\documentclass[12pt]{article}

\usepackage{graphicx}
\usepackage{amssymb}

\usepackage{amsmath}
\usepackage{amsfonts}
\usepackage{enumerate}
\usepackage{amsthm}
\usepackage{lineno}

\usepackage{csquotes}

\usepackage{lscape}

\newcommand{\bB}{\mbox{\boldmath $B$}}

\newcommand{\bX}{\mbox{\boldmath $X$}}

\newcommand{\bE}{\mbox{\boldmath $E$}}

\newcommand{\bQ}{\mbox{\boldmath $Q$}}
\newcommand{\bzero}{\boldmath 0}

\newcommand{\bbeta}{\mbox{\boldmath $\beta$}}

\newcommand{\bPsi}{\mbox{\boldmath $\Psi$}}

\usepackage{color}

\usepackage[default]{jasa_harvard}    
\usepackage{JASA_manu}

\newtheorem{thm}{Theorem}[section]

\begin{document}

\title{{\Large \bf  Regression Analyses of Distributions using Quantile Functional Regression} 
\author{Hojin Yang$^{\dagger}$, Veerabhadran Baladandayuthapani$^{\dagger}$, Arvind U.K. Rao$^{\ddagger}$,  \\ and Jeffrey S. Morris$^{\dagger}$ \\
Department of Biostatistics$^{\dagger}$\\ 
Department of Bioinformatics and Computational Biology$^{\ddagger}$\\ 
 The University of Texas MD Anderson Cancer Center, Houston, TX 77030  \\ 
 \texttt{hyang6@mdanderson.org},  
 \texttt{Veera@mdanderson.org},  \texttt{ARUppore@mdanderson.org},  \\ and 
 \texttt{jefmorris@mdanderson.org} 
 }
 } 
\date{} 

\maketitle

\mbox{}
\vspace*{-0.4 in}


\thispagestyle{empty}
\begin{center}
\textbf{Abstract}
\end{center}

\noindent Radiomics involves the study of tumor images to identify  quantitative markers explaining cancer heterogeneity.  The predominant approach is to extract hundreds to thousands of image features, including histogram features comprised of summaries of the marginal distribution of pixel intensities, which leads to multiple testing problems and can miss out on insights not contained in the selected features.  In this paper, we present methods to model the entire marginal distribution of pixel intensities via the quantile function as functional data, regressed on a set of demographic, clinical, and genetic predictors to investigate their effects of imaging-based cancer heterogeneity.  We call this approach \textit{quantile functional regression},  regressing subject-specific marginal distributions across repeated measurements on a set of covariates, allowing us to assess which covariates are associated with the distribution in a global sense, as well as to identify distributional features 
characterizing these differences, including mean, variance, skewness, heavy-tailedness, and various upper and lower quantiles.
To account for smoothness in the quantile functions and gain statistical power, we introduce custom basis functions we call \textit{quantlets} that are sparse, regularized, near-lossless, 
and empirically defined, adapting to the features of a given data set and containing a Gaussian subspace so {non-Gaussianness} can be assessed.   
We fit this model using a Bayesian framework that uses nonlinear shrinkage of quantlet coefficients to regularize the functional regression coefficients and provides fully Bayesian inference after fitting a Markov chain Monte Carlo.  
We demonstrate the benefit of the basis space modeling through simulation studies, and apply the method to  Magnetic resonance imaging (MRI) based radiomic dataset from Glioblastoma Multiforme to relate imaging-based quantile functions to various demographic, clinical, and genetic predictors, finding specific differences in tumor pixel intensity distribution between males and females and between tumors with and without DDIT3 mutations. 

\vspace*{.3in}

\thispagestyle{empty}

\clearpage
\pagenumbering{arabic}


\section{ {\bf Introduction} }

Glioblastoma multiforme (GBM), also known as glioblastoma and grade IV astrocytoma,
is the most common and most aggressive cancer that begins within the brain. 
Studying GBM is difficult in that  the cause of most cases is unclear, there is no known way to prevent the disease, and most people diagnosed with GBM survive only  12 to 15 months, with less than $3\%$ to $5\%$ surviving longer than five years \cite{tutt2011glioblastoma}.
Most GBM diagnoses are made by medical imaging such as  computed tomography (CT), magnetic resonance imaging (MRI), and positron emission tomography (PET). MRI is frequently chosen because it offers a wide range of high-resolution image contrast that can   serve as indicators for clinical decision making or for tumor progression in GBM studies. 
A GBM tumor, which usually originates from a single cell,  demonstrates heterogeneous physiological and morphological features  as it proliferates \cite{marusyk2012intra}.  Those heterogeneous features make it difficult to predict treatment impacts and outcomes for patients with GBM.  
 Investigating tumor heterogeneity is critical in cancer research since inter/intra-tumor differences have stymied the systematic development of targeted therapies for cancer patients \cite{felipe2013cancer}.
 
 \begin{figure}[!htb]
 \centering
 \caption{\small Characterizing tumor heterogeneity from distributional summaries of MRI pixel intensities: 
the two graphs include kernel density estimates and the raw empirical quantile functions as representations of tumor heterogeneity (pixel intensities within the tumor); black line: female patient without DDIT3 mutation; red line: male patient without DDIT3 mutation; blue line: female patient with DDIT3 mutation; and green line: male patient with DDIT3 mutation. The images in other columns represent the T1-post contrast MRIs of the brains, with tumor boundaries indicated by black lines. 
  \label{S5_Figure_Intro}}
\includegraphics[height=4.2in,width=5.6in]{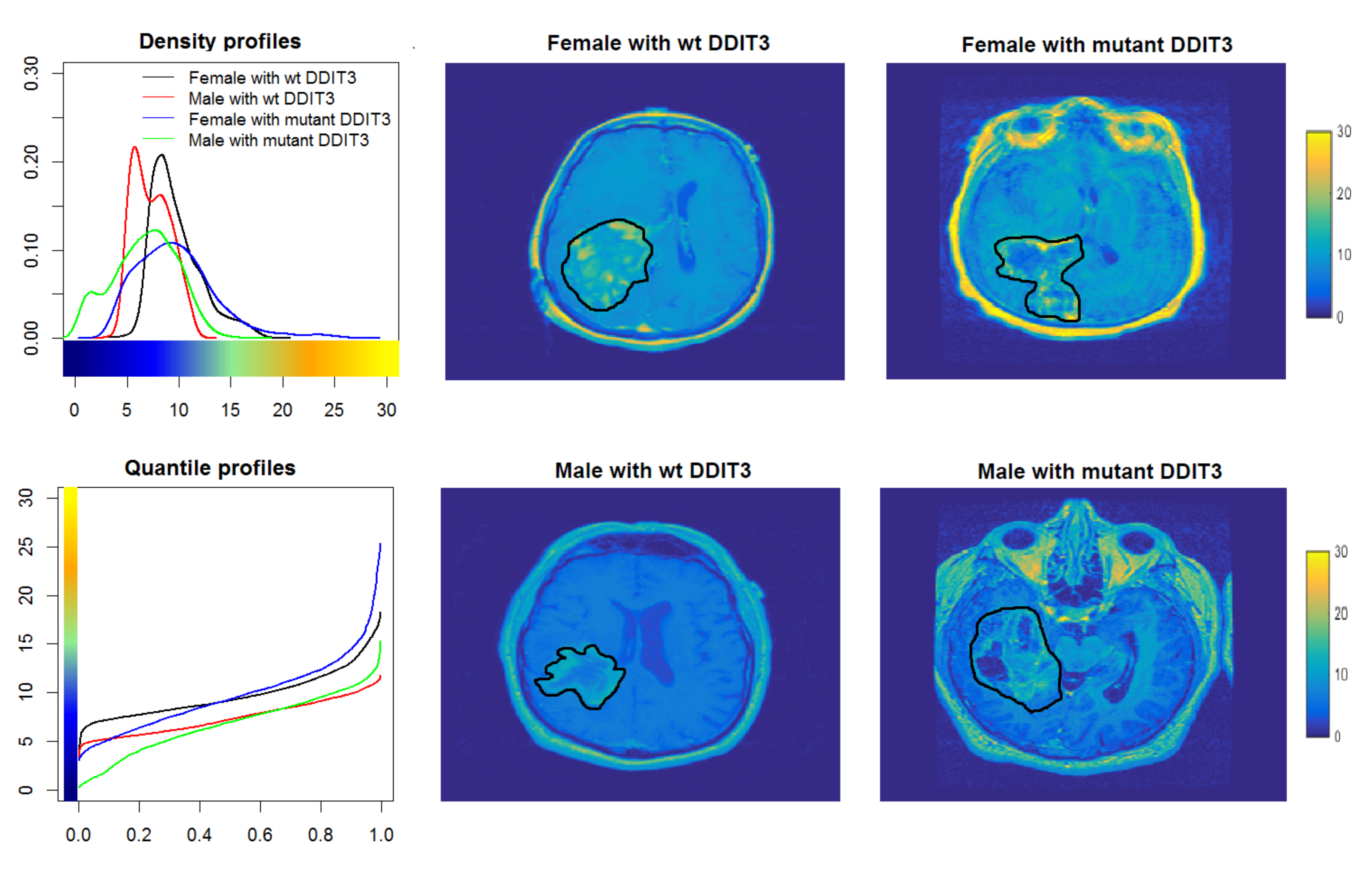}
\end{figure}

Our motivating dataset comes from The Cancer Imaging Archive (TCIA, cancerimagingarchive.net) -- a comprehensive archive of biomedical images of various cancer types along with associated clinical and genomic data (described in detail in Section 4). As an illustration, the rightmost four plots of Figure \ref{S5_Figure_Intro} display MRI images for 4 patients with GBM, two males and two females, and with and without mutations in the DDIT3 gene, an important gene associated with GBM development, with tumor boundaries indicated by the black lines.  The upper left plot contains smoothed density estimates of the pixel intensities while the bottom left plot contains the empirical quantile functions for these tumors.  
Features of these images may comprise clinically useful biomarkers since these pixel intensities denote the amount of contrast enhancement (or vascularization) on T1-weighted sequence; or extent of infiltration into neighboring tissue (in T2-weighted or fluid-attenuated inversion recovery (FLAIR) MR sequence).   It is of scientific interest to study the pixel intensity distributions for a set of 64 GBM tumors of which these four are a subset and investigate their associations with various covariates including age, sex, tumor subtype, DDIT3 mutation status, EGFR mutation status, and survival status ($>12$ months, $\le 12$ months), to assess how various aspects of GBM tumor heterogeneity are reflected in the tumor images.

\textit{Radiomics} is a field of study to identify quantitative biomarkers from biomedical imaging data.  The typical approach is to extract various features of the images and then relate them to various clinical and genetic outcomes.  While some of these features characterize various spatial relationships among the pixel intensities, an important subset called \textit{histogram features}  \cite{just2014improving} extracts information from the marginal distribution of pixel intensities within the tumor, such as the mean and variance.
While the feature extraction strategy that is typical in radiomics is reasonable and often can yield meaningful results, it has numerous drawbacks.   The exploratory regression analysis of numerous different summaries raises multiple testing problems, and if the key distributional differences are not contained in the pre-defined summaries, then this approach can miss out on important insights.

\begin{figure}[!htb]
\centering
\caption{\small Differences in density distributions: 
panel A reveals four densities
black Normal($\mu=1, \sigma=5$), blue Normal($\mu=1, \sigma=10$),
green Normal($\mu=10, \sigma=10$), and red  Skewed Normal ($\mu=10, \sigma=10, \xi=-0.8$)  and 
their corresponding cumulative density functions and quantile functions are shown in panels B and C, respectively. 
  \label{S0_Figure}}
\includegraphics[height=2.2in,width=5.6in]{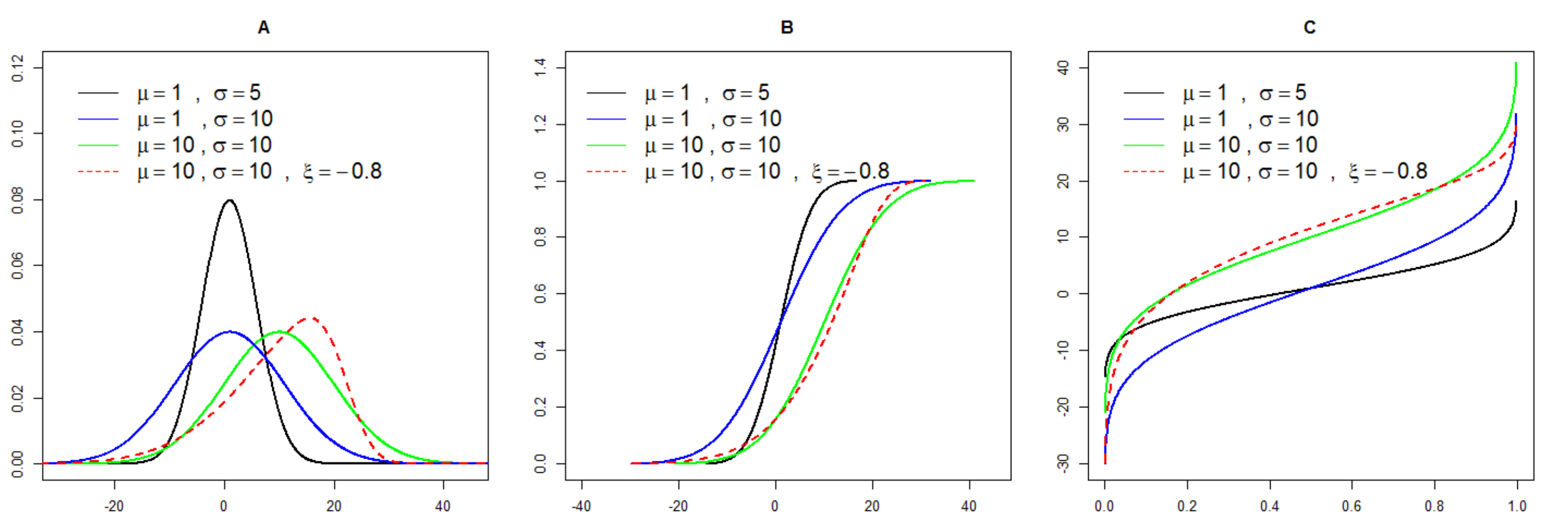}
\vspace{0.5cm} 
\end{figure}

To illustrate this point, consider the densities plotted in panel (a) of Figure \ref{S0_Figure}. 
We see that just extracting the mean from the entire density function cannot distinguish two distributions for which  
the means are identical but one is more variable than the other (black solid line and blue solid  line).
Similarly,  note the red dashed line and green solid lines mark densities with identical means and variances, so only considering these 
summaries would miss out on their difference in skewness.  Also, solely looking at the center of a distribution via the mean or median can miss out
on differences in the tails, which in some settings can be the most scientifically relevant parts of the distributions.
It would be preferable to model the entire distribution, thus retaining the information in the data and potentially finding any differences.

There are various choices for representing the subject-specific distributions, including the distribution function, the quantile function, or the density function (if it exists).  The three panels of Figure \ref{S0_Figure} show all three for the example distributions.  In this paper, we choose to represent and model the distribution through the quantile function, which has numerous advantages as described in Section 2.6, including a fixed, common domain $[0,1]$, their ease of estimation by order statistics without any need for smoothing parameter specification, and the ability to readily compute distributional moments.  
  Thus, our approach is to represent each subject's data via their empirical quantile function $Q_i(p), p\in {\cal P}= [0,1]$, computed from the order statistics, and then treat these as functional responses regressed on a set of scalar covariates $x_{ia}; j=1,\ldots,A$ through $Q_i(p) = B_0(p) + \sum_{a=1}^J x_{ia} \beta_a(p) + E_i(p)$.  This models the \textit{distribution of subject-level distributions} as a function of subject-level covariates.  We call the fitting of this model \textit{quantile functional regression}, which is \underline{fundamentally different and distinguished
  from other models } for quantile regression in existing literature in Section 2.1.  Regression analysis using the quantile function as the response is based upon the Wasserstein metric between distributions \cite{dobrushin1970definition}, which can be shown to be equivalent to an L2 distance between the corresponding quantile functions.


One simple approach to fitting this model would be to interpolate each subject's data onto a common grid of ${\cal P}$ and then perform independent regressions for each interior point $p$.  This would lead to estimators that are unbiased but inefficient, as they would not borrow strength across nearby $p$, which should be similar to each other.  We refer to this strategy as \textit{naive quantile functional regression}.   As is typically done in other functional regression settings (see review article by \cite{morris2015functionalreg}), alternatively one could borrow strength across $p$ using basis representations, with common choices including splines, principal components, and wavelets.  In this paper, we will introduce a new strategy for construction of a custom basis set 
we call 
 \textit{quantlets} that is sparse, regularized, near-lossless, 
and empirically defined, adapting to the features of the given data set and containing the Gaussian distribution as a prespecified subspace so non-Gaussianness can be assessed.  Representing the quantile functions with a quantlet basis expansion, we propose a Bayesian modeling approach for fitting the quantile functional regression model that utilizes shrinkage priors on the quantlet coefficients to induce regularization of the regression coefficients $\beta_a(p)$, and leading to a series of global and local inferential procedures that can first determine whether $\beta_a(p)\equiv 0$ and then assess which $p$ and/or distributional summaries (e.g. mean/variance/skewness/Gaussianness) characterize any such difference.  While based on quantile functions, our model will also be able to provide predicted distribution functions and densities for any set of covariates to use as summaries for users more accustomed to interpreting densities than quantile functions.

While developed in the context of our GBM motivating case study, the methods we develop are general and can be applied to a wide range of contexts in which multiple observations are obtained per subject and one wishes to associate subject-specific distributions to explanatory variables. 
This paper is organized as follows. In Section 2, we introduce the general quantile function regression model, introduce \textit{quantlets}, describe how to construct a set of quantlet basis functions for a given data set, and describe our Bayesian approach to fitting the model.  In Section 3, we describe simulation studies conducted to evaluate the finite-sample performance of our method and demonstrate the benefit of incorporating quantlet bases in the modeling. In Section 4, we apply our method to data in our GBM case study and perform various investigations to obtain insightful scientific results. We provide concluding remarks in Section 5.


\section{{\bf Models and Methods}}

\subsection{{\bf Quantile Functions and Empirical Quantile Functions}}  \label{sec:QF}

Let $Y$ be a real valued random variable which in our context, represents the pixel intensity from a tumor image in our GBM application, 
and $F_{Y}(y)$ be its 
cumulative distribution function (right-continuous) such that $F_{Y}(y)=\text{P}(Y\le y)$,
and $p=F_{Y}(y)$ be the percentage of the population less than or equal to $y$.
The quantile function of $Y$, defined for $p \in [0,1]$,  is defined as
 \begin{align*}
Q(p)=Q_{Y}(p)=F_{Y}^{-1}(p)=\inf{(y: F_{Y}(y) \ge p     )}.
 \end{align*}
Distributional moments are easily computable as simple functions of quantile function, for example with mean, variance, skewness, and kurtosis given by 
 \begin{align}  \label{Quant_formula} 
&\mu_{Y}  =  \text{E}(Y)= 
\int_{0}^{1} Q_{Y}(p)  dp     \notag \\ 
&{\sigma}_{Y}^2 =  \text{Var}(Y)=  
 \int_{0}^{1}(Q_{Y}(p) -\text{E}(Y) )^2 dp,   \notag \\ 
 &{\xi}_{Y}=  \text{Skew}(Y)= 
\int_{0}^{1}(Q_{Y}(p) -\text{E}(Y))^3/\text{Var}(Y)^{3/2} dp, ~~\text{and}     \notag  \\ 
&{\varphi}_{Y}=    \text{Kurt}(Y)= 
\int_{0}^{1}(Q_{Y}(p) -\text{E}(Y))^4/\text{Var}(Y)^2 dp.   
\end{align}

Given a sample of $m$ repeated observations for a given subject, intensities for multiple pixels for the subject's tumor in our GBM application, 
let $Y_{(1)}\le \cdots\le Y_{(m)}$ be the corresponding order statistics.  For $p \in [1/(m+1),m/(m+1)]$, a subject-specific empirical quantile function of $Y$ can be computed, e.g. using linear interpolation across order statistics,    \begin{align*}
\widehat{Q}(p) =  (1-w)Y_{([(m+1)p])} + wY_{([(m+1)p]+1)},
 \end{align*}
 where 
 $[x]$ is an integer less than or equal to $x$
 and  $w$ is a weight such that $(m+1)p=[(m+1)p]+w$.
This empirical quantile function is an estimate of the true quantile function.

 As shown in \cite{parzen2004quantile}, for a fixed $p$, the empirical estimator  
is consistent and is asymptotically equivalent to a Brownian bridge 
when the density function $f_{Y}(y)$ exists and is positive.  
This can serve as a summary of the subject-specific distribution that does not require specification of any smoothing parameter, that in this paper we regress on outcomes to assess how they vary across covariates.
In this paper, we are interested in studying outcomes $Y$ that are absolutely continuous, meaning that the corresponding quantile functions are continuous and smooth, without jumps that would occur for discretely valued random variables.
 For brevity, we omit the estimator notation for the empirical quantile functions and just refer to them as 
 $Q(p)$.

\begin{table}
\caption{Types of regression based on response type and objective function. 
      \label{objective}}
\resizebox{1.0\columnwidth}{!}{    
\begin{tabular}{   c| c c } \hline
& $\text{{\bf  Objective function}}$   &  $\text{{\bf  Objective function}}$ 
  \\ 
    $\text{{\bf Response}}$ $(\cdot)$
& $E( (\cdot) |X)$ &  $F_{(\cdot)}^{-1}(p|X)$         \\    \hline \hline
$\text{{\bf scalar}}$ $Y$	&$\text{{\bf classic regression}}$ 	
&$\text{{\bf quantile regression}}$	  \\
$\text{{\bf function}}$ $Y(t)$	&$\text{{\bf functional  regression}}$ 	
&$\text{{\bf functional quantile regression}}$	   \\
$\text{{\bf quantile function}}$ $Q(p)$
&$\text{{\bf quantile functional regression}}^{\ast}$	 	
&$\text{{\bf quantile functional quantile regression}}$ \\
	\hline	
		\hline	
\end{tabular}
}
\end{table}      

\subsection{ {\bf Quantile Functional Regression Model}} \label{liter}

Suppose that for a series of subjects $i=1,\ldots,n$ we observe a sample of $m_i$ observations from which we construct a subject-specific empirical quantile function $Q_i(p_j)$ for $p_j=j/(m_i+1); j=1,\ldots, m_i$, along with a set of $A$ covariates  $\bX_i =(x_{i1},\dots, x_{iA})^{T}$, which are the demographic, clinical, and genetic factors described in the introduction for our GBM application.
Note that by construction all subject-specific empirical quantile functions $Q_i(p_j)$ are non-decreasing in $p$.  See Section 4 of the supplement for further discussion of monotonicity issues in this framework.

The quantile functional regression model is given by
 \begin{equation}
Q_i(p)= \sum_{a=1}^{A}x_{ia}\beta_{a}(p) +  E_i(p)= \bX_i^{T}\bB(p) +  E_i(p),   
 \label{p5_qfm_i}  
\end{equation}
where  $\bB(p)=( \beta_1(p), \dots, \beta_A(p) )^{T}$ is a column vector of length $A$ 
containing unknown fixed functional coefficients for the quantile $p$ and 
   $E_i(p)$ is a residual error process, assumed to follow a mean zero Gaussian process with   
  the covariance surface, $\Sigma(p_1, p_2)=\text{cov}\{ E_i(p_1), E_i(p_2) \}$
  and
  to be independent of $\bX_i$. 
 The structure of
  $\Sigma(\cdot,\cdot)$ captures the variability across subject-specific quantiles, and the diagonals capture
  the intrasubject covariance across $p$.   
In practice, we will focus our modeling on $p \in \mathcal{P} = [\delta, 1- \delta]$, with  $\delta=\max_{i\le n}\{1/(m_i+1)\}$ 
being the most extreme quantile estimable from the subject with the fewest observed data points.  In this paper,
we are primarily interested in settings with at least moderately large numbers of observations per subject, i.e. $m_i$ not too small, 
and in later studies will extend our work to sparse data settings with few observations per subject.

To place our model in the proper context within the current literature on quantile and functional regression, Table \ref{objective} lists various types of regression in terms of response and objective function. 
 In contrast to classical regression, which specifies the mean of the response conditional on a set of covariates, \textit{quantile regression} \cite{he2000quantile,koenker2005quantile,yang2015quantile} 
 works by estimating a pre-specified $p$-quantile of the response distribution conditional on the covariates, either with independent  \cite{koenker2004quantile,hao2007quantile,davino2013quantile} 
  or spatially/temporally correlated errors \cite{koenker2004quantile,reich2012bayesian,reich2012spatiotemporal}.  Most existing methods fit independent quantile regressions for each desired $p$, which can lead
  to crossing quantile planes, although recent methods (e.g. \citeasnoun{yang2017joint}) jointly model all quantiles, borrowing strength across $p$ using Gaussian process priors.  Parallel to these efforts are methods to perform \textit{Bayesian density regression} \cite{dunson2007bayesian}, in which the density of the response variable is modeled as a function of covariates via dependent Dirichlet processes  \cite{muller1996bayesian,maceachern1999dependent,griffin2006order,dunson2006bayesian}.
   These quantile regression models are inherently different from the setting of this paper, as they are modeling the quantile of the \textit{population} given covariates, while our framework is modeling the quantile function of each \textit{subject} as a function of subject-specific covariates.  Another difference is that, in general, these methods do not model intrasubject correlation in settings for which there is more than one observation per subject.

Other regression methods have been designed for functional responses.  There is a subset of the functional regression literature (see \citeasnoun{morris2015functionalreg} for an overview)
 that involve regression of a functional response on a set of covariates, with classical functional regression focusing on the mean function conditional on covariates 
 \cite{faraway1997regression,wu2000kernel,guo2002functional,ramsay2006functional,morris2006wavelet,reiss2010fast,goldsmith2011functional,goldsmith2011penalized,scheipl2015functional,meyer2015bayesian}, 
 and \textit{functional quantile regression} that computes
 the quantile of functional response conditional on covariates, using
  the check function as the objective function
   \cite{brockhaus2015functional,brockhaus2015fdboost}
   or the asymmetric Laplace likelihood as a Bayesian analog  \cite{morris2018tech}.
   Again these methods are not modeling subject-specific, but rather population-level quantiles.
Other recent works on functional quantile regression have focused on the quantile of the scalar response distribution regressed on a set of functional covariates   
 \cite{ferraty2005conditional,cardot2005quantile,chen2012conditional,kato2012estimation,kato2012asymptotics,li2016inference}, which is also an inherently different problem from the one addressed here.

All of these methods differ, fundamentally, from the quantile functional regression framework described in this paper.  For these methods, the quantile regression is computing the $p^{th}$ quantile of the population given covariates X, while in our case, we are interested in modeling the $p^{th}$ quantile of an individual subject's distribution given X.  In our case, we are modeling the empirical quantile function for each subject as the response, and using a classical (mean) regression of these subject-specific quantile functions onto a set of scalar covariates, i.e. estimating the expected quantile function for a subject given a set of covariates.   Note that this regression problem is based upon the Wasserstein metric between distributions \cite{dobrushin1970definition}, which can be shown to be equivalent to an L2 distance between the corresponding quantile functions.   It would also be possible to compute the $q^{th}$ quantile of the distribution
of specific empirical quantile functions for each $p$ conditional on covariates, which could be dubbed \textit{quantile functional quantile regression}, but this model is not addressed in the current paper.

 
\subsection{{\bf Quantlet Basis Functions}} \label{sec:quantlets}

If all empirical quantile functions are sampled on (or interpolated onto) the same grid (i.e. $m_i \equiv m \forall i=1,\ldots,n$), then a simple way to fit model (\ref{p5_qfm_i}) would be to fit separate linear regressions for each $p$.  However, this naive approach would treat observations across $p$ as independent.  This leads to a regression model that fails to borrow strength across $p$, and thus is expected to be inefficient for estimation of the functional coefficients $\beta_a(p)$, and ignores correlation across $p$ in the residual error functions $E_i(p)$, which would adversely affect any subsequent inference.  We call this approach \textit{naive quantile functional regression} in our comparisons below.

Basis function representations can be used to induce smoothness across $p$ in $\beta_a(p)$ and capture intra-subject correlation in the residual error functions $E_i(p)$.  In existing functional regression literature, common choices for basis functions include splines, Fourier, wavelets, and principal components, and smoothness is induced across $p$ by regularization of the basis coefficients via L1 or L2 penalization \cite{morris2015functionalreg}.  Here, we introduce a strategy to construct a custom basis set called \textit{quantlets} for use in the quantile functional regression model that have many desirable properties, including regularity, sparsity, near-losslessness, interpretability, and empirical determination allowing them to capture the salient features of the empirical quantile functions for a given data set.

\begin{figure}[!htb]
\centering
\caption{Graphical illustration of the entire procedure for constructing the quantlets.
  \label{diagram}}
\includegraphics[height=2.5in,width=4.5in]{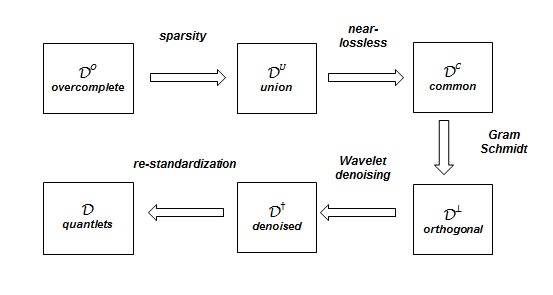}
\vspace{0.5cm} 
\end{figure}

We empirically construct the quantlets for a given data set as a common near-lossless basis that can nearly perfectly represent each subject's empirical quantile function, and then we use these basis functions as building blocks in our quantile functional regression model as described later.  Given a sample of subject-specific empirical quantile functions, we construct a quantlet basis set by the following steps:
\begin{enumerate}
\item Construct an overcomplete dictionary that contains bases spanning the space of Gaussian quantile functions plus a large number of Beta cumulative density functions.  For each subject, use regularization to choose a sparse set among these dictionary elements.  

\item Take the union of all selected dictionary elements across subjects, and find a subset that simultaneously preserves the information in each empirical quantile function to a specified level, as measured by the cross-validated concordance correlation coefficient.

\item Orthogonalize this subset using Gram-Schmidt, apply wavelet denoising to regularize the orthogonal basis functions, and then re-standardize.
\end{enumerate}
We refer to the set of basis functions resulting from this procedure as \textit{quantlets}.   
We describe these steps in detail and then discuss their properties. 
See Figure~\ref{diagram} for an overview of the entire procedure, for which each step is given as follows.

\textbf{Form overcomplete dictionary:}    Suppose that $L^2(\Pi(\mathcal{P}))$ is a Banach space such that $\{ Q: p \in \mathcal{P} \rightarrow \mathbb {R}~~ \text{measurable s.t.}~~ 
\|Q\|_2=\bigr ( \int |Q(p)|^2 d\Pi(p)  \bigl )^{1/2} < \infty \}$, where
$\Pi$ is a uniform density with respect to the Lebesgue measure. We define the first two basis functions to be a constant basis $\xi_1(p)=1$ for $p \in [0,1]$ and standard normal quantile function $\xi_2(p) = \Phi^{-1}(p)$.  These orthonormal bases span the space of all Gaussian quantile functions, with the first coefficient corresponding to the mean and the second coefficient the variance of the distribution.  
We form an overcomplete dictionary that includes these along with a large number of dictionary elements constructed from Beta cumulative density functions (CDF).  The shape of the Beta CDF is able to follow a 
``steep-flat-steep" shapes that we have observed characterize the features of empirical quantile functions in a wide array of applications, so has the potential for efficient representation.

The individual dictionary elements $\xi_k(p)$ are given by
 \begin{equation} \label{p5_phi}
\xi_k(p) =    P_{N^{\perp}} \biggr (    \frac{ F_{\theta_k}(p) -   \mu_{\theta_k} } { \sigma_{\theta_k} }  \biggl ) 
=  P_{N^{\perp}}  \biggr (  \int_{0}^{1} ( I(u \le p)  -\mu_{\theta_k})/\sigma_{\theta_k}dF_{\theta_k}(u)  \biggl ) , 
\end{equation}
where $F_{\theta_k}(p)$ is the CDF of a Beta($\theta_k$) distribution for some positive parameters $\theta_k=\{a_k, b_k\}$, $\mu_{\theta_k}= \int_{0}^1 F_{\theta_k}(u)  du$ and 
$\sigma_{\theta_k}^2= \int_{0}^1 (F_{\theta_k}(u)-\mu_{\theta_k})^2du$ are the centered   
and scaled values of these distributions for standardization, respectively, 
and $P_{N^{\perp}}$
indicates the projection operator onto the orthogonal complement to the Gaussian basis elements $\xi_1(p)$ and $\xi_2(p)$, with $P_{N^{\perp}}\{f(p)\}= f(p) - \xi_1(p) \int_0^1 f(p) \xi_1(p) dp - \xi_2(p) \int_0^1 f(p) \xi_2(p) dp$ .  Put together, the set
$\mathcal{D}^{O}=\{ \xi_1, \xi_2 \} \cup \{  \xi_k:    \theta_k  \in \Theta \}$  comprises an {\it overcomplete dictionary}  family
  on $\Theta \subset \mathbb {R}_{+}^{2}$. 
  In practice, to fix the number of dictionary elements, we choose a grid on the parameter space to obtain $\Theta=\{ \theta_k= (a_k,b_k)  \}_{k=3}^{K^{O}}$ by uniformly sampling on 
 $\Theta \subset (0, J )^{2}$  for some sufficiently large $J$, 
and choosing $K^{O}$ to be a large integer (e.g. we use $K^{O}=12,000$ in this paper).
Details of how to select $\Theta$ can be found in the Supplementary materials.   If desired, this dictionary can be arbitrarily expanded to include any other basis functions on $[0,1]$ that one might think could capture salient features of the given data set.

The use of a large dictionary of Beta CDF in this step is supported by the following theorem, that demonstrates that any quantile function whose first derivative is 
absolutely continuous can be represented by a conical combination of Beta CDFs.
\begin{thm} \label{betabasis}
Let $Q(p)$ be a quantile defined on $p\in [0,1]$,  
 $F_{k,n}(p)$ be a beta cumulative distribution function defined as
$F_{k,n}(p)= \int_{0}^{p}  \frac{ \Gamma(n+2)}{\Gamma(k+1)\Gamma(n-k+1)} x^{k}(1-x)^{n-k} dx$, and
 $q(p)$ be the first derivative of $Q(p)$.
Define
$$Q_n(p)= \sum_{k=0}^{n} c_{k,n}  \int_{0}^{p}  \frac{ \Gamma(n+2)}{\Gamma(k+1)\Gamma(n-k+1)} x^{k}(1-x)^{n-k} dx
= \sum_{k=0}^{n}c_{k,n}  F_{k,n}(p),$$
where $c_{k,n}=\alpha_{k}/(n+1) $, $\alpha_{k}=q(k/n)$ and $0 \le x \le 1$.
Suppose that $q(x):[0,1]\rightarrow \mathbb{R}$ be continuous function 
for the sufficiently small $\delta>0$, that
there exists a constant $C$ such that $\|q\|_{\infty}=\sup_{ x\in [0,1]}|q(x)| \le C$, and
that $c_{k,n} \rightarrow c_k$  for each $k$, where $c_k$ is some constant.
Then for any $p\in [0,1]$ 
$$ \lim_{n\rightarrow \infty}  Q_n(p) = Q(p).$$
\end{thm}
{\bf Proof of Theorem  } {   
We consider the relation between beta and Binomial distributions
\begin{align} \label{rbp}
\sum_{k=0}^{n}c_{k,n}  \frac{ \Gamma(n+2)}{\Gamma(k+1)\Gamma(n-k+1)} x^{k}(1-x)^{n-k}= \sum_{k=0}^{n} \alpha_{k}\binom{n}{k}x^{k}(1-x)^{n-k}
\end{align}
Notice that the formula in right hand side in (\ref{rbp})
is the random Bernstein polynomial of order $n$.
Since $q(x)$ is continuous on the closed and bounded interval $[0,1]$, 
it is uniformly continuous and thus for any positive $\epsilon$,
there exists a $\delta(\epsilon)$ such that
$|x-y|<\delta(\epsilon)$ implies $|q(x)-q(y)|<\epsilon$.
Fix $x\in [0,1]$.
Let $X_1, \dots, X_n$ be a sample from Bernoulli$(x)$ distribution. 
Let $\hat x_n$ be the sample average, $\hat x_n=\sum_{i=1}^{n}X_i/n$. Then, 
it is easy to see $Eq(\hat x_n)=q_n(x)$. Hence, 
\begin{align*}
|q(x) - q_n(x)| &\le E | q(\hat  x_n) - q(x)| +  | Eq(\hat  x_n) - q_n(x)|  \\
&= E \biggl \{ | q(\hat  x_n) - q(x)|I(|\hat x_n-x|<\delta(\epsilon))   \biggr \} 
+  E \biggl \{ | q(\hat  x_n) - q(x)|I(|\hat x_n-x|\ge\delta(\epsilon))   \biggr \}   \\
&\le  \epsilon + 2\|q\|_{\infty}P(|\hat x_n-x|\ge\delta(\epsilon)),
\end{align*}
where  $\|q\|_{\infty}=\sup_{ x\in [0,1]}|q(x)|$.
It follows from Chebychev's inequality that
$$P(|\hat x_n-x|\ge\delta(\epsilon)) \le \frac{x(1-x)}{n\delta(\epsilon)^2} \le \frac{1}{4n\delta(\epsilon)^2} ~~\text{for all}~~ x\in [0,1].$$
Hence, we have $\sup_{ x\in [0,1]}|q(x) - q_n(x)|\le \epsilon + 2\|q\|_{\infty} \frac{1}{4n\delta(\epsilon)^2}$. Letting $n \rightarrow \infty$ and then  $\epsilon \rightarrow 0$
 yields
  \begin{align} \label{ucq}
 \lim_{n\rightarrow \infty} \sup_{ x\in [0,1]}|q(x) - q_n(x)| =0,
  \end{align}
 where this implies that $q_n(x)$ converges $q(x)$ uniformly (over $[0,1]$).
It follows from the integrability of $q(x)$  by continuity  and well known  triangle inequality that 
 \begin{align*}
 \biggl |\int_0^{p} q(x) - q_n(x)dx  \biggr | \le 
 \int_0^{p}  |q(x) - q_n(x) |dx .
 \end{align*}
From the result (\ref{ucq}),
we already know that given $\epsilon$, there exists $N$ such that $|q(x) - q_n(x)|<\epsilon $ for $n>N$ (not depend on $x$).
Therefore, when $n>N$,
 \begin{align*}
 \biggl |\int_0^{p} q(x) - q_n(x)dx  \biggr | \le 
 \int_0^{p}  |q(x) - q_n(x) |dx  \le p\epsilon.
 \end{align*}
Since $\max_{p\in[0,1]}p=1$, this implies that for any $p \in [0,1]$,
$$ \lim_{n \rightarrow \infty}Q_n(p)= \lim_{n \rightarrow \infty}\int_{0}^{p}q_n(x) = \int_{0}^{p}q(x) dx = Q(p),$$
which is what we want to  prove.  \hfill $\Box$
}

This theorem provides justification for using a dictionary containing many Beta CDF to represent the empirical quantile functions, and supports the notion that given a large enough dictionary,
the linear combination of beta CDFs should be sufficient for representing each individual's empirical quantile function. 


\textbf{Sparse selection of dictionary elements:} For each $i$, we use regularization via penalized likelihood to obtain a sparse set of dictionary  elements to represent each subject's empirical quantile function.  While other choices of penalty could be used, here we use the Lasso \cite{tibshirani1996regression}, minimizing
 \begin{equation}   \label{p5_lasso}
 \| Q_{i}(p) - \sum_{k \in \mathcal{D}^{O}   } \xi_k(p)Q_{ik}^{O  }\|^2_{2}+
 \lambda_i \sum_{k \in \mathcal{D}^{O}    }\|Q_{ik}^{O}\|_{1}, 
 \end{equation}
for a fixed positive constant $\lambda_i$, where the choice of each $\lambda_i$ 
is determined by cross validation and  $Q_{ik}^{O}$ are basis coefficients for the elements of $\mathcal{D}^{O} $.  
The standardization of the basis functions  ensures they are on a common scale which is important for the regularization method.  
By using the regularization methods, we obtain different sets of  selected {\it dictionary} elements for each 
subject,  denoted by 
$\mathcal{D}_i=\{ \xi_k \in  \mathcal{D}^{O}  :  Q_{ik}^{O} \ne 0 \}$.  Taking the union across subjects, we obtain a unified set of  {\it dictionary} elements denoted by 
 $\mathcal{D}^U= \cup_{i=1}^{n} \mathcal{D}_i$, which we construct to always include the Gaussian basis functions $\xi_1$ and $\xi_2$.

\textbf{Finding near-lossless common basis:}  
The above sparse selection is done for each subject $i$, however, we would like to use a common 
basis across all subjects to fit the quantile functional regression model.  The unified set of dictionary elements $\mathcal{D}^U$ is likely to be very redundant, with some of the dictionary elements selected for many subjects' empirical quantile functions and many others selected for only a few subjects, and not all necessary.  We would like to  find a common basis set $\mathcal{D}^\mathcal{C}$ that is as sparse as possible while retaining virtually all of the information in the original empirical quantile functions.  
We call such a basis \textit{near-lossless}, which we define more precisely below.

As a measure of losslessness, we use the leave-one-out concordance correlation coefficient (LOOCCC), $\rho_{(i)}$.   This quantifies the ability of a basis set $\mathcal{D}^U_{(i)}$ that has been empirically constructed using all samples except the $i$th one to represent the observed quantile function 
 \begin{equation}  \label{p5_rhoi}  
\rho_{(i)}=\frac{ \text{Cov}( Q_i(\cdot), \sum_{k \in   \mathcal{D}^U_{(i)} }  \xi_k(\cdot)  Q_{ik}^{U} ) }
{ \text{Var}( Q_i(\cdot) )  + \text{Var}( \sum_{k \in   \mathcal{D}^U_{(i)} }  \xi_k(\cdot)  Q_{ik}^{U})
+  [ \text{E}(Q_i(\cdot))-  \text{E}(\sum_{k \in   \mathcal{D}^U_{(i)} }  \xi_k(\cdot)  Q_{ik}^{U})]^2    }, 
\end{equation} 
where 
$\text{Cov}$, $\text{Var}$ and $\text{E}$ are taken with respect to ${\Pi}$ and
$Q_{ik}^{U}$ are basis coefficients corresponding to the elements $\xi_k$ contained in the set $\mathcal{D}^U_{(i)}$. 

This measure $\rho_{(i)} \in [0,1]$, with $\rho_{(i)}=1$ indicating the basis set $\mathcal{D}^U_{(i)}$ is sufficiently rich such that there is no loss of information about $Q_i(p)$ in its corresponding projection.  One advantage of this measure over other choices such as mean squared error is that it is scale-free, in the sense that it is invariant to the scale of the quantile functions $Q_i$ and the basis functions $\xi_k$.  Aggregating across subjects, we can compute $\rho^0=\text{min}_i\{\rho_{(i)}\}$ or $\overline{\rho}=\text{mean}_i\{\rho_{(i)}\}$ to summarize the ability of the chosen basis to reconstruct the observed data set in its entirety, with $\overline{\rho}$ the average across all subjects and $\rho^0$ the worst case.  If $\rho^0=1$, we say this basis is \textit{lossless}, and if $\rho^0>1-\epsilon$ for some small $\epsilon$ then we say this basis is \textit{near-lossless}.

\begin{figure}[!htb]
\centering
\caption{Construction of Quantlet Bases.  The concordance correlation for the GBM application: (A) minimum concordance
($\rho^{0}$, red) and average ($\bar\rho$, blue) across samples as function of $K^\mathcal{C}$,
(B) $\rho^{0}$ and $\bar\rho$ for {\it quantlets} basis and principal components, varying with the number of basis coefficients.  
  \label{S5_Figure_qant}}
\includegraphics[height=2.5in,width=5.0in]{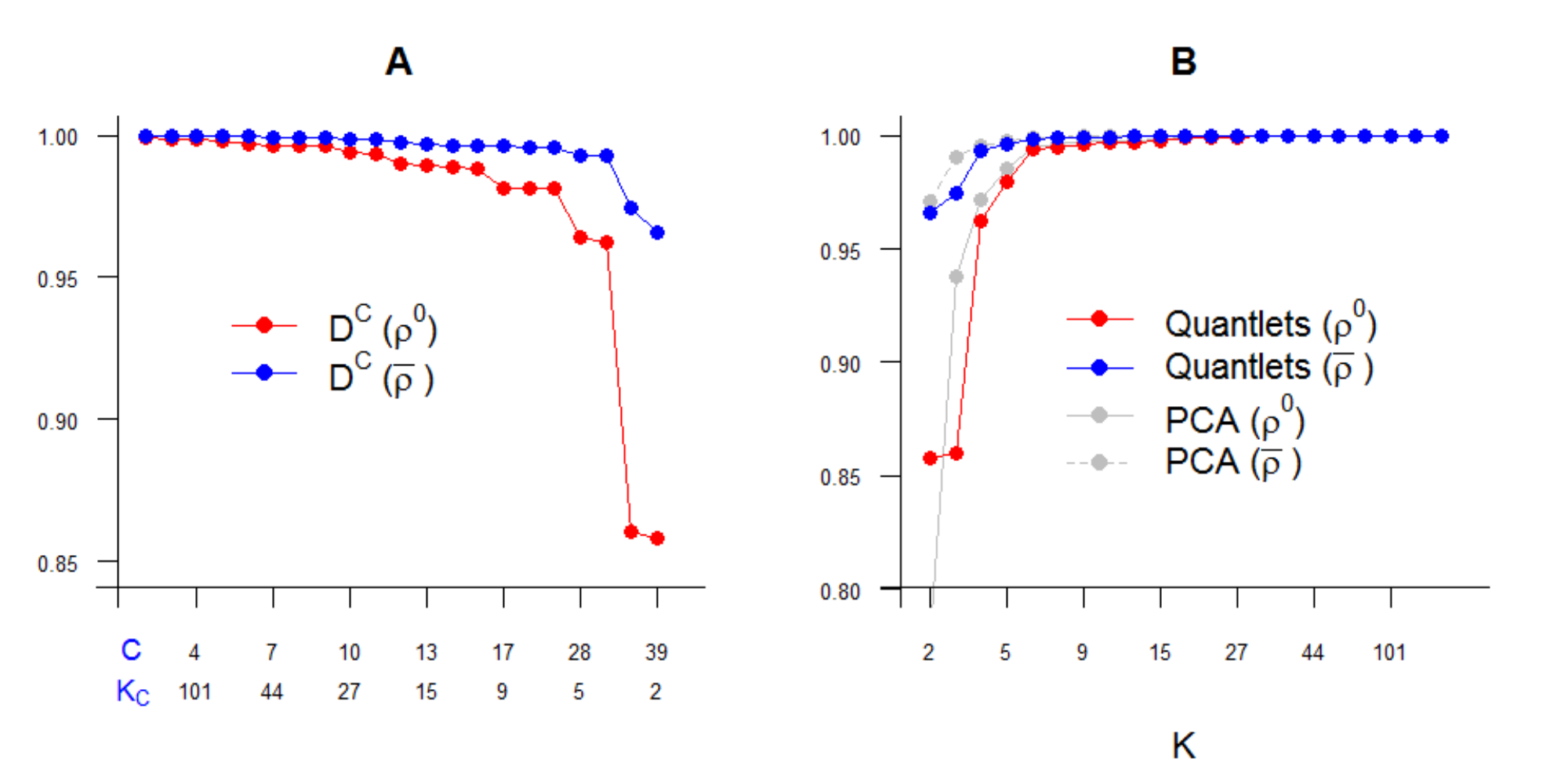}
\end{figure}

To find a sparse yet near-lossless basis set, we define a sequence of reduced basis sets $\{\mathcal{D}^U_{(i)\mathcal{C}}, \mathcal{C}=1, \ldots, n-1\}$ that contain the Gaussian basis functions $\xi_1$ and $\xi_2$ plus all dictionary elements $\xi_k(p)$ that are selected for at least $\mathcal{C}$ of the $n-1$ empirical quantile functions, excluding the $i$th one, i.e. $\mathcal{D}^U_{(i)\mathcal{C}}=\{\xi_k, k:\sum_{i' \ne i = 1}^n I(Q^O_{i'k} \ne 0) \ge \mathcal{C}\}.$ 
We can construct plots of $\rho^0$ or $\overline{\rho}$ vs. $\mathcal{C}$ to choose a value of $\mathcal{C}$ that leads to a sparse basis that can recapitulate the observed data at the desired level of accuracy (as shown below).  Given this choice, we next compute the corresponding reduced basis set using all of the data $\mathcal{D}^\mathcal{C}=\{\xi_k, k:\sum_{i = 1}^n I(Q^O_{ik} \ne 0) \ge \mathcal{C}\}$ containing $K=K_\mathcal{C}$ basis coefficients.  The left panel of Figure \ref{S5_Figure_qant} contains this plot for our GBM data set.  From this, we select $\mathcal{C}=10$ which leads to $K_\mathcal{C}=27$ basis functions
as this number of basis preserves a concordance of at least $\rho^0=0.990$ for each subject ($\epsilon=0.01$) and an average concordance of $\overline{\rho}=0.998$.

\textbf{Orthogonalization and Denoising:} 
Next, we use Gram-Schmidt to orthogonalize the basis set $\mathcal{D}^\mathcal{C}$ to generate an orthogonalized basis set $\mathcal{D}^{\perp}=\{\psi^\perp_k(p), k=1,\ldots,K\}$, where $\{\psi^\perp_1(\cdot), \psi^\perp_2(\cdot)\}=\{\xi_1(\cdot), \xi_2(\cdot)\}$ comprise the Gaussian basis and $\{\psi^\perp_k(\cdot), k=3,\ldots,K\}$ are orthogonalized basis functions computed from and spanning the same space as the remaining bases in $\mathcal{D}^\mathcal{C}$, indexed in descending order of their total percent variability (total energy) explained for the given data set. 
Specifically, suppose that $Q_{ik}^{\perp}$, $k= 1, \dots, K$ and $i=1, \dots, n$  are the empirical coefficients corresponding to the elements of $\mathcal{D}^\perp$, ordered as in $\mathcal{D}^{\mathcal{C}}$.
We compute the percent total energy for basis $k$ as  $\mathcal{E}_k=\sum_{i=1}^{n}Q_{ik}^{ \perp 2}/\sum_{i=1}^{n}\sum_{k=1}^{K}Q_{ik}^{\perp  2}$, and then relabel $\psi_k, k=3, \ldots, K$ to be in descending order of $\mathcal{E}_k$.

  In practice, we have observed that the first number of orthogonal basis functions are relatively smooth, but the later basis functions can be quite noisy, sometimes with high-frequency oscillations.  As we do not believe these oscillations capture meaningful features of the empirical quantile functions, we regularize the orthogonal basis functions using wavelet denoising to adaptively remove these oscillations.  
For a choice of mother wavelet function 
  $\varphi_{j,l}(p)=2^{j/2}\varphi(2^j p - l)$ with integers $j, l$, we construct
  the wavelet shrunken and denoised basis function $\psi^{\dagger}_k(p)$ \cite{donoho1995wavelet}, given by
 \begin{equation}
\psi^{\dagger}_k(p) = \sum_{j= 0}^J \sum_{l= 1}^{L_j} d_{k,j,l}^{\dagger}\varphi_{j,l}(p),
  \end{equation}
where $L$ is a grid of size $L=2^{10}=1024$ for our GBM data,
$d_{k,j,l}=\int \psi^\perp_k(p)\varphi_{j,l}(p) dp=\langle \psi^\perp_k, \varphi_{j,l} \rangle$,
 $d_{k,j,l}^{\dagger}=d_{k,j,l}$ if $|d_{k,j,l}|>\sigma \sqrt{2 \log L}$ and $d_{k,j,l}^{\dagger}=0$ If $|d_{k,j,l}|\le \sigma \sqrt{2 \log L}$.
 We use an empirical estimator for $\sigma$
that is the median absolute deviation of the wavelet coefficients at the highest frequency level $J$.
Details for denoising are described in Section 1.1 of the supplement.

  After applying the denoising method to all of the orthogonal basis functions in the set $\mathcal{D}^{\perp}$ to get $\mathcal{D}^\dagger=\{\psi^\dagger_k(p), k=1,\ldots,K\}$, we re-standardize these basis functions  by $\psi_k(p)=(\psi_k^{\dagger}(p) - \mu^\dagger_k)/\sigma^\dagger_k$ for $k=3, \ldots, K$ with $\mu^\dagger_k = \int_0^1 \psi_k^{\dagger}(p) dp$ and $\sigma^\dagger_k=\sqrt{\int_0^1 \{\psi_k^{\dagger}(p)-\mu^\dagger_k\}^2  dp}$ such that $\int_0^1 \psi_k(p) dp=0$ and $\int_0^1 \psi_k(p) \psi_k(p) dp = 1$ for $k=3,\ldots,K$.  

We refer to the resulting basis set $\mathcal{D}=\{\psi_k(p), k=1,\ldots, K\}$ as the \textbf{{\it quantlets}}, which we use as the basis functions in our quantile functional regression modeling.     Figure \ref{S5_QBE} contains the first 16 quantlet basis functions from the GBM data set.

   \begin{figure}[!htb]
\centering
\caption{First 16 quantlet basis functions for GBM data set.
  \label{S5_QBE}}
\includegraphics[height=5.3in,width=5.7in]{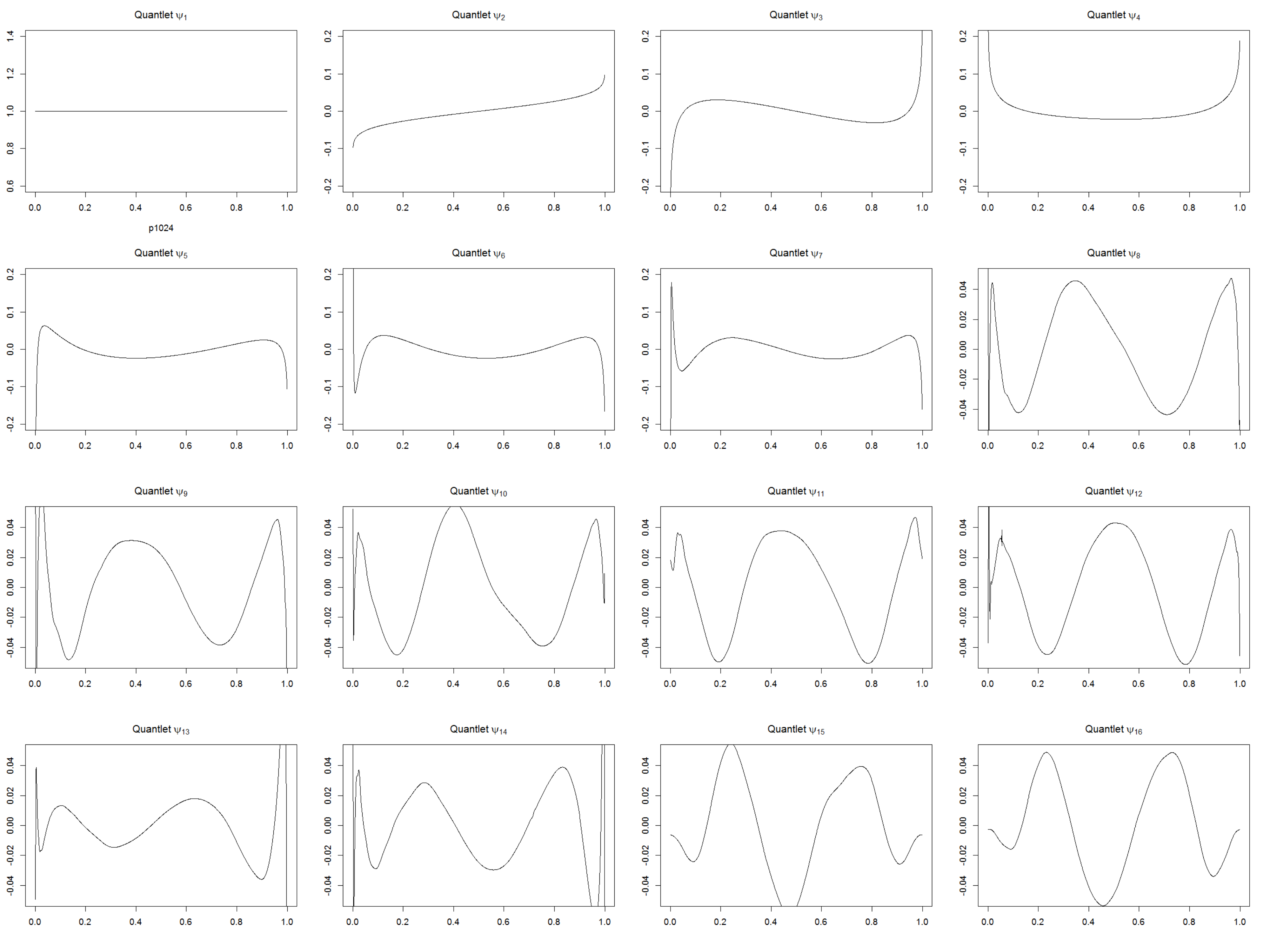}
\vspace{0.5cm} 
\end{figure}

\textbf{Properties of quantlets:}  These quantlets have numerous properties that makes them useful for modeling in our quantile functional regression framework.
\begin{itemize}
\item \textbf{Empirically defined}:  The empirical quantile functions for different applications can have very different features and characteristics.  Given their derivation from the observed data, the quantlets are customized to capture the features underlying the given data set, giving them advantages over pre-specified bases like splines, wavelets, or Fourier series.

\item \textbf{Near-losslessness}: By construction, the set of quantlets are at least \textit{near-lossless} in the sense that the basis is sufficiently rich to almost completely recapitulate the empirical quantile functions $Q_i(p)$.  As a result, we can project the empirical quantiles into the space spanned by the quantlets with negligible error, and thus it is reasonable to consider modeling the quantlet coefficients for the empirical quantile functions as observed data.

\item \textbf{Regularity}: The denoising step tends to remove any wiggles or high frequency noise from the orthogonal basis functions $\psi^\perp_k(p)$, leading to visually pleasing yet adaptive basis functions that are relatively smooth and regular.  We have found these tend to be more regular looking than other empirically determined basis functions like principal components  (compare Figure \ref{S5_QBE} to Supplementary Fig 5).

\item \textbf{Sparsity}: The procedure we have defined to construct the quantlets tends to also produce a basis set that is relatively low dimensional and thus a sparse representation.  We have found these basis functions to have similar sparsity to principal component bases, measured by computing the average LOOCCC $\overline{\rho}$ for quantlets and analogously for principal components (i.e. computing the principal components leaving out the $i$th sample, and then estimating $\rho_{(i)}$ measuring the losslessness of the resulting basis set) -- see Figure \ref{S5_Figure_qant}B and Figure \ref{S2_Figure_1}C.  Using a low dimensional basis enhances the computational speed of our procedure and reduces the uncertainty in the quantile functional regression coefficients $\beta_a(p)$, as can be seen in our sensitivity analyses (Supplementary Table 5).

\item \textbf{Interpretability}:  Unlike principal components, the quantlets have some level of interpretability in that the first two basis functions define the space of all Gaussian quantile functions (see Figure \ref{S5_QBE}).  
For Gaussian data, only the first two basis functions will be needed, while comparing with dimensions $k=3, \ldots, K$ provides a measure of the degree of \textit{non-Gaussianity} in the distribution. 
The remaining quantlets for $k\ge3$ are not necessarily interpretable since they are empirically determined, but by our observation for many data sets the next two quantlets capture some sense of skewness and some sense of heavy-tailedness like kurtosis.

\end{itemize}

\subsection{ \bf{Quantlet-based Modeling in Quantile Functional Regression}} \label{qbmiqfr}

 
Given the $i$th empirical quantile function $Q_i(p_j)$ evaluated at $p_j=j/(m_i+1), j=1,\ldots, m_i$, constructed from the order statistics $Y_{i(j)}, j=1,\ldots, m_i$, and a quantlet basis set $\mathcal{D}=\{\psi_k(p), k=1,\ldots,K\}$ derived as described in Section \ref{sec:quantlets}, we write a quantlet basis expansion $Q_i(p_j)=\sum_{k=1}^K Q^*_{ik} \psi_k(p_j)$ with $Q^*_{ik}$ being the $k$th empirical quantlet basis function for subject $i$.  For this paper, we will assume that $K<\text{min}_i(m_i)$, with the understanding that $K \ll \text{min}_i(m_i)$ for an extremely large number of applications, including our GBM data.  Extensions of this framework to sparse data settings for which $m_i<K$  for some $i$ are tractable and of interest, but given the length and complexity of this paper and the additional challenges raised by this sparse case, we will leave it to future work.

 With $\bQ_i=[Q_i(p_1), \dots, Q_i(p_{m_i}) ]$ a row vector containing the $i$th empirical quantile function and $\bPsi_i$ a $K \times m_i$ matrix with element $\bPsi_i(k,j)=\psi_k(p_j)$, we can compute the $1 \times K$ vector of empirical quantlet coefficients $\bQ^*_i=[Q^*_{i1}, \ldots, Q^*_{iK}]$ by $\bQ^*_i = \bQ_i \bPsi_i^-$, where $\bPsi_i^-=\bPsi_i^T (\bPsi_i \bPsi_i^T)^{-1}$ is the generalized inverse of $\bPsi_i$.  
 Based on the \textit{near-lossless} property of the quantlets by design, $\bQ^*_i$ contains virtually all of the information in the raw data $\bQ_i$, and thus we model these as our data.
Concatenating $\bQ^*_i$ across the $n$ subjects, we are left with a $n \times K$ matrix $\bQ^*$,
and consider obtaining  estimates and inference on the quantiles and parameters of model (\ref{p5_qfm_i}) on any desired grid of $p$ of size $J$, by $\bQ(\mathcal{P})=\bQ^*\bPsi$ and $\bB(\mathcal{P})=\bB^*\bPsi$ with $\bPsi$ a $K \times J$ matrix with elements $\psi_k(p_j)$, where $\bB^*$ an $A \times K$ matrix of corresponding quantlet-space regression coefficients.

The Wasserstein distance between cumulative distribution functions \cite{bickel1981some}  
 is defined as $L(F, G)=\inf_{U,V}\|U-V\|_m$ for $F$ and $G$ two distribution,  where
 all pairs of random variables $(U,V)$ are followed from $F$ and $G$, respectively.
Following \citeasnoun{bellemare2017distributional}, the infimum is attained by the inverse transformation of a random variable $\mathcal{P}$ uniformly distributed on $[0,1]$. i.e.,
$L(F, G)= \int_{0}^{1} |F^{-1}(p)-G^{-1}(p) |^m dp.$ 
The quantile functional regression model of (\ref{p5_qfm_i}) is a framework for the Wasserstein distance with $m=2$, 
minimizing the empirical risk 
$\sum_{i=1}^{n}\int_{0}^{1} |Q_i(p)-\sum_{a=1}^{A}x_{ia}\beta_{a}(p) |^2 dp. $
Based on the matrix notation,
 we rewrite the empirical risk for the Wasserstein loss function as
\begin{equation} \label{mat}
\text{tr}[( \bQ^*\bPsi -\bX \bB^*\bPsi  )( \bQ^*\bPsi-\bX \bB^*\bPsi  )^{T}],
\end{equation}
where $\bX$ is an $n \times A$ matrix with $\bX(i,a)=x_{ia}$.
It follows from the corresponding normal equation  $\bX^{T}\bQ^{*}\bPsi\bPsi^{T}= \bX^{T}\bX\bB^{*}\bPsi\bPsi^{T}$ that
 the minimizers  $\widehat{\bB^*}=(\bX^{T}\bX)^{-1}\bX^{T}\bQ^*$  is seen to be a point estimator in a quantile functional regression framework like ours.  This partially motivates our approach of performing the regressions on the quantile scale.

We can also consider regressing on the covariates in the \textit{quantlet space model} 
\vspace{-2mm}
\begin{equation}
\bQ^*=\bX \bB^* + \bE^*, \\ \label{eq:quantreg}
\end{equation}
where  $\bE^*$ an $n \times K$ matrix of quantlet space residuals.  
From (\ref{eq:quantreg}),
we can relate this quantlet-space model back to the original quantile functional regression model (\ref{p5_qfm_i}) through the quantlet basis expansions $\beta_a(p)=\sum_{k=1}^K B^*_{ak} \psi_k(p)$ and $E_i(p)=\sum_{k=1}^K E^*_{ik} \psi_k(p)$.
The rows of $\bE^*$ are assumed to be independent and identically distributed mean-zero Gaussians, with $\bE^*_{i.} \sim N(\bzero, \Sigma^*)$, where $A_{i.}$ or $A_{.i}$ denotes the $i$th row or column of the matrix A.
  Here, we assume $\Sigma^*=\text{diag}_k\{\sigma^2_k\}$, which enables us to fit in parallel the models for each column, $\bQ^*_{.k} = \bX \bB^*_{.k} + \bE^*_{.k}, k=1,\ldots,K$, and yet accommodate correlation across $p$ since modeling in the quantlet space induces correlation in the original data space, with the covariance operator for $E_i(p)$ given by $\Sigma(p,p')=\text{cov}\{E_i(p), E_i(p')\} = \bPsi(p) \Sigma^* \bPsi(p')$, where $\bPsi(p)=(\psi_1(p), \ldots, \psi_K(p))^T$.     The empirical nature of the derived quantlets makes this structure well-equipped to capture the key correlations across $p$ in the observed data, as shown for our 
real data set (See Supplementary Figure 9).
If desired, 
one could model $\Sigma^*$ as an unconstrained $K \times K$ matrix, which would provide additional flexibility in the precise form of $\Sigma$ but at a potentially much greater computational cost.


\subsection{ {\bf Bayesian Modeling Details}}

This model could be fit using vague conjugate priors for the regression coefficients, $B^*_{ak} \sim N(0, \tau^2)$ for some extremely large $\tau^2$.  This could be called a \textit{quantlet-no sparse regularization} approach.  It would result in virtually no smoothing of $\beta_a(p)$ relative to the naive (one-$p$-at-a-time) quantile functional regression model, but it would still account for correlation across $p$ in the residual errors, so may have inferential advantages over the naive approach.   We can further improve performance by inducing regularity and smoothness in the quantile functional regression parameters $\beta_a(p)$, which we accomplish through regularization or shrinkage priors, as is customary for Bayesian functional regression models.

Motivated by a belief that the covariate effects should be more regular than the empirical quantile functions themselves, we assume sparsity-inducing priors on the $B^*_{ak}$ coefficients.
We use a spike-Gaussian slab \cite{lempers1971posterior,mitchell1988bayesian} distribution.  The spike at 0 induces sparsity while the Gaussian prior applies a roughness penalty.
Motivated by the belief that certain quantlets are \textit{a priori} more likely to be important for representing covariate effects, we partition the set of $K$ quantlet dimensions into $H$ clusters of basis functions, each with their own set of prior hyperparameters.  This allows us, for example, to allow a higher prior probability for certain quantlet dimensions to be important  such as the 
the Gaussian basis levels $\{\psi_1, \psi_2\}$ and the quantlets explaining a high proportion of the relative variability in the empirical quantile functions.  Recalling that quantlets are indexed in descending order of their proportion of relative variability explained, we can group together the Gaussian coefficients as one cluster, and then split the rest sequentially into $H$ clusters each containing sets of basis functions whose relative variability explained are of similar order of magnitude.  

Let $\Pi$ be a $K \times J$ matrix with element $\Pi(k,j)=\xi_k(p_j)$  and 
$\Phi$ be a $K \times J$ matrix with element $\Phi(k,j)=\psi^\perp_k(p_j)$ 
for $k=1,\dots, K$ and $J=1,\dots,J$, where $\xi_k(p_j)$ and $\psi^\perp_k(p_j)$ are elements in $\mathcal{D}^{\mathcal{C}}$
and $\mathcal{D}^{\perp}$. Note that we use Gram-Schmidt to orthogonalize the basis set $\mathcal{D}^\mathcal{C}$ to generate an orthogonalized basis set $\mathcal{D}^{\perp}$. Then, we measure the variability for each element as $\text{diag}(\Lambda)$ based on the spectral decomposition structure, 
$\Lambda = \Phi \Pi ^{T}\Pi\Phi ^{T}$.
We can split the elements in $\psi^\perp_k(\cdot) \in \mathcal{D}^{\perp}$  into $H$ clusters each containing sets of basis functions sharing a similar variability explained, where
the  hierarchical cluster algorithm on the variability is used to determine it.
Although we use the quantlets basis function $\psi_k(\cdot)$ which is obtained by denoising and re-standardizing for the corresponding element $\psi^\perp_k(\cdot)$, the identical clustering information is used in the estimation procedure.

Specifically,
let $\mathcal{K}$ be a set of indices $\mathcal{K} =\{1,\dots,K \}$
and 
$\mathcal{H}$ be a set of indices $\mathcal{H} =\{1,\dots,H \}$ such that 
$\mathcal K=\cup_{h=1}^{H} \mathcal K_{h}$ with $\mathcal K_{h}\cap \mathcal K_{h'}=\emptyset$ for  $h\not=h'$, where
  $\mathcal{K}_h= \{1\le k\le K| f(k)=h\}$ for the clustering map, $f(k)=h$.
 Let   $\mathcal{H}_h=\{ \pi(k)  |  k \in \mathcal K_{h} \} \equiv \{1,\dots, |\mathcal K_{h}| \}$ be  the ordered set consisting of the integers
  such that for all $k$ and $k' \in \mathcal K_{h}$,
 if $k<k'$ implies   $\pi(k)  <\pi(k')$.
By defining
the index $h_{k,l}$ to indicate the quantlets $\psi_k$ as the $l^{\text{th}}$ component of $\mathcal{H}_{h}$
 within the $h$ cluster,
the prior on $B^*_{ak}$ is given by
  \begin{align}    \label{p5_prior_B} 
 &B_{ak}^* \equiv B_{ah_k,l}^*  \sim  \gamma_{ah_k,l} N(0, \tau^2_{ah_kl}) + (1-\gamma_{ah_k,l})I_{0}       \\
 &\gamma_{ah_k,l}  \sim \text{Bernoulli}(\pi_{ah}) ,      \notag 
  \end{align}
  where $I_{0}$ is a point mass distribution at zero, and $\gamma_{ah_k,l}$ 
  is an indicator of whether the $k$th quantlet basis coefficient is important for representing the effect for the $a$th covariate within 
  the $h$ cluster as the $l^{\text{th}}$ component. 
  The hyperparameter $\pi_{ah}$ indicates the prior probability that a quantlet coefficient in set 
  $\mathcal{K}_h$ is important, and $\tau^2_{ah_k,l}$ the prior variance, and regularization factor, for coefficient $B^*_{ak}$ conditional on it being chosen as important.

  In order to fit model (\ref{eq:quantreg}) using a Bayesian approach, we also need to specify priors on the variance components $\{\sigma^2_k, k=1,\ldots,K\}$.  
We place a vague proper inverse gamma prior on each diagonal element $\sigma_{k}^2$ given by
      $\sigma_{k}^2 \sim \text{inverse-gamma}( \nu_0/2,  \nu_0/2 )$,
where $\nu_0$ is some relatively small positive constants. Other relatively vague priors could also be used. 
If one wanted to allow $\Sigma^*$ to be unconstrained, an Inverse Wishart prior could be assumed for the $K \times K$ matrix.
The likelihood funciton is given
$\bQ^*_{.k} \sim  N( \bX \bB^*_{.k} , \sigma_{k}^2 I )$  in the projected space for each $k=1,\ldots,K$, .

\subsection{\bf{Details of MCMC}}

 The parameters $\pi_{ah}$ and $\tau^2_{ah_k,l}$ can be estimated using an empirical Bayes method, 
  assuming $\tau_{ak}^2\equiv\tau_{ah_k,l}^2=V_{ah_k,l}\Gamma_{ah}$ for some parameters $\Gamma_{ah}$,
   which allows for full flexibility in these regularization parameters within the $\mathcal K_{h}$ group  for the $a$th covariate.  This structure also enables us to integrate out the {\it quantlets} coefficients and compute the marginalized  likelihood for  $\Gamma_{ah}$ and $\pi_{ah}$  as: 
\begin{align*}
l(\pi_{ah}, \Gamma_{ah})  \propto &(1 +  \Gamma_{ah})^{-\sum_{  l \in  \mathcal H_{h}    }^{ |\mathcal K_{h} |}\gamma_{ah_k,l} /2 }
\exp{\biggl \{ -1/2 \sum_{l \in  \mathcal H_{h} }^{ |\mathcal K_{h} | }\zeta_{ah_k,l}^2\gamma_{ah_k,l}/(1 +  \Gamma_{ah})
\biggr \}}  \notag \\
&\qquad 
\times
(\pi_{ah})^{\sum_{  l \in  \mathcal H_{h}    }^{ |\mathcal K_{h} |}\gamma_{ah_k,l} }
(1-\pi_{ah})^{|\mathcal K_{h}| -\sum_{  l \in  \mathcal H_{h}    }^{ |\mathcal K_{h} |}\gamma_{ah_k,l}  }.
\end{align*}
On the marginalized likelihood, the MLEs  of $\Gamma_{ah}$ and $\pi_{ah_k}$  can be obtained by 
  \begin{align*}  
& \hat \pi_{ah}  =  \sum_{l \in  \mathcal H_{h} }^{ |\mathcal K_{h} | }\gamma_{ah_k,l}/|\mathcal K_{h}| , 
~~~ \widehat{\Gamma}_{ah}  =  \max \biggl( 0,     \sum_{l \in  \mathcal H_{h} }^{ |\mathcal K_{h} | }\zeta_{ah_k,l}^2\gamma_{ah_k,l}/
\sum_{l \in  \mathcal H_{h} }^{ |\mathcal K_{h} | }\gamma_{ah_k,l}-1   
  \biggr ) \\
  & \widehat{O}_{ah_k,j}=  \frac{\hat \pi_{ah} }{1 - \hat \pi_{ah} }(1 +  \widehat{\Gamma}_{ah} )^{-1/2}
  \exp{\biggl \{ 
  \frac{1}{2}\zeta_{ah_k,l}^2 \frac{ \widehat{\Gamma}_{ah} }{1+ \widehat{\Gamma}_{ah} } 
  \biggr \}}, ~~~
    \hat{\gamma}_{ah_k,l}=  \frac{\widehat{O}_{ah_k,l}}{1+\widehat{O}_{ah_k,l}}.
  \end{align*}
These empirical Bayes estimates can be computed for each iteration of the MCMC procedure. 
The Bayes estimates of $\pi_{ah}$ and $\tau_{ah_k,l}$ are given by
$\hat \pi_{ah}$ and $\widehat{V}_{ah_k,l} \widehat{\Gamma}_{ah}$.

We fit the quantlet space model (\ref{eq:quantreg}) using Markov chain Monte Carlo (MCMC).
Let ${\bf Q}_{.k}^*$ and ${\bB}_{.k}^{\ast}$  
  be the $k$th column vector of ${\bQ}^{\ast}$ and ${\bB}^{\ast}$, respectively.
  For each quantlet basis $k=1,\ldots,K$, we sample the $a$th covariate effect from $f(B_{ak}^{\ast} | {\bf Q}^{\ast}, {\bB}_{(-a)k}^{\ast},\sigma^2_k)$, where ${\bB}_{(-a)k}^{\ast}$ is a vector of length $A-1$ containing all covariate effects except the $a$th of $\bB^*$  in  model (\ref{eq:quantreg}) 
  for the $k$th quantlet coefficient.
 We repeat this procedure for all covariates,  $a=1, \dots, A$ and quantlet basis function $k=1,\ldots,K$.
This distribution is a mixture of  a point mass at zero and a normal distribution, with normal mixture proportion $\alpha_{ak}$ and the mean and variances of the normal distribution $\mu_{ak}$ and $v_{ak}$ given by 
   \begin{align*}
   B_{ak}^{\ast} \equiv B_{ah_k,l}^*  \sim 
   \alpha_{ah_k,l}
    N( \mu_{ah_k,l}, v_{ah_k,l}) +
  (1- \alpha_{ah_k,l})I_{0}
\end{align*} 
where  $\alpha_{ah_k,l}$, $\mu_{ah_k,l}$ and  $v_{ah_k,l}$ are given by 
   \begin{align*}
&\alpha_{ah_k,l}
=     \text{P}( \gamma_{ah_k,l} =1 |  {\bf Q}^{\ast}_{.k}, {\bB}_{(-a)k}^{\ast},\sigma^2_k)   
=  \widehat{O}_{ah_k,l}/( \widehat{O}_{ah_k,l}+1),\\
&\mu_{ah_k,l}= \widehat{B}_{ah_k,l}^{\ast}(1+ V_{ah_k,l}/\tau_{ah_k,l})^{-1},~~
~~v_{ah_k,l}= V_{ah_k,l}(1+ V_{ah_k,l}/\tau_{ah_k,l} )^{-1} ,  \notag \\
& \widehat{O}_{ah_k,l}=  \frac{\hat \pi_{ah} }{1 - \hat \pi_{ah} }(1 + V_{ah_k,l}/\tau_{ah_k,l} )^{-1/2}
  \exp{\biggl \{ 
  \frac{1}{2}\zeta_{ah_k,j}^2 \frac{ V_{ah_k,l}/\tau_{ah_k,l} }{1+V_{ah_k,l}/\tau_{ah_k,l} } 
  \biggr \}}, \\
 & \zeta_{ah_k,l}= \widehat{\beta}_{ah_k,l}^* /V_{ah_k,l}^{1/2},~~ 
V_{ah_k,l}=\left(\sum_{i=1}^{n} x_{ia}^2/\sigma_{k}^2\right)^{-1}, 
\end{align*}   \label{qbmiqfr}
and $\widehat{B}_{ah_k,l}^{\ast}$ is frequentist estimator mentioned in Subsection~\ref{qbmiqfr}.
For each quantlet basis $k=1,\ldots,K$, we sample $\sigma_k^2$ from 
its complete conditional
 \begin{align*}
\text{P}( \sigma_{k}^{2}|{\bB}^{\ast}_{.k}, {\bf Q}^{\ast}_{.k},  {\bX})
\sim  \text{Inverse Gamma}\{  (\nu_0+n)/2,
(\nu_0+ \text{SSE}({\bB}^{\ast}_{.k}) )/2\},
\end{align*}
where $\text{SSE}({\bB}^{\ast}_{.k})= {\bf Q}_{.k}^{\ast  T}({\bf I} -{\bX}
( {\bX}^{T}{\bX}   )^{-1}{\bX}^{T} ){\bf Q}_{.k}^{\ast}$.

 \subsection{ \bf{Posterior Inference}} \label{sec:BayesInference}

After obtaining posterior samples for all quantities in the {\it quantlet space} model (\ref{eq:quantreg}),
these posterior samples are transformed back to the {\it data space} using 
${\beta}_a^{(m)}(p) = \sum_{k=1}^{K} B_{ak}^{ \ast (m)}\psi_k(p), m=1,\ldots,M$ where $M$ is the number of MCMC samples after burn in and thinning.
From these posterior samples, various Bayesian inferential quantities can be computed, including point wise and joint credible bands, global Bayesian p-values, and multiplicity-adjusted probability scores, as detailed below.  These can be computed for $\beta_a(p)$ itself or any transformation, functional, or contrast involving these parameters.

\textbf{Point and joint credible bands:}  
Pointwise credible intervals for $\beta_a(p)$ can be constructed for each $p$ by simply taking the $\alpha/2$ and $1-\alpha/2$ quantiles of the posterior samples.  Use of these local bands for inference does not control for multiple testing, however.  Joint credible bands have global properties, with  the $100(1-\alpha)\%$ joint credible bands for $\beta_a(p)$ satisfying $  \text{P}(L(p) \le   \beta_a(p) \le U(p) ~~ \forall p \in \mathcal{P} ) \ge 1-\alpha$.  Using a strategy as described in  \cite{ruppert2003semiparametric},  we can construct joint bands by
   \begin{align} \label{p5_jointci}
J_{a,\alpha}(p)= \hat{\beta}_a(p) \pm q_{(1-\alpha)}  \bigl [ \widehat{\text{St.Dev}}\{\hat \beta_a(p)\} 
\bigr ],
\end{align} 
where $\hat \beta_a(p)$ and $\widehat{\text{St.Dev}}\{\hat \beta_a(p)\}$ are the mean and standard deviation
for each fixed $p$ taken over all MCMC samples. 
Here the variable $q_{(1-\alpha)}$ is the $(1-\alpha)$
quantile taken over all MCMC samples of the quantity
   \begin{align*}
Z_a^{(m)}= \max_{p \in \mathcal{P}}\left| \frac{\beta_a^{(m)}(p)-\hat \beta_a(p)}{\widehat{\text{St.Dev}}\{\hat  \beta_a(p)\} }  \right|.
\end{align*} 

\textbf{SimBaS and GBPV:}  
Following \citeasnoun{meyer2015bayesian}
we can construct  $J_{a,\alpha}(p)$ for multiple levels of $\alpha$ and
determine for each $p$ the minimum $\alpha$ such that $0$ is excluded 
from the joint credible band, which we call {\it Simultaneous Band Scores (SimBaS)}, $\text{P}_{a,SimBaS}(p)=\min{\{\alpha: 0\not\in  J_{\alpha}(p) \}}$, which can be directly estimated by
   \begin{align*}
\text{P}_{a,SimBaS}(p)=  M^{-1}\sum_{m=1}^{M}I\biggl \{ 
\biggl |
\frac{ \hat\beta_a(p) }{\widehat{\text{St.Dev}}\{\hat \beta_a(p)\}}
\biggr |
\le Z_a^{(m)}
\biggr \}.  
\end{align*}  
These can be used as
local probability scores that have global properties, effectively adjusting for multiple testing.  For example, we can flag all $\{p:\text{P}_{a,SimBaS}(p)<\alpha\}$ as significant.
From these we can compute $P_{a,Bayes}=$min$_p\{P_{a,SimBaS}(p)\}$, which we call {\it global
Bayesian p-values} (GBPV) such that we  reject the global hypothesis that $\beta_a(p) \equiv 0$ whenever 
$P_{a,Bayes}<\alpha$.

\textbf{Probability scores for distributional moments:}  
As mentioned in Section \ref{sec:QF}, distributional moments can be constructed as straightforward
functions of the quantile function, and thus from posterior samples of quantile functional regression
parameters one can construct posterior samples of these moments for various levels of covariates
$\bX$.
Denoting ${\beta}^{(m)}(p)=({\beta}_1^{(m)}(p), \dots, {\beta}_A^{(m)}(p))^{T}$ 
for each MCMC sample $m=1,\ldots,M$, posterior samples of distributional moments conditional
on $\bX$ are given by
 \begin{align}  \label{conmoments}
&{ \mu}^{(m)}_{\bX}  =  \int_{0}^{1} \bX^{T}\beta^{(m)}(p)  dp   \notag \\ 
&{\sigma}^{2(m)}_{\bX} =  \int_{0}^{1}(\bX^{T}\beta^{(m)}(p) - { \mu}^{(m)}_{\bX}  )^2 dp,   \notag \\ 
&{\xi}^{(m)}_{\bX}=  \int_{0}^{1}(\bX^{T}\beta^{(m)}(p) -{ \mu}^{(m)}_{\bX}  )^3/{\sigma}^{3(m)}_{\bX} dp, ~~\text{and}\notag \\  
&{\varphi}^{(m)}_{\bX}=   \int_{0}^{1}(\bX^{T}\beta^{(m)}(p) - { \mu}^{(m)}_{\bX} )^4/{\sigma}^{4(m)}_{\bX} dp.  
\end{align}
The conditional expectations of other basic statistics are similarly derived.   
We can construct posterior probability scores to assess differences of moments between groups or specific levels of continuous covariates as follows.  For each posterior sample, we compute the appropriate moment from the formulas in (\ref{conmoments}) for two covariate levels, $\bX_1$ and $\bX_2$, and compute the difference, e.g. for the mean $\Delta_m = \mu_{1m} - \mu_{2m}$.  Then, we 
define the posterior probability score for the comparison as:
$$P_{\mu_1-\mu_2}=2\min{\{  M^{-1} \sum_{m=1}^{M}I(\Delta_m>0) \},  M^{-1} \sum_{m=1}^{M}I(\Delta_m<0) \}}$$
In assessing a dichotomous covariate $x_{a}$, we compare $x_a=0$ and $x_a=1$ while holding all other covariates at the mean, while when assessing a continuous covariate we compute differences for two extreme values of $x_a$, with the corresponding probability scores for the respective moments denoted $P_{a,\mu}$, $P_{a,\sigma}$, $P_{a,\xi}$, or $P_{a,\varphi}$.

\textbf{Summarizing Gaussianity:}  
As mentioned above, the first two quantlets form a complete basis for the space of Gaussian quantile functions, so by comparing the first two coefficients to the remainder one can obtain a rough measure of ``Gaussianity'' of the predicted
distribution for a given set of covariates $\mathbf{X}$.  One measure that can be computed is $\sum_{k=1}^2 (\mathbf{X} \hat{\beta}_{ak})^2/\sum_{k=1}^K (\mathbf{X}\hat{\beta}_{ak})^2$, which will be on $[0,1]$, with a value of 1 precisely when the predicted quantile function is completely determined by the first two (Gaussian) bases and smaller scores indicating greater degrees of non-Guassianity.

\textbf{Predicted PDF and CDF:}  
To some researchers, distribution functions or probability density functions are more intuitive than quantile functions, and given their one-to-one relationship, it is possible to construct CDF or PDFs from the posterior samples as follows.  CDFs can be constructed by simply plotting $p$ vs. $\text{E}\{\hat{Q}(p)|\mathbf{X},\mathbf{Y}\}$, and given posterior samples of the predicted quantile functions on an equally spaced grid $0<p_1, \ldots, p_J<1$, one can estimate predicted pdf for a set of covariates as described in Section 1.3 of the supplement.

 \noindent Following is our recommended sequence of Bayesian inferential procedures.
 \begin{enumerate}[1.]
 \item Compute the global Bayesian p-value $P_{a,Bayes}$ for each predictor or contrast. 
\item For any covariates for which $P_{a,Bayes}<\alpha$, characterize the differences:
\vspace{-0.4cm}
 \begin{enumerate}[2a.]
\item Flag which probability grid points $p$ are different using $P_{SimBas}(p)<\alpha$.
\item  Compute moments; assess which moments differ according to the covariates.
\item  Assess whether the degree of Gaussianity appears to differ across covariates.
 \end{enumerate}
 \vspace{-0.4cm}
\item If desired, compute the predicted densities or CDFs for any set of covariates.
 \end{enumerate}

\section{ {\bf Simulation Study}}
We conducted a simulation study to evaluate the performance of the quantile functional modeling framework and the use of quantlet basis functions.

\subsection{\bf{Skewed Normal Scenario}}

We generated random samples for four groups of subjects whose mean quantile function was assumed to be from a skew normal distribution
 \begin{equation}
f(x)=\frac{2}{\omega} \phi \biggl ( \frac{x-\eta}{\omega} \biggr )  
\Phi \biggl ( \alpha   \biggl ( \frac{x-\eta}{\omega} \biggr )  \biggr )  
  \label{skewnormal} 
 \end{equation}
with the respective values of $(\eta, \omega, \alpha)$  being
$(1, 5, 0)$, $(3, 5, 0)$, $(1, 6.5, 0)$, and $(9.11, 7.89, -4)$, which correspond to a $N(1,5)$, $ N(3,5), N(1,6.5)$, and a skewed normal with mean $1$, variance $5$, and skewness $-0.78$ denoted by $SN(1,5,-0.78)$.    
Panels A and E of Figure \ref{S2_Figure_1} below show the densities
and quantile functions, respectively, corresponding to these distributions.

For each group $j=1,\ldots, 4$, we generated the random process $Q_{ij}(p)$ for $i=1,\ldots n$ subjects, taking 1024 samples from the corresponding skewed normal distribution, with $p\in \mathcal{P} = [1/1025, \ldots, 1024/1025]$, and some correlated
noise $\epsilon_{ij}(p)$ added to allow some random biological variability in the individual subjects' distributions.  That is, $Y_{ij}(p) = \beta_j(p)+\epsilon_{ij}(p)$,  where $\epsilon_{ij}(p)$ follows an Ornstein-Uhlenbeck process such that  $\text{Cov}(\epsilon_{ij}(p), \epsilon_{ij}(p') )=0.9^{|p-p'|}$.

After constructing the empirical quantile function $Q_{ij}(p)$ by reordering $Y_{ij}(p)$ in $p$, the quantile functional regression model we fit to these data was
 \begin{equation}
 Q_{ij}(p)= \sum_{a=1}^4 X_{ija} \beta_a(p) + \epsilon_{ij}(p),
  \label{mgroups} 
 \end{equation}
with covariates defined such that $X_{ij1}=1$ is for the intercept and $X_{ija}=\delta_{j=a}$ for $a=2, 3, 4$ group indicators for groups 2-4.  Note that with this parameterization, the means of the four groups are, respectively,
$\beta_1(p)$, $\beta_1(p)+\beta_2(p)$, $\beta_1(p)+\beta_3(p)$, and $\beta_1(p)+\beta_4(p)$, and by construction
$\beta_2(p)$ represents a location offset, $\beta_3(p)$ a scale offset, and $\beta_4(p)$ a skewness offset.  Panel E of Figure \ref{S2_Figure_1} displays the true mean quantiles for each group and panel F the true values for these quantile functional regression coefficients.

 \begin{figure}[!htb]
 \caption{ Simulated data in the skewed normal scenario and their {\it quantlet} representations:  (A) density functions of the population, (B) the near-lossless criterion varying with the different number of basis functions, (C) the concordance correlation varying with the cumulative number of the {\it quantlets}, and compared with principal components  (D) the relation between empirical quantile functions and {\it quantlet} fits, (E) mean quantile functions by group and (F) quantile functional regression coefficients.
  \label{S2_Figure_1}}
\centering
\includegraphics[height=3.5in,width=5.5in]{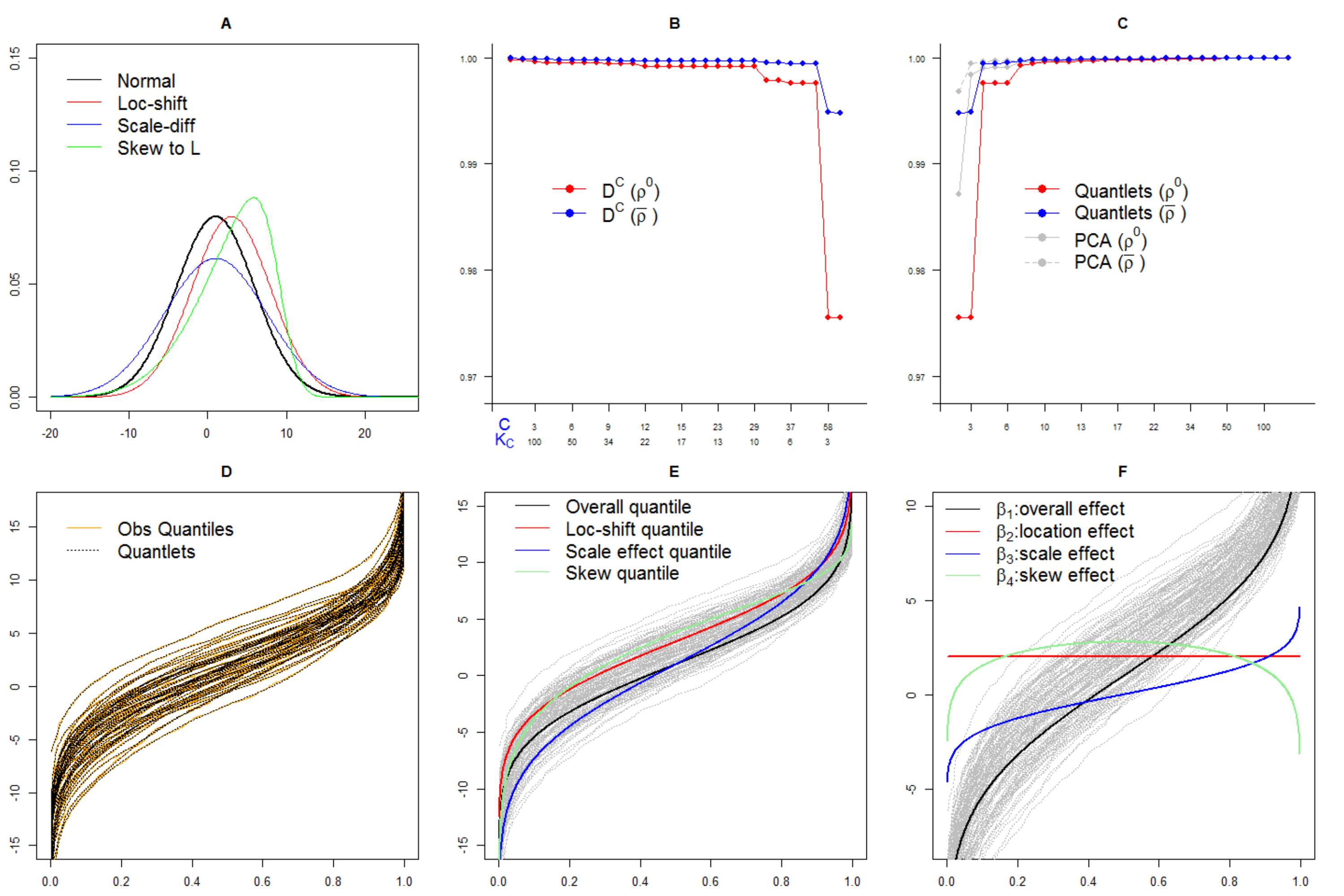}
\end{figure}

We constructed a quantlet basis set for this data set as described above, with some results summarized in panels B, C, and D of Figure \ref{S2_Figure_1}.  
The union set $\mathcal{D}^{U}= \cup_{i=1}^{n} \mathcal{D}_i$ included $2,868$ basis functions, and 
we chose a {\it common} set, $\mathcal{D}^{\mathcal{C}}$, that retained $10$ basis functions, which
resulted in a near-lossless basis set with $\rho^0=0.997$ (see $K_\mathcal{C}=10$ in  panel B).  After orthogonalization, denoising, and re-standardization, the
set of quantlets had sparsity properties similar to principal components (see panel C),
 and the fitted {\it quantlet} projection almost perfectly coincided with the observed data for all of the empirical quantile functions (panel D).  Supplemental Figure 4 contains a plot of these $10$ quantlet basis functions.

We applied several different approaches to these data: 
(A) naive quantile regression method (separate classical quantile regressions for each $p$ 
 by using  \textit{rq} function in {\it quantreg} R package \cite{koenker2005quantile}
),
 (B) naive quantile functional regression approach (separate functional regressions for each subject-specific quantile $p$),
    (C) principal components method (quantile functional regression using PCs as basis functions),
(D) {\it quantlet} without sparse regularization, (E) {\it quantlet} with sparse regularization, and 
(F) Gaussian model (quantlet approach but keeping only the first two coefficients).
The naive quantile regression method (A) ignores all intrasubject correlation in the data and estimates the \textit{population} quantile conditional on covariates, not the \textit{subject-specific} quantile conditional on covariates desired in this quantile functional regression setting, but it is included here since it is an approach some researchers might try in this setting.
In each case, the MCMC was run for $2,000$ iterations, keeping every one after a burn-in of $200$.
The results are shown in Supplementary Figure 8. 
 We compared the methods in terms of the area within the joint credible region and the 
corresponding integrated coverage rate, defined respectively as 
$\mathcal{A}(a)= \int_{0}^{1} | J_{a,\alpha}^{upper}(p) -  J_{a,\alpha}^{lower}(p) |^2 dp$ and
$ \mathcal{C}(a)= \int_{0}^{1} I( J_{a,\alpha}^{lower}(p) \le {\bbeta}_a(p)
  \le J_{a,\alpha}^{upper}(p) ) dp$,
where $J_{a,\alpha}^{upper}(p)$ and $J_{a,\alpha}^{lower}(p)$ are
the upper and lower joint credible bands, respectively.

To investigate the degree of monotonicity afforded by the model, we constructed predicted quantile functions for a broad range of covariate values, and computed the degree of $\epsilon$-monotonicity, defined to be $P^\mathcal{M}_\epsilon(X)=\int_0^1 I[\widehat{Q}(p|X)-\text{max}_{p'<p}\{\widehat{Q}(p|X)\}>\epsilon] dp$ for some $\epsilon$ considered negligibly small in the context of the scale of $Y$ in the current data set.  
 We report the empirical rates of the $\epsilon$-monotonicity as $1-n^{-1}\sum_{i=1}^{n}P^\mathcal{M}_\epsilon(X_i)$.  This empirical summary measure can be used to assess if a given model produces predictors with significant non-monotonicities across $p$ or not.

Table \ref{S2_CI} reports $\mathcal{A}(a)$ and  $\mathcal{C}(a)$ for all quantile functional coefficients.  Methods A-E all had good coverage properties, but use of the basis
functions in modeling (C, D, E) clearly led to tighter joint credible bands than the naive quantile regression and naive quantile function regression methods that did not borrow strength across $p$, as expected, and the use of sparse regularization (E) led to tighter bands than the quantlet method with no shrinkage (D).  Supplementary Figure 8 demonstrates the wiggliness and extremely wide joint credible bands of the naive methods.
Note also that for the coefficient with significant skewness $\beta_4(p)$, the Gaussian model (F) had extremely poor coverage, while for the coefficients corresponding to the Gaussian groups, the quantlet model (E) had performance no worse than the Gaussian method.   This is encouraging, suggesting that when the quantile functions are Gaussian there is not much loss of efficiency from using a richer quantlet basis set.  

Supplementary Figure~10 
depicts the simultaneous band scores $P_{SimBaS}(p)$ for the two contrast functions associated with the scale effect $\beta_3(p)$ and skewness effect
$\beta_4(p)$, with regions of $p$ for which $P_{SimBaS}(p)<0.05$ are flagged as significantly different.  As seen in  Supplementary Figure 10, 
we expect to flag the tails in the scale effect and a broad region in the middle and in the extreme tails for the skewness effect.  Note how the quantlet method with sparse regularization (E) flagged a larger set of regions than the other approaches, especially (B).  In all cases, the global adjusted Bayesian p-values $\text{P}_{Bayes}=\min{\{\text{P}_{map}(p)\}}$ were less than $0.0005$; hence, 
the null hypothesis  $\beta_a(p)\equiv 0$ was rejected in all models.

\begin{table}
\caption{Results for Simulation 1: Area and coverage for the joint 95\% confidence intervals:
 (A) naive quantile regression approach,
(B) naive quantile functional regression approach,
(C) principal component method, (D) {\it quantlet} space without sparse regularization, 
(E) {\it quantlet} space with sparse regularization, and (F) Gaussian {\it quantlet} space approach.
      \label{S2_CI}}
\begin{center}
\resizebox{0.98\columnwidth}{!}{  
\begin{tabular}{ c | c c c c  c c } \hline
 $\text{ {\bf Type} }$     
  & $\text{ {\bf A} }$ &  $\text{{\bf B}}$ & $ \text{{\bf C}}$  & $ \text{{\bf D}}$ &$ \text{{\bf E}}$  &$ \text{{\bf F}}$   \\ \hline
${\beta_1(p)}$ 	
&2.448 (1.000) &	1.603 (1.000)	&	1.092 (0.999)	&	1.186  (1.000)  & 1.069 (1.000)  & 1.071 (1.000)  \\
${\beta_2(p)}$ 	
& 3.487 (1.000) &	2.246 (1.000)	&	1.551 (1.000)	&	1.706 (1.000)  & 1.465 (1.000) & 1.551 (1.000)   \\
${\beta_3(p)}$ 
& 3.581 (1.000)
&	2.242 (1.000)	&	1.599 (1.000)	&	1.717  (1.000)   & 1.457 (1.000) & 1.599 (1.000)  \\ 	
${\beta_4(p)}$ 	& 3.658 (1.000)
&	2.281 (1.000)	&	1.583 (1.000)	&	1.651  (1.000)  & 1.499 (1.000)  & 1.520 (0.421)  \\
	\hline	
\end{tabular}
}
\end{center}
\end{table}      
\begin{table}

\caption{Simulation 1: Testing for conditional moment statistics in simulation:
 (A) naive quantile regression approach,
(B) naive quantile functional regression approach,
(C) principal component method, (D) {\it quantlet} space without sparse regularization, 
(E) {\it quantlet} space with sparse regularization, (F) Gaussian {\it quantlet} space approach,
   and  (G) feature extraction approach,
  where the values in this table are the posterior probability scores derived by its corresponding method for each test (the first column).
      \label{S2_Test}   
    }
\begin{center}
\resizebox{0.72\columnwidth}{!}{   
\begin{tabular}{c   c| c c c  c c c c} \hline
  $\text{{\bf H}}_0$    &$\text{ {\bf True} }$   
 & $\text{ {\bf A} }$ &  $\text{{\bf B}}$ & $ \text{{\bf C}}$    & $ \text{{\bf D}}$   
  & $ \text{{\bf E}}$        & $ \text{{\bf F}}$             & $ \text{{\bf G}}$       
    \\ \hline
$\mu_{1}=\mu_{3}$   &	$\mu_{1}=\mu_{3}$	 &0.205
&	0.001	&	0.193	&	0.211	&	0.217	&	0.212	&	0.205 \\
$\mu_{2}=\mu_{4}$   &	$\mu_{2}=\mu_{4}$	 &0.438
&	0.001	&	0.447	&	0.465	&	0.445	&	0.462	&	0.438 \\
$\sigma_{1}=\sigma_{3}$   &	$\sigma_{1}\neq\sigma_{3}$	 	 &0.001
&	0.001	&	0.001	&	0.001	&	0.001	&	0.001	&	0.001  	\\
$\sigma_{2}=\sigma_{4}$   &	$\sigma_{2}=\sigma_{4}$		&0.187
&	0.002	&	0.420	&	0.334	&	0.331	&	0.016	&	0.187  	\\
$\xi_{1}=\xi_{3}$   &	$\xi_{1}=\xi_{3}$	 &0.389  
	&	0.374	&	0.498	&	0.488	&	0.479	&	0.493	&	0.389  \\
$\xi_{2}=\xi_{4}$   &	$\xi_{2}\neq\xi_{4}$	 &0.001	
&	0.001	&	0.001	&	0.001	&	0.001	&	0.505	&	0.001 \\
\hline
\end{tabular}
}
\end{center}
\end{table}      

We computed posterior probabillity scores to compare the mean, standard deviation, and skewness for each pair of distributions (Table \ref{S2_Test}), and
Supplemental Table 2 
contains the posterior means and credible intervals for each summary.  We see that the basis function methods (C-E) all flagged the
correct differences, while the naive quantile functional regression approach (B) had major type I error problems in the moment tests and the 
Gaussian method (F) unsurprisingly was unable
to detect differences in skewness.  As an additional comparison, we also applied the so-called {\it feature extraction} approach (G), which involved first
computing the moments from the set of values for each subject and then performing statistical test comparing these across the groups. 
 Encouragingly, we found
these results were near identical to those found using our quantile functional regression with quantlets (E), suggesting that our unified functional
modeling approach does not lose power relative to feature extraction approaches when the distributional differences are indeed contained in the
moments. 

Constructing predicted quantile functions for a wide range of predictors and assessing $\epsilon$-monotonicity, we found that the all predicted quantile functions from the quantlet-based methods were monotone, while the naive quantile functional regression method had $\epsilon$-monotonicity of 25.8\% and 96.8\% for $\epsilon=0.001$ and $0.01$, respectively, demonstrating that quantlet basis functions encouraged the predicted quantile functions to be monotone in $p$.

\subsection{\bf{Multi Modality Scenario}}

We also conducted the additional simulation based on multi modality scenario, 
  in order to see the performance of our method  as a balanced assessment.  
  Specifically, we generated random samples for four groups of subjects whose mean quantile function was assumed to be from four mixture distributions, where 
  two mixture skewed normal distributions consist of 
$SN(-3.06,3.67,0)$ and $SN(9.11,7.89, -4)$  with $0.5$ and $0.5$  probabilities for one while
$SN(-7.1,2.4,0)$ and $SN(-3.11,7.89, 4)$  with $0.3$ and $0.7$  probabilities for the other,
and 
two mixture normal distributions consist of  $N(-2.5,2.5)$, $N(4, 3)$ and $N(9.5, 2.1)$ with $0.3$, $0.5$ and $0.2$ probabilities for one while
$N(-2.5,1.5)$, $N(4, 3.56)$ and $N(9.5, 1.1)$ for the other 
with the same probabilities.

\begin{figure}[!htb]
 \caption{ Simulated data in the multi modality scenario and their {\it quantlet} representations:  (A) density functions of the population, 
 (B) mean quantile functions by group and (C) quantile functional regression coefficients,
 (D) the near-lossless criterion varying with the different number of basis functions,
 (E) the relation between empirical quantile functions and {\it quantlet} fits, and (F) posterior estimates for each $\beta(p)$.
  \label{re_sim}}
\centering
\includegraphics[height=3.5in,width=5.5in]{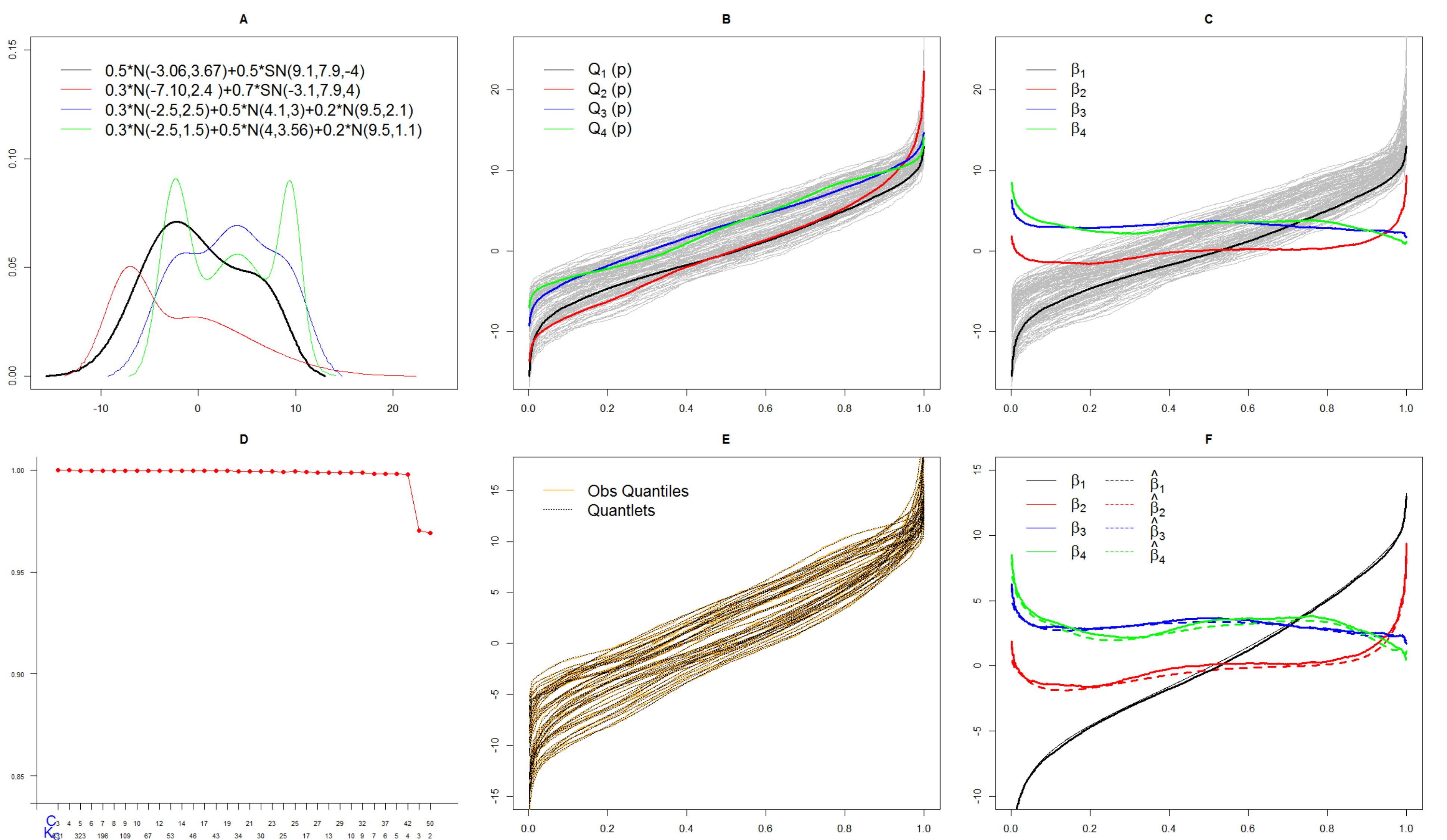}
\end{figure}

Panels A, B and C of Figure~\ref{re_sim} 
 show the densities,   mean quantile functions by group, and 
quantile functional regression coefficient estimates, respectively, corresponding to these distributions, where
all observed quantile functions are depicted with
 the gray-lines in each panel.
 All other settings in this simulation are the same to those of the first simulation
 for the model equation, covariates, noise process, and sample size.
We chose a {\it common} set, $\mathcal{D}^{\mathcal{C}}$, that retained $17$ basis functions, which
resulted in a near-lossless basis set with $\rho^0=0.997$ (see $K_\mathcal{C}=17$ in  panel D).  After orthogonalization, denoising, and re-standardization, 
 and the fitted {\it quantlet} projection almost perfectly coincided with the observed data for all of the empirical quantile functions (panel E). 
After running the MCMC algorithm, posterior estimate for each $\beta_a(p)$  is contained  as the dashed-line in
panel F of Figure~\ref{re_sim}.

As  expected, Table~\ref{S3_CI} shows that our method (E)
outperforms all other competing methods in that it leads to tighter bands with good coverage.  
Also, From Table~\ref{S3_Test}, we see that 
test results found using our method (E), were near identical to true results and 
our method does not lose power relative to feature extraction approaches (G) when the distributional differences are indeed contained in the moments, where $(\mu, \sigma, \xi)$ for each group is set to be
$(-0.06,  5.30,   0.02)$, $(-0.07,  6.40,   0.39)$, $(3.05,  5.05,  -0.05)$, and  $(3.05,  5.03,   0.07)$, respectively.

\begin{table}
\caption{Results for Simulation 2: Area and coverage for the joint 95\% confidence intervals:
 (A) naive quantile regression approach,
(B) naive quantile functional regression approach,
(C) principal component method, (D) {\it quantlet} space without sparse regularization, 
(E) {\it quantlet} space with sparse regularization, and (F) Gaussian {\it quantlet} space approach.
      \label{S3_CI}}
\begin{center}
\resizebox{0.98\columnwidth}{!}{  
\begin{tabular}{ c | c c c c  c c } \hline
 $\text{ {\bf Type} }$     
  & $\text{ {\bf A} }$ &  $\text{{\bf B}}$ & $ \text{{\bf C}}$  & $ \text{{\bf D}}$ &$ \text{{\bf E}}$  &$ \text{{\bf F}}$   \\ \hline
${\beta_1(p)}$ 	
&	3.528 (0.160)	&	2.041 (0.094)	&	0.296 (0.001)	&	2.821 (0.062)&	1.012 (1.000)	&2.813 (0.047) \\
${\beta_2(p)}$ 	
&	3.505 (0.363)	&	1.981 (0.229)	&	2.360 (0.062)	&	2.648 (0.059)&	1.492 (1.000)	&2.918 (0.127)   \\
${\beta_3(p)}$ 
&	3.367 (0.593)	&	1.954 (0.529)	&	2.855 (0.003)	&	4.282 (0.101)&	1.488 (1.000)	&0.536 (0.052)
  \\ 	
${\beta_4(p)}$
&	3.384 (0.485)	&	1.988 (0.332)	&	2.292 (0.023)	&	5.039 (0.630)	&1.556 (1.000)	&0.601 (0.001)
 \\
	\hline	
\end{tabular}
}
\end{center}
\end{table}      

\begin{table}
\caption{Simulation 2: Testing for conditional moment statistics in simulation:
 (A) naive quantile regression approach,
(B) naive quantile functional regression approach,
(C) principal component method, (D) {\it quantlet} space without sparse regularization, 
(E) {\it quantlet} space with sparse regularization, (F) Gaussian {\it quantlet} space approach,
   and  (G) feature extraction approach,
  where the values in this table are the posterior probability scores derived by its corresponding method for each test (the first column).
      \label{S3_Test}   
    }
\begin{center}
\resizebox{0.72\columnwidth}{!}{   
\begin{tabular}{c   c| c c c  c c c c} \hline
  $\text{{\bf H}}_0$    &$\text{ {\bf True} }$   
 & $\text{ {\bf A} }$ &  $\text{{\bf B}}$ & $ \text{{\bf C}}$    & $ \text{{\bf D}}$   
  & $ \text{{\bf E}}$        & $ \text{{\bf F}}$             & $ \text{{\bf G}}$       
    \\ \hline
$\mu_{1}=\mu_{2}$   &	$\mu_{1}=\mu_{2}$	 &	0.000 	&	0.000 	&	0.000 	&	0.000 	&	0.102 	&	0.000 	&	0.096  \\
$\mu_{3}=\mu_{4}$   &	$\mu_{3}=\mu_{4}$	 &	0.000 	&	0.000 	&	0.352 	&	0.000 	&	0.274 	&	0.000 	&	0.260  \\
$\sigma_{1}=\sigma_{2}$   &	$\sigma_{1}\neq\sigma_{2}$	 	
&	0.000 	&	0.000 	&	0.000 	&	0.000 	&	0.000 	&	0.000 	&	0.000  	\\
$\sigma_{3}=\sigma_{4}$   &	$\sigma_{3}=\sigma_{4}$		&	0.000 	&	0.000 	&	0.347 	&	0.000 	&	0.381 	&	0.352 	&	0.074  	\\
$\xi_{1}=\xi_{2}$   &	$\xi_{1}\neq\xi_{2}$	 &	0.000 	&	0.000 	&	0.328 	&	0.120 	&	0.004 	&	0.438 	&	0.000  \\
\hline
\end{tabular}
}
\end{center}
\end{table}      

\section{{\bf Quantile Functional Regression Analysis of GBM Data}}


In our GBM case study, radiologic images consisting of pre-surgical T1-weighted post contrast MRI sequences from 64 patients were obtained from from the Cancer Imaging Archive (cancerimagingarchive.net), along with measurements of certain covariates, including sex (21 females, 43 males), age (mean 56.5 years), DDIT3 gene mutation (6 yes, 58 no), EGFR gene mutation (24 yes, 40 no), GBM subtype (30 mesenchymal, 34 other), and survival status (25 less than 12 months, 39 greater than or equal to 12 months),
where  \citeasnoun{tutt2011glioblastoma}
has pointed out that most people diagnosed with GBM survive only 12 to 15 months, so that we followed this and used 12 moments as the cutoff in our context. 
This cut-off is commonly referred to as an extreme discordant phenotype design \cite{nebert2000extreme} and is a well-established grouping to enhance signals relevant to survival \cite{tyekucheva2011integrating}.

Following \citeasnoun{saha2016demarcate},
registration and inhomogeneity correction were conducted using Medical Image Processing and Visualization (MIPAV) software. Inhomogeneity correction known as nonparametric, nonuniform intensity normalization (N3) correction was conducted to remove the shading artifacts in MRI scans. Then,
 tumors were segmented in 3-D by clinical experts using the Medical Image Interaction Toolkit. Images and their 3-D tumor masks were subsequently re-sliced for isotropic pixel resolution using the NIFTI toolbox in MATLAB. From these re-sliced images, the slice with largest tumor area in the T1-post contrast image was selected as the Regions of Interest (ROI) for analysis. 
  We extracted the set of $m_i$ pixel intensities within the ROI for each patient $i=1,\ldots,n=64$, where
  the number of pixels within the tumor ranged from 371 to 3421.


\textbf{Model:}. We sorted the pixel intensities for each patient, yielding an empirical quantile function $Q_i(p_{ij})$ on a grid of observational points $p_{ij}=j/(m_i+1), j=1, \ldots, m_i$.  We related these to the clinical, demographic, and genetic covariates using the following quantile functional regression model:
 \begin{eqnarray}
 \label{eq:GBM}
Q_i(p)=&\beta_{\text{overall}}(p) + x_{\text{sex},i}\beta_{\text{sex}}(p) 
 +x_{\text{age},i}\beta_{\text{age}}(p) +x_{\text{DDIT3},i}\beta_{\text{DDIT3}}(p)     \notag \\
&  +x_{\text{EGFR},i}\beta_{\text{EGFR}}(p)  + x_{\text{Mesenchymal},i}\beta_{\text{Mesenchymal}}(p)  \notag \\
& 
+ x_{\text{survival},i}\beta_{ survival_{12}}(p) 
  +  E_i(p).
\end{eqnarray}

We constructed quantlets for these data using the procedure described in Section \ref{sec:quantlets}.  After the first step, we were left with a union basis set $\mathcal{D}^U$ containing 546 basis functions.   The first panel of Figure \ref{S5_Figure_qant} plots the near-losslessness parameters $\rho_0$ and $\overline{\rho}$ 
against the number of basis coefficients $K_\mathcal{C}$ in the reduced set.  Based on this, we selected the combined basis set $\mathcal{D}_\mathcal{C}$ for $\mathcal{C}=10$, which contained $K_\mathcal{C}=27$ basis functions and was near-lossless, with $\rho^0=0.990$ and $\overline{\rho}=0.998$.  We then orthogonalized, denoised, and re-standardized the resulting basis to yield the set of quantlets, the first 16 of which are plotted in Figure \ref{S5_QBE}.  As shown in panel 2 of Figure \ref{S5_Figure_qant}, these quantlets yielded a basis with similar sparsity property as principal components computed from the empirical quantile functions.

After computing the quantlet coefficients for each subject's empirical quantile function, we fit the quantlet-space version of model (\ref{eq:GBM}) as described above, obtaining $2,000$ posterior samples after a burn-in of $200$, after which the results were projected back to the original quantile space to yield posterior samples of the functional regression parameters in model (\ref{eq:GBM}).  MCMC convergence diagnostics were computed, and suggested that the chain mixed well (Supplementary Figure 17).  From these, we constructed $95\%$ point wise and joint credible bands for each $\beta_a(p)$ and computed the corresponding simultaneous band scores $P_{a,SimBaS}(p)$ and global Bayesian p-values $P_{a,Bayes}$ as described in Section \ref{sec:BayesInference}.

 \begin{figure}[!htb]
 \caption{Posterior inference for functional coefficients for T1-post contrast image:
for each covariate (6), the left panel includes posterior mean estimate, point and joint credible bands, GBPV in heading along with
SimBas less then $.05$ (orange line), 
 and the right panel includes predicted densities for the two levels of the covariate along with the posterior probability scores
   for the moment different testings.
     \label{S5_Figure_2}}
\centering
\includegraphics[height=5.3in,width=5.7in]{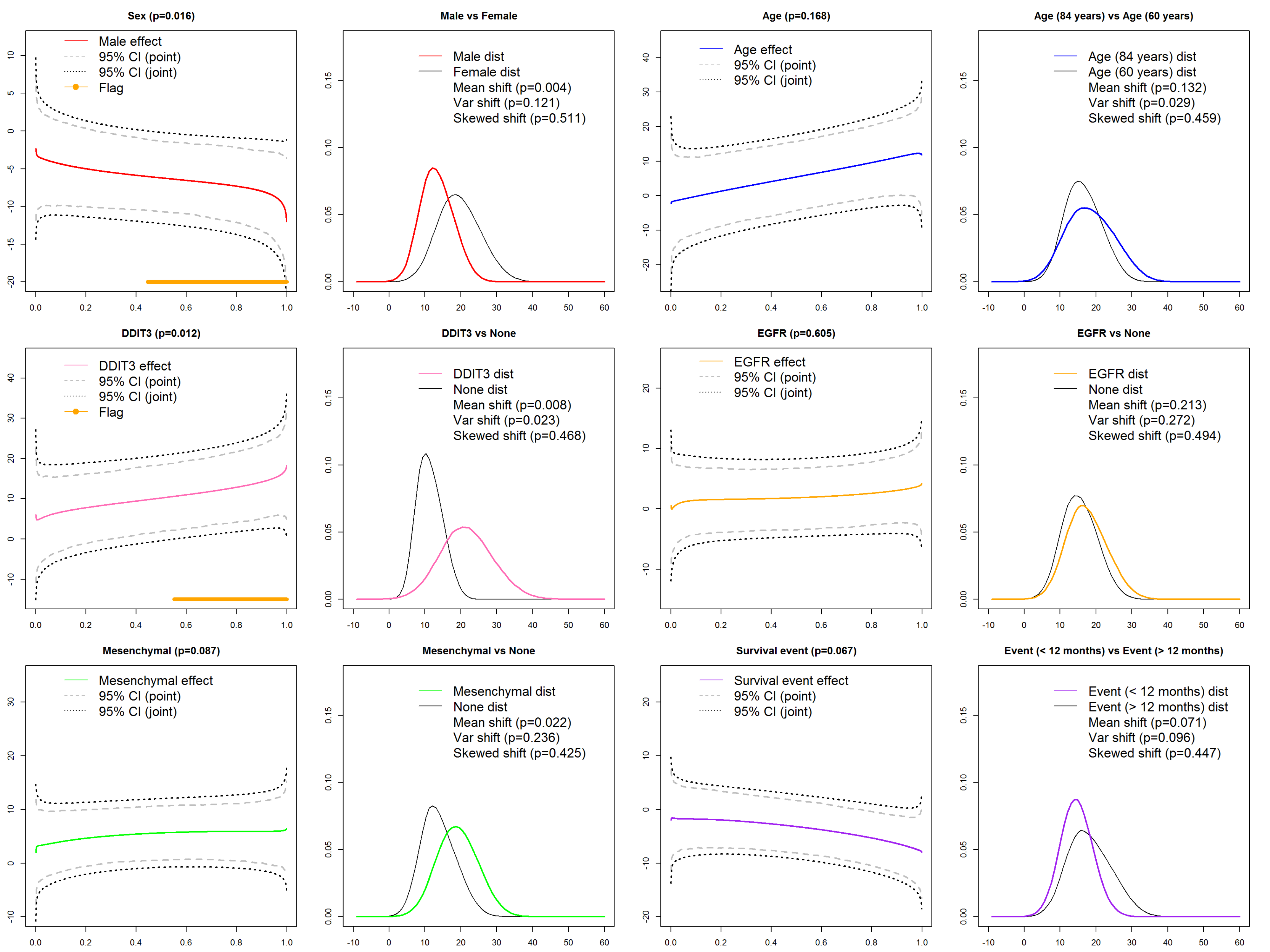}
\end{figure}

\textbf{Results:}. Figure  \ref{S5_Figure_2} summarizes the estimation and inference for each of the covariates in the model.  For each covariate there is one panel presenting the functional predictor $\beta_a(p)$ along with the point wise (grey) and joint (black) credible bands, and an indicator of which $p$ are flagged such that $\beta_a(p)\ne 0$ (orange lines indicating $P_{a,SimBaS}(p)<0.05$).   The other panel contains density estimates for each covariate level (holding all others at the mean), computed as outlined in the supplementary materials,  along with posterior probability scores summarizing whether the mean, variance or skewness appeared to differ across these groups.  Supplementary Table %
3
contains measures of the relative Gaussianness of the distributions for the various groups along with 95\% credible intervals.

The global Bayesian p-values for testing $\beta_a(p)\equiv0$ for each covariate are in the corresponding figure panel headers, and reveal that for sex (p=0.016) and DDIT3 (p=0.012), the functional covariates are flagged as significant, and for the mesenchymal subtype (p=0.087) and survival (p=0.067) endpoints, there was some indication of a possible trend.  We see that for sex, there was evidence of a mean shift (p=0.004) with females tending to have higher pixel intensities than males, especially in the upper tails of the distribution, and the female distribution appearing to be slightly more Gaussian than the males.  For DDIT3, we see evidence of a mean and variance shift, with tumors with DDIT3 mutation tending to have higher intensities and greater variability than those without, especially in the upper tail of the distribution.  The mesenchymal subtype, while not flagged as statistically significant in the global test, shows some tendency for a mean shift with the mesenchymal subtype tending to have higher distributional values and perhaps slightly more non-Gaussian characteristics.
 Follow-up studies can assess the significance of this upward shift in distribution of pixel intensities for female patients and DDIT3 mutated tumors.

 One cause of higher pixel intensities in MRI images of tumors is greater accumulation of fluid in body tissues, called edema, which can be an indicator of poor prognosis 
 \cite{zinn2011radiogenomic}.  Thus, it may be true that female patients and patients with DDIT3 mutations have tumors with greater edema, which is plausible given results in the literature showing DDIT3 mutation is associated with shorter survival time \cite{saha2016demarcate}.  
 Follow-up studies can assess the significance of this upward shift in distribution of pixel intensities 
 (or the extent of tumor vascularisation)
 for female patients and DDIT3 mutated tumors.
Since the gender has the specific effect on GBM 
\cite{colen2014glioblastoma}  and 
 DDIT3 also plays a key role in resistance to therapy, due to its hypoxia-related activity 
\cite{ragel2007identification},
 and 
 in GBM tumorogenesis \cite{ping2015identifying}, where
  DDIT3 is a p53 driven gene \cite{tivnan2016resistance}
suggesting that this radiographic observation might be associated with p53 associated cell death (showing as lower T2, FLAIR or T1c signal), 
 our findings have a strong connection with the results in the existing literature.
In addition, we notice that the longer survival time tends to be shifted to the right, representing  higher intensity
  and higher vascularisation.
  As pointed out by  \citeasnoun{gilbert2016antiangiogenic},
  one of the only effective therapeutic strategies for GBM is antiangiogeneic therapies, and it would make sense that patients with greater baseline vasculature would be more likely to respond to therapy, and thus experience improved survival times.  

\textbf{Sensitivity Analysis and Comparison:}  Our results are presented for $K=27$ basis functions, but to assess sensitivity to choice of K we also ran our model for a wide range of possible values of $K$, with Supplementary Table 5 showing global Bayesian p-values for the entire range
of potential values for K (from 546 to 2), along with run time.  The run time tracks linearly with $K$.  Note that we get the same
substantive results over the range of basis sizes, so results are quite robust to choice of number of quantlets.  However, keeping more quantlets than necessary clearly adds to the uncertainty of parameter estimates, as indicated by the larger joint band widths. Also, keeping too few basis functions can lead to some missed results and also wider joint band widths.  Moderate basis sets that are as parsimonious as possible while retaining the near-lossless property seem to give the tightest credible bands and thus the greatest power for global and local tests.
 We also performed a sensitivity analysis on the parameter $\nu_0$ (inverse gamma prior) indicating the prior strength for the variance components and found that results for slightly larger or smaller values yielded nearly identical results. 
We lastly conducted  a sensitivity analysis  for lasso 
to see how selection of more or fewer dictionary elements via larger or smaller lasso parameters effects the ultimate number of quantlets.
Choice of greater or fewer dictionary elements via larger or smaller lasso parameters still resulted in sparse sets of quantlet basis functions using the near-lossless criterion
as can be seen in Figures 23 in Supplementary material. 
 Also, from Figures 11, 21 and 22 in Supplementary material, we see that there are not dramatic changes on the final results.

 \begin{figure}[!htb]
 \caption{Comparison between quantlet and naive approaches for DDIT3 status for (A) quantlet approach with sparse regularization and (B) the naive \textit{one-p-at-a-time} quantile functional regression approach. 	
  \label{S5_Figure_2_g}}
\centering
 \includegraphics[height=3.1 in,width=5.7in]{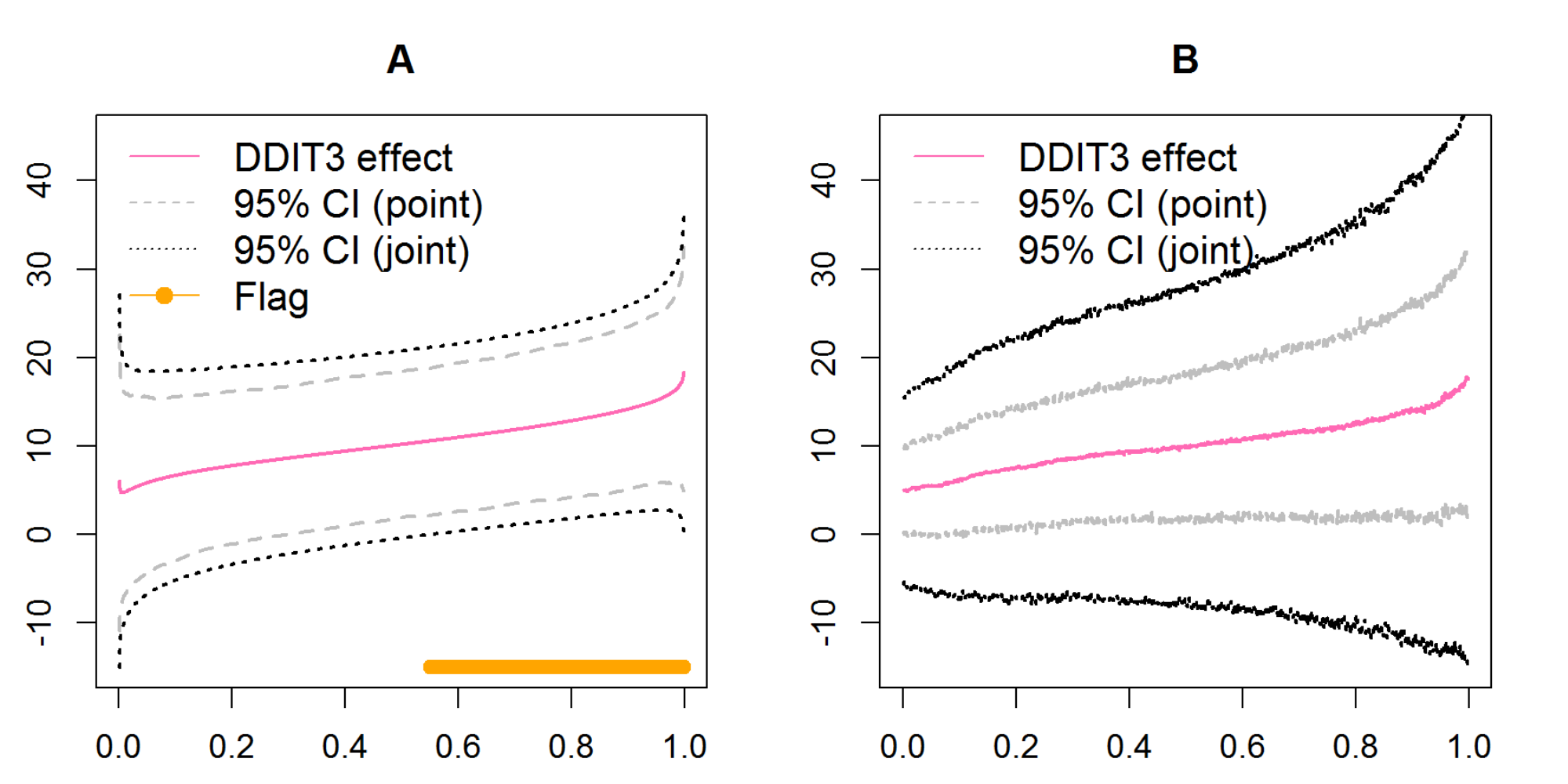}
\end{figure}
To compare different methods with our quantlet with sparse regularization approach, we also applied to these data a quantlet approach with no sparse regularization and 
a naive quantile functional regression method modeling independently for each $p$ (after interpolating onto a common grid). 
  Posterior mean estimates, credible intervals, and other inferential summaries are given in Supplementary Figure 11 and Table 6.  
    Note that the quantlets method with sparse regularization tends to yield estimates that are smoother and with tighter joint credible bands than either the naive or the quantlets-no sparse regularization runs.   As we can see in Figure~\ref{S5_Figure_2_g} the differences between the quantlet and naive methods are substantial, and demonstrate the significant power gained by borrowing strength across $p$ using the quantlet-based modeling approach.  The completely naive quantile functional regression approach gave nonsensical results for this application (Supplementary Figure 19).
    
     Supplementary Figure~18 contains the predicted quantiles functions over a grid of covariate combinations for this model.  Although the quantile functional regression using quantlets does not explicitly impose monotonocity in the predicted quantile functions, we see that the predicted quantile functions are all monotone non-decreasing. 
See Section 4 of the supplement for further details and discussion of monotonicity issues.

   \begin{table}
\caption{Posterior probability score of difference tests for the GBM data set: 
(B) naive quantile functional regression approach, 
(E) {\it quantlet} space with sparse regularization, and (G) feature extraction approach,
where the values in this table are the posterior probability scores derived by its corresponding method for each different test between treatment and reference groups in the top row.
      \label{S5_tab_pval} }
      \vspace{-0.1cm} 
\begin{center}
\resizebox{0.88\columnwidth}{!}{   
\begin{tabular}{c   |  c    c c | c c c | c c c} \hline
   $\text{{\bf Test}}$    
    &$$   &$\mu_{T}=\mu_{R}$      &$$ 
    &$$   &$\sigma_{T}=\sigma_{R}$     &$$   
    &$$    &$\xi_{T}=\xi_{R}$      &$$ 
    \\    \hline
   $\text{{\bf Method}}$     
    &$\text{{\bf B}}$   &$\text{{\bf E}}$      &$\text{{\bf G}}$ 
    &$\text{{\bf B}}$   &$\text{{\bf E}}$     &$\text{{\bf G}}$   
    &$\text{{\bf B}}$   &$\text{{\bf E}}$      &$\text{{\bf G}}$ 
    \\    \hline
 $\text{{\bf Sex}}$   &0.000     &0.004 &0.028
  &0.000     &0.121 &0.067
   &0.342     &0.511 &0.548
  \\
 $\text{{\bf Age}}$   &0.000     &0.132  &0.326
  &0.000    &0.029 &0.014
   &0.181     &0.459 &0.003
  \\
   $\text{{\bf DDIT3}}$   &0.000    &0.008  &0.020
  &0.000    &0.023 &0.046
   &0.347   &0.468 &0.442
  \\
   $\text{{\bf EGFR}}$   &0.000    &0.213  &0.470
  &0.000    &0.272 &0.391
   &0.365     &0.494 &0.470
  \\
   $\text{{\bf Mesenchymal}}$   &0.000     &0.022 &0.043
  &0.000    &0.236 &0.458
   &0.071     &0.425 &0.189
  \\
     ${\bf Survival_{12}}$   &0.000   &0.071 &0.160
  &0.000    &0.096 &0.034
   &0.309    &0.447 &0.941
  \\  \hline
\end{tabular}
}
\end{center}
\end{table}      



    Table \ref{S5_tab_pval} contains posterior probability scores assessing differences in moments for these three methods, plus a feature extraction approach in which moments were first calculated from each subject's samples and then statistically compared with a Bayesian regression fit.
As in the simulations, we see that the naive quantile functional regression method appears to have type I error problems in the mean and variance.  
While our method (E) does not yield additional power when the distributional differences are captured by the moments (G), our approach does not give much power away in these settings and yet can detect distributional differences that are not contained in the moments, e.g. differences in specific extreme quantiles. 
      Specifically, by the estimation and inference of our method, we can thoroughly understand the pixel intensity distribution for each of the covariates. For instance, for the male, (E) provide the insight that males have lower pixel intensities than females because the male effect, $B(p)$ along with the point wise and joint credible band has a decreasing tendency from the first panel of Figure~\ref{S5_Figure_2}. 

\section{{\bf Discussion }}

In this paper, motivated by a clinical imaging application in cancer, we have introduced a strategy for regressing the distribution of repeated samples for a subject on a set of covariates through a model we call \textit{quantile functional regression}.  We distinguish this model from other types of quantile regression and functional regression  methods in existing literature, in that it is regressing the \textit{subject-specific} quantile, not the \textit{population-level} quantile, on covariates, and accounts for intrasubject correlation.  We describe how it serves as a middle ground between two commonly-used strategies of (1) performing a series of regressions on arbitrary summaries of the distribution such as mean or standard deviation and (2) independent regression models for each quantile $p$ in a chosen set.  Our approach models a subject's entire quantile function as a functional response, building in dependency across $p$ in the mean and covariance using custom basis functions called \textit{quantlets} that are empirically defined, near-lossless, regularized, sparse, and with some of the individual bases being interpretable.  These basis functions have sparsity properties similar to principal components, but appear more regular and interpretable.  They provide a flexible representation of the underlying quantile functions while containing a sufficient Gaussian basis as a subspace.  The {\it quantlets} basis function that is successfully utilized to capture distinct characteristics of the quantile function, consisting of the subspace spanned by the normal quantile and the subspace spanned by the mixture beta distributions.   
These quantlets are constructed based on a dictionary of Beta CDF, which can be shown to be sufficient for representing any quantile function with a first derivative that is uniformly continuous, and has numerous useful statistical properties including near-lossless representation, sparsity, regularity, and some interpretablity.

We fit the quantile functional  regression model using a Bayesian approach with sparse regularization priors on the quantlet space regression coefficients that smooths the regression coefficients and yields a broad array of Bayesian inferential summaries computable from the posterior samples of the MCMC procedure.  
For example, we can construct global tests of signficance for each covariate using global Bayesian p-values, and then characterize these differences by flagging regions of $p$ while adjusting for multiple testing, and obtaining probability scores for any moments or other summaries of the distributions.   

In this paper, we have presented the quantile functional regression framework using a standard linear model with scalar covariates and independent Gaussian residual error functions, but as in other functional regression contexts the model can be extended to include other complex structures that extend the usability of the modeling framework.  This includes functional covariates, nonparametric effects in the covariates $x_{ia}$, random effects and/or spatially/temporally correlated residual errors to accommodate correlation between subjects induced by the experimental design, and the ability to perform robust quantile functional regression to downweight outlying samples using heavier-tailed likelihoods. These types of flexible modeling components are available as part of the Bayesian functional mixed model (BayesFMM) framework that has been developed in recent years
\cite{morris2006wavelet,zhu2011robust,zhu2012robust,meyer2015bayesian,zhang2016functional,Zhu2017robustf,lee2016semiparametric}.
By linking the software developed here to generate the quantlets and fit quantile functional regression models with the BayesFMM software, it will be possible to extend the quantile functional regression framework to these settings and thus analyze an even broader array of complex data sets generated by modern research tools.

Our approach has been designed with relatively high dimensional data in mind, i.e. data for which there are at least 
a moderately large number of observations per subject (at least 50 or 100). 
  We are currently working on extensions of this method to handle lower dimensional data with fewer observations per subject,
   which requires a careful propagation of uncertainty in the estimators of the empirical quantile functions into the quantile functional regression.  This propagation of uncertainty could also be done in larger sample cases like the one presented here, but given the substantial complexity and length already in this paper we leave this for future work.  
As mentioned in Section 5, we are currently working on extensions of this method to handle the empirical quantile estimator established by fewer/massive observations because of the different tumor size and the imperfection of the image segmentation per subject and leave this for future work in this paper

Also, in settings with enormous numbers of observations per subjects, e.g. millions to billions or more, the procedure described in this paper to construct the quantlets basis would be too computationally burdensome.  Given that in those settings, it is unlikely that so many observations are needed to quantify the subject-specific quantile function, we have worked out algorithms to down-sample the empirical quantile functions in these cases in a way that engenders computational feasibility but is still near-lossless.  This also will be reported in future work.   Other data have measurements on many 1000s to 100,000s of subjects, which can be accommodated by computational adjustments of the procedure reported herein, but again we leave this for future work.   In this paper, we focused on absolutely continuous random variables that have no jumps in the quantile functions.  It is also possible to adapt our quantlet construction procedure to allow jumps at a discrete set of values, thus accommodating discrete valued random variables, but again this extension will be left for future work.



\newpage

\appendix

\makeatletter   
 \renewcommand{\@seccntformat}[1]{APPENDIX~{\csname the#1\endcsname}.\hspace*{1em}}
 \makeatother

\section{\bf{Details of Estimation}}

\subsection{\bf{Details of Denoising}}

In practice, we have observed that the first number of orthogonal basis functions are relatively smooth, but the later basis functions can be quite noisy, sometimes with high-frequency oscillations.  As we do not believe these oscillations capture meaningful features of the empirical quantile functions, we regularize the orthogonal basis functions using wavelet denoising to adaptively remove these oscillations.

Given a choice of mother wavelet function $\varphi(p)$, wavelets are formulated by the operations of dilation and translation given by  
  $$\varphi_{j,l}(p)=2^{j/2}\varphi(2^j p - l)$$ with integers $j, l$ indicating scale and location, respectively.
We can decompose any arbitrary function $\psi^\perp_k(p) \in L^2(\Pi(\mathcal{P}))$ into the generalized Fourier series as 
 \begin{equation}
 \psi^\perp_k(p) = \sum_{j= -\infty}^{\infty} \sum_{l= -\infty}^{\infty} d_{k,j,l}\varphi_{j,l}(p),
  \end{equation}
where $d_{k,j,l}=\int \psi^\perp_k(p)\varphi_{j,l}(p) dp=\langle \psi^\perp_k, \varphi_{j,l} \rangle$ are the wavelet coefficients corresponding to $\psi^\perp_k$.
Wavelet coefficient $d_{k,j,l}$ describes features of the function $\psi^\perp_k$ at the spatial locations indexed by $l$ and scales indexed by $j$.
 A fast algorithm, the discrete wavelet transform (DWT), can be used to compute these wavelet coefficients in linear time for data sampled on an
 equally spaced grid whose size $L$ is a power of two, yielding a set of $L$ wavelet coefficients, with $L_j$ wavelet coefficients at each of $J$ wavelet scales and $L_0$
 scaling coefficients at the lowest scale. 
 We apply this wavelet transform to the the basis functions $\psi^\perp_k(p)$ sampled on an equally-spaced fine
 grid on $p$, for example using a grid of size $L=2^{10}=1024$ for our GBM data.

  Functions can be adaptively denoised by shrinking these wavelet coefficients nonlinearly towards zero \cite{donoho1995wavelet}.  Various shrinkage/thresholding rules can be used to accomplish this, 
such as hard thresholding with a threshold of $\sigma \sqrt{2 \log L}$ introduced by \citeasnoun{donoho1995wavelet}, which yields a risk within a log factor of 
the ideal risk.   In that case, the wavelet shrunken and denoised basis function $\psi^{\dagger}_k(p)$ can be constructed as
 \begin{equation}
\psi^{\dagger}_k(p) = \sum_{j= 0}^J \sum_{l= 1}^{L_j} d_{k,j,l}^{\dagger}\varphi_{j,l}(p),
  \end{equation}
such that  $d_{k,j,l}^{\dagger}=d_{k,j,l}$ if $|d_{k,j,l}|>\sigma \sqrt{2 \log L}$ and $d_{k,j,l}^{\dagger}=0$ If $|d_{k,j,l}|\le \sigma \sqrt{2 \log L}$.
When $\sigma$ is unknown, it is often replaced by an empirical estimator that is the median absolute deviation of the wavelet coefficients at the highest frequency level $J$.

\subsection{\bf{Details of Predicted PDF}}

 If desired, one can construct the estimate of the conditional probability density function
  given covariates $\bX$ from 
   \begin{align*}
\hat{ f}(x| \bX)  =  M^{-1} \sum_{m=1}^{M}
\delta / \biggr (  \bX^{T}\hat\beta^{(m)}(p)  -  \bX^{T}\hat\beta^{(m)}(p-\delta)\biggl ) , ~~~   
  \end{align*}
  where $\delta$ is a fixed positive constant 
  and $x=\inf{(y: y \ge  \bX^{T}\hat\beta^{(m)}(p)     )}$. 
 Note that the above formula is derived from 
  $f(Q(p))) d Q(p)/dp =1$ 
  by changing the variable. 
Remark that the conditional quantile function $X_i^{T}\hat\beta^{(m)}(p)$ for some samples may not 
   enforce strict monotonicity, which leads to the negativity density value for its computation. However, when we take the coarse grid points for $x$  with the sufficient gap between any two adjacent points,  equivalently
   $\delta$ to be the large value,
    which  would allow 
   $X_i^{T}\hat\beta^{(m)}(p)  -  X_i^{T}\hat\beta^{(m)}(p-\delta)>0$, the valid probability density function
  is available. In practice, we use 
  $\max{(0,  X_i^{T}\hat\beta^{(m)}(p)  -  X_i^{T}\hat\beta^{(m)}(p-\delta) )}$
as the differential for $Q(p)$ for an arbitrary $\delta$.

\section{ \bf{Other results from Simulation }}

\subsection{ \bf{Simulation for Basis Representation}} \label{sim1}

\begin{figure}[!htb]
\caption{ Density functions in simulation 1: panel (A) contains
normal (skyblue), $t_{(1)}$ (black), shifted gamma (3,1) (purple), 
and dirichlet (gray dot) and panel (B) contains
 mixtures of $SN(-3.06,3.67,0)$ and $SN(9.11,7.89, -4)$  with $0.5$ and $0.5$ probabilities (black), 
$SN(-7.1,2.4,0)$ and $SN(-3.11,7.89, 4)$  with $0.3$ and $0.7$ probabilities (red),
 $N(-2.5,2.5)$, $N(4, 3)$ and $N(9.5, 2.1)$ 
and  $N(-2.5,1.5)$, $N(4, 3.56)$ and $N(9.5, 1.1)$  
  with $0.3$, $0.5$ and $0.2$ probabilities (blue and green), denoted by $E$, $F$, $G$, and $H$, respectively.
  \label{S1_Figure_Exs}}
\centering
\includegraphics[height=2.5in,width=5.0in]{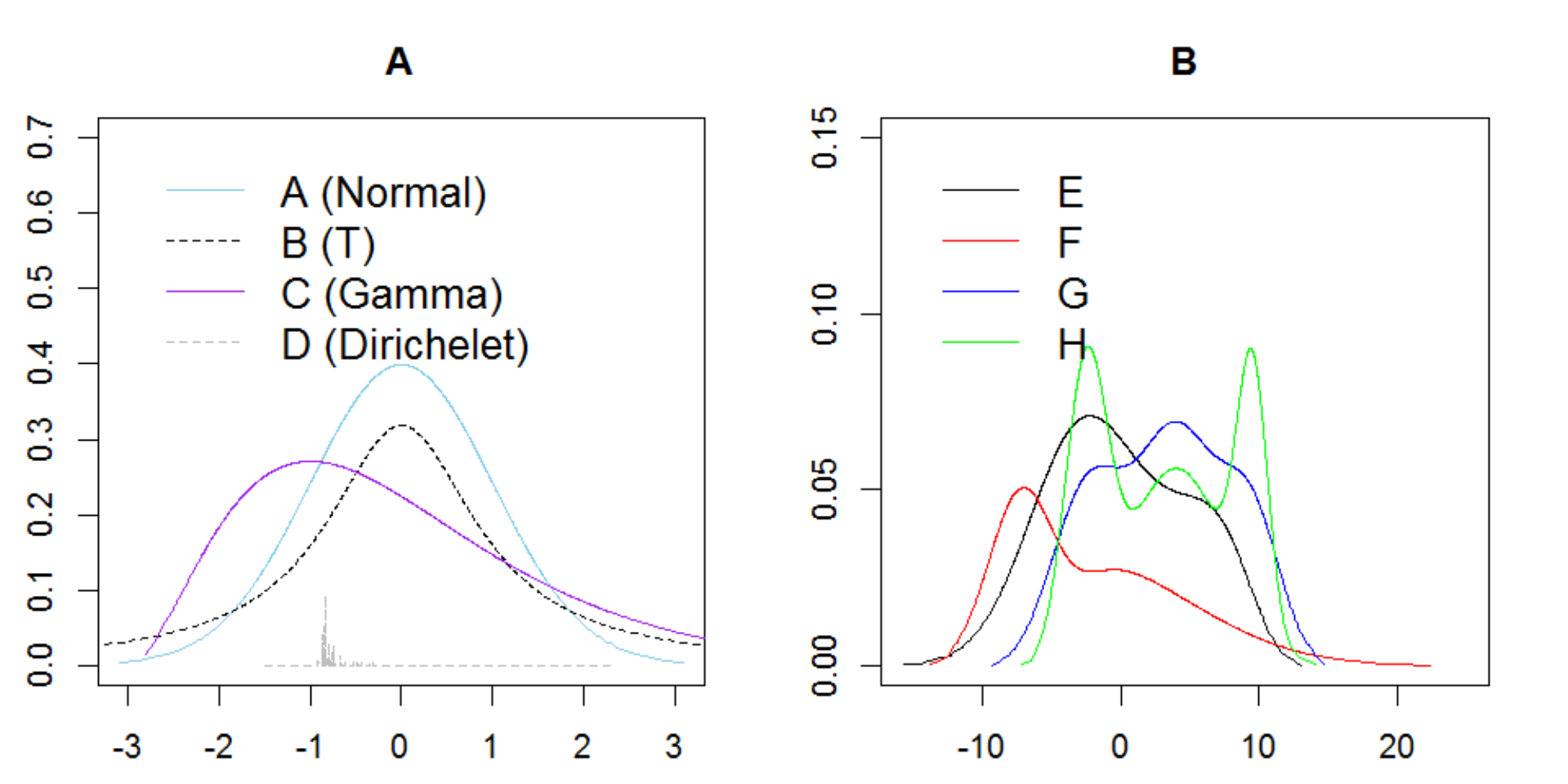}
\end{figure}

 We let $N$, $T$, $G$, $D$, $M_1$, $M_2$, $M_3$, and $M_4$
 be random variables from the standard normal, student-$t$, gamma, dirichlet distribution, and  
 mixture distributions, respectively.
We first generated four profiles of quantile functions, $Q_{N}(p)$, $Q_{T}(p)$, $Q_{G}(p)$, 
$Q_{D}(p)$, $Q_{M_1}(p)$, $Q_{M_2}(p)$, $Q_{M_3}(p)$,
and $Q_{M_4}(p)$ 
defined on $\mathcal{P} = [\delta, 1- \delta]$ and 
 $J$ fixed grid points
 $\{p_1, \dots, p_J \} \in \mathcal{P}$, 
where 
$\delta$ and $J$ were set to be $1/1000$ and $999$, respectively.  
Specifically,
we independently simulated each $Q_{(\cdot)}(p)$ according to the generating process  
 \begin{align}
Q_{(\cdot)}(p) =  \inf{(y: F_{(\cdot)}(y) \ge p     )},
 \label{s1_Q} 
 \end{align}
 following the normal distribution $N(0,1)$, student $t$ distribution with one degree of freedom,    
shifted gamma distribution with shape $3$ and scale $1$ to $-3$,   
dirichlet distribution with base measure to be the kernel density estimator of the first observation in GBM data, 
two mixture skewed-normal distributions for which the mixture components were 
$SN(-3.06,3.67,0)$ and $SN(9.11,7.89, -4)$  with $0.5$ and $0.5$  probabilities for one while
$SN(-7.1,2.4,0)$ and $SN(-3.11,7.89, 4)$  with $0.3$ and $0.7$  probabilities for the other,
and 
two mixture normal distributions for which the mixture components were $N(-2.5,2.5)$, $N(4, 3)$ and $N(9.5, 2.1)$ for one while
$N(-2.5,1.5)$, $N(4, 3.56)$ and $N(9.5, 1.1)$ for the other with $0.3$, $0.5$ and $0.2$ probabilities, respectively.
To illustrate, two panels of Figure \ref{S1_Figure_Exs} show eight probability density trajectories for $F_{N}(y)$, $F_{T}(y)$, $F_{G}(y)$, $F_{D}(y)$,
$F_{M_1}(y)$, $F_{M_2}(y)$, $F_{M_3}(y)$, and $F_{M_4}(y)$.
We see that each distribution has different characteristic in that 
 $F_{T}(y)$ has heavy tails,  $F_{G}(y)$ is skewed to the right,
 $F_{D}(y)$ has high frequency, 
   and $F_{M_k}(y)$ has multiple peaks for $k=1, \dots 4$,
 compared to  $F_{N}(y)$, which is standard in this scenario.

 We constructed a {\it quantlets} representation for each distribution as follows. 
We generated the parameter space as a set of a sequence pairs, $\Theta=\{ \theta_k= (a_k,b_k)  \}_{k=1}^{11,881}$, uniformly sampled on $[0.1,1000]^{2}$ and     
an {\it overcomplete dictionary}, $\mathcal{D}^{O} =\{  \xi_k:    \theta_k  \in \Theta \}$, where $\xi_2$ (Gaussian) is not included in $\mathcal{D}^{O}$ to allow for a fair comparison in this scenario so
that $P_{N^{\perp}}$ is the  identity operator as the orthogonal complement to the empty space.  
 We restrict the maximum value of the parameter space as the total number of the probability gird points $J$
 and the minimum value of the probability grid points $p_1$. 
   We prefer to have a large number of $K^{O}$ but restrict its minimum as $K^{O}> J$.
 This setting is motivated by 
  the structure of the random Bernstein polynomials \cite{petrone1999random,bornkamp2009bayesian}.
We used the Lasso method to find the individual {\it dictionary}, $\mathcal{D}_i$ and used it
as {\it quantlets} from $\mathcal{D}^{O}$  for each distribution case.
Note that we did not need to find $\mathcal{D}^{U}$ or $\mathcal{D}^{C}$  because there was only a single observation for each distribution case.

We compared our {\it quantlets} with other competing representations such as 
 B-spline \cite{schumaker2007spline}, 
 integrated spline \cite{ramsay1988monotone} denoted by I-spline and  convex spline \cite{meyer2008inference} denoted by C-spline basis representations.
In Figure \ref{S1_basis}, we see the shape of I-spline and C-spline basis functions.
To compare the different methods, we examined  one type of performance measure for prediction accuracy
and computed the empirical mean integrated squared error of the test data set as 
$n^{-1}\sum_{i=1}^n( Q_{(\cdot)}(p_i)^{test}-\widehat {Q}_{(\cdot)}(p_i)^{test} )^2$, where 
$\widehat {Q}_{(\cdot)}(p_i)$ is the predicted basis representation 
built from the aforementioned training set of 300 grid points 
and $( Q_{(\cdot)}(p_1)^{test},\dots, Q_{(\cdot)}(p_n)^{test})$s' are observations in the test set of 999 grid points for each distribution case. We used a B-spline basis of order 4, degree 3 on 10 knots uniformly spaced in  
$\mathcal{P}$ to generate I-spline and C-spline in this simulation.
Also, to investigate the degree of monotonicity, we compute the degree of $\epsilon$-monotonicity, defined to be $P^\mathcal{M}_\epsilon =\int_0^1 I[\widehat{Q}_{(\cdot)(p)}-\text{max}_{p'<p}\{\widehat{Q}_{(\cdot)}(p)\}>\epsilon] dp$ for some $\epsilon$ considered negligibly small in the context of the scale of $Y$ in the current data set. 
 When $P^\mathcal{M}_\epsilon \approx 0$, it says a strong monotonicity and we use the 
  monotonicity measure $1-P^\mathcal{M}_{0.01}$ for each distribution.

\begin{figure}[!htb]
 \caption{\small
  Basis representations based on different methods:
representations for {\it quantlets} (red), B-spine (dashed)
 I-spline (dot) and C-spline (dashed-dot) 
are given for the following quantiles, where the gray dot line is the true quantile function in each panel: 
(A) normal, (B)  t, (C) gamma and (D) dirichlet,
(E) mixture of $SN(-3.06,3.67,0)$ and $SN(9.11,7.89, -4)$  with $0.5$ probability, 
(F) mixture of $SN(-7.1,2.4,0)$ and $SN(-3.11,7.89, 4)$  with $0.3$ and $0.7$ probabilities,
(G) mixture of $N(-2.5,2.5)$, $N(4, 3)$ and $N(9.5, 2.1)$
and (H) mixture of $N(-2.5,1.5)$, $N(4, 3.56)$ and $N(9.5, 1.1)$  
  with $0.3$, $0.5$ and $0.2$ probabilities, respectively.
  \label{S1_Figure_Fit}}
\centering
\includegraphics[height=5.2in,width=6.4in]{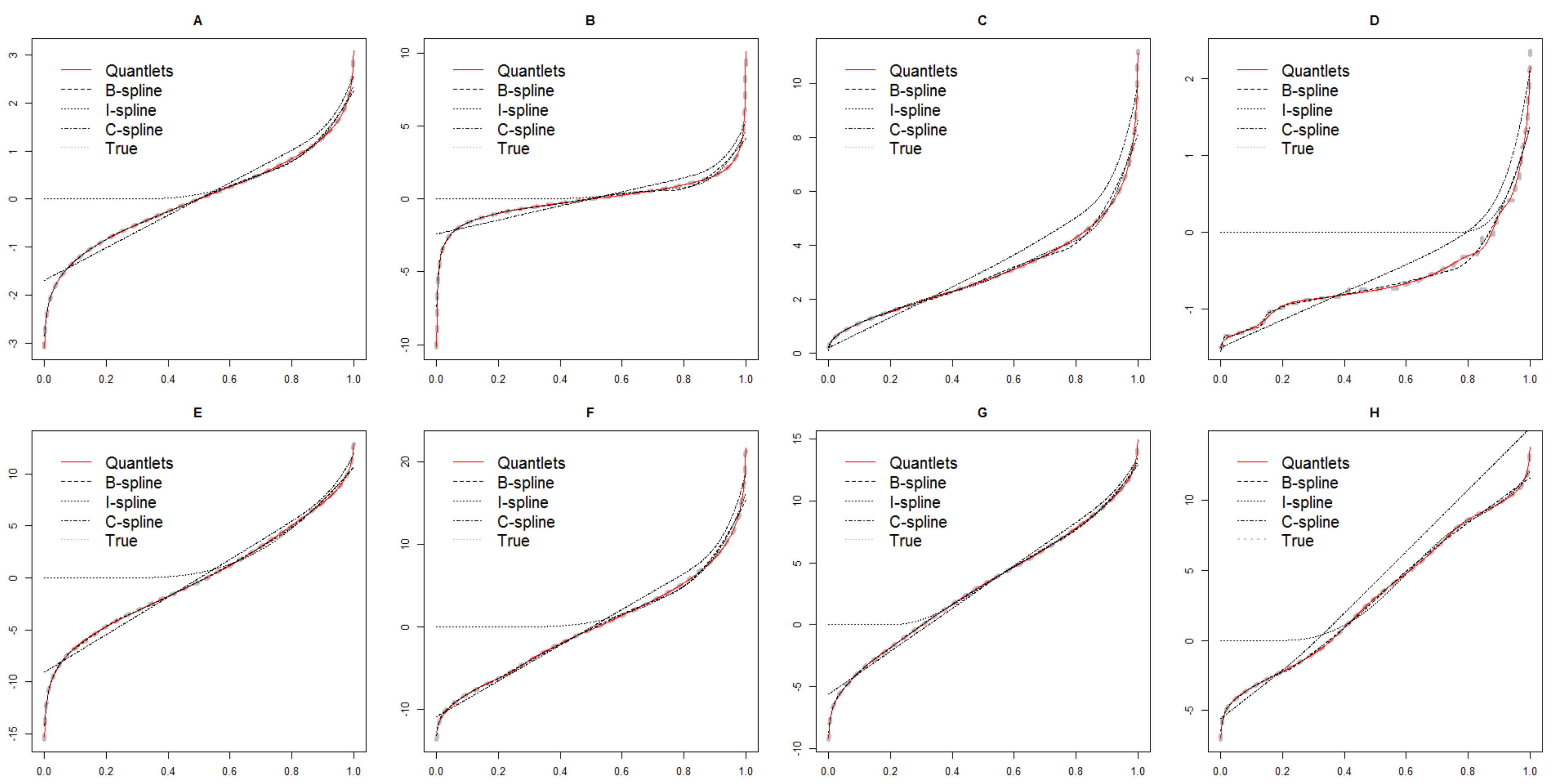}
\vspace{0.5cm} 
\end{figure}

\begin{figure}[!htb]
 \caption{\small I-spline and C-spline basis functions: (A) I-spline and (B) C-spline.
  \label{S1_basis}}
\centering
\includegraphics[height=2.5in,width=5.0in]{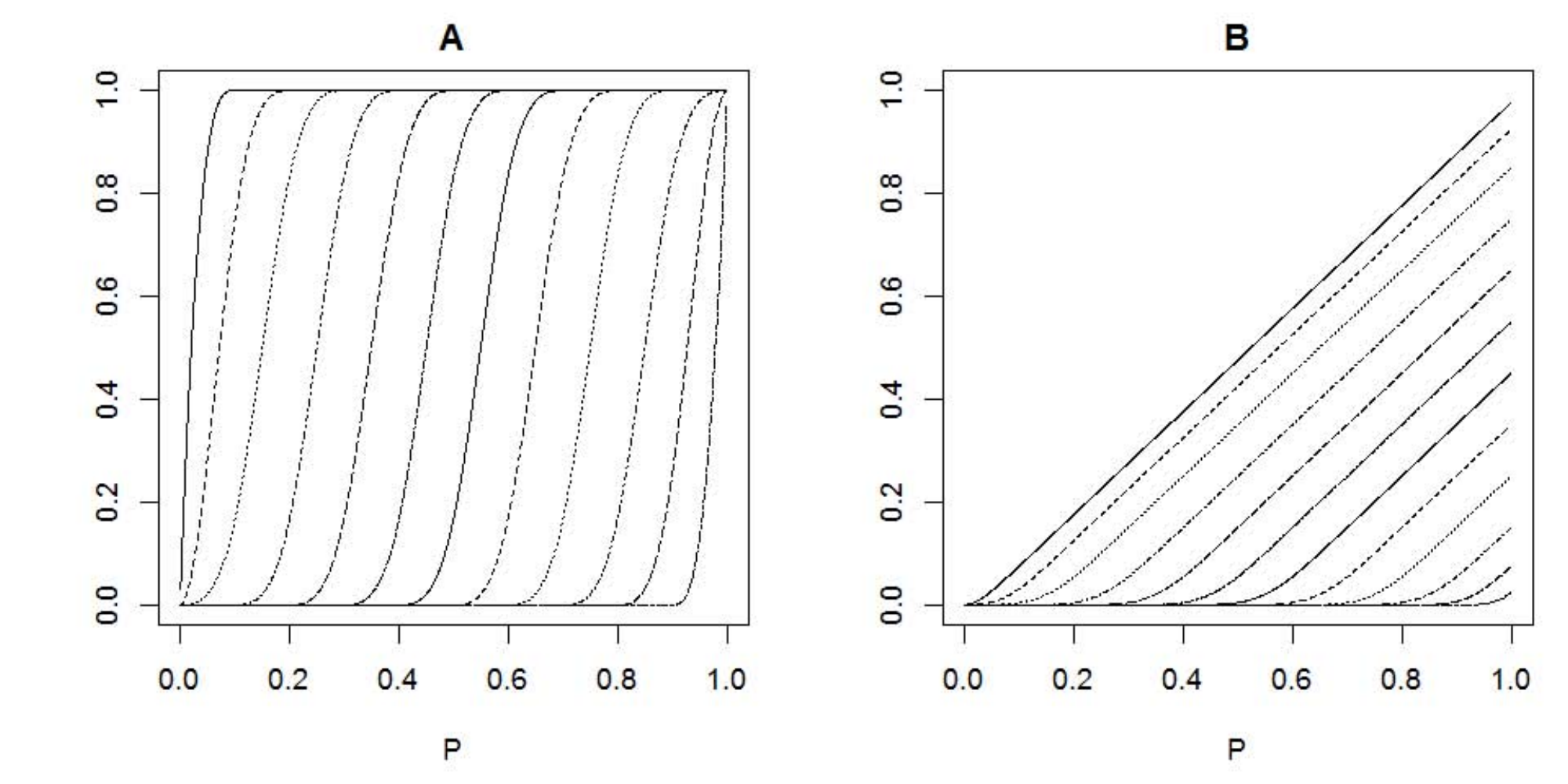}
\end{figure}

  Table \ref{Basis_table} presents
the empirical mean integrated squared error  for each method  calculated 
and $\epsilon$-monotonicity
  from the test data sets
of 999 grid points on $\mathcal{P}$.  
 We see that our method
 significantly outperformed the competing methods 
for all distribution cases.
Although B-spline method showed lower IMSE compared to other competing methods,
it does not show the monotone property in that it shows wiggly fitted regions near the curvature points. Hence,
 we conclude that our method is better than the existing spline approaches. 
Figure \ref{S1_Figure_Fit}
plots the true quantile functions (gray dot line) along with the fits for quantlets (red), B-spline (dashed),
 I-spline (dot) and C-spline (dashed-dot) and shows that the quantlets can be the best representation for the shapes of these quantile functions compared to all others.

\subsection{ \bf{Other Additional Results from Simulations}} \label{add}

There are additional results for the simulation conducted in the main paper.
We ran the MCMC algorithm  for $2,000$ iterations, keeping every one after a burn-in of $200$
and then, transformed all the estimates in  the {\it quantlets} space into the {\it data} space.
The results are shown in Figure \ref{S2_Figure_2}
based on each method: 
(A) naive quantile regression method (separate classical quantile regressions for each $p$ 
 by using  \textit{rq} function in {\it quantreg} R package \cite{koenker2005quantile}),
 (B) naive quantile functional regression approach (separate functional regressions for each subject-specific quantile $p$),
    (C) principal components method (quantile functional regression using PCs as basis functions),
(D) {\it quantlet} without sparse regularization, (E) {\it quantlet} with sparse regularization, and 
(F) Gaussian model (quantlet approach but keeping only the first two coefficients).
Compared to the other methods,
 our method (E) provided  smoother estimators and tight $95\%$ joint confidence intervals for all the parameters. 
Figure   \ref{S2_Figure_3} depicts the simultaneous band scores $P_{SimBaS}(p)$ for the two contrast functions associated with the scale effect $\beta_3(p)$ and skewness effect
$\beta_4(p)$, with regions of $p$.
  Since the two true contrast functions  $\beta_3(p)$ and $\beta_4(p)$ have one and two zero points while
  the null hypothesis is $\beta_a(p)=0$,
respectively, as shown in Figure   \ref{S2_Figure_2}, 
those points need to be detected with the higher $P_{SimBaS}(p)$ at their zero points (not reject the null hypothesis). Compared to the other methods, our method (E) showed lower type II error at the level of significance, $\alpha=0.05$ (solid black line).

Table \ref{S2_Stat} includes  true conditional moment parameters such as the mean, standard deviation and skewness, and 
corresponding point and interval estimators for the four groups derived from the closed form of the formulas in (see Section 2.7) for each method. 
Although the performance of the point estimators seems to be similar for all cases,   
the performance of the interval estimators is clearly better when using the {\it quantlets} basis approaches compared to the naive approach because 
those intervals contain true parameters for the four groups.
Summarizing Gaussianity score for the four groups, assessed by the relative energy,
is reported in Table \ref{S2_Norm}.
We see that the first three groups not involved with the skew parameter $\alpha$ 
 can be explained as Gaussianity with the higher score whereas
 the  fourth group, which is involved with the skew parameter, is hard to explain by Gaussianity,
where values for the $95\%$ confidence intervals are reported in parentheses. 
 Figure~\ref{S2_QBE} reveals the quantlets basis functions in the simulation.

Because our method  involves a lot of computational burden, we investigated  
  the computational aspect of our method.
  Figure~\ref{runtime} depicts the run time for computing the basis set as the function of
 sample size ($N$) and probability grid size ($m_i$) from the simulated data in the multi modality scenario.
We see that for the grid size, $m_i\le 2^8$, the number of the subjects does not yield the heavy computation.
However, for the grid size, $m_i\ge 2^{10}$, the amount of the computations is dramatically increased as the number of the subject increase. Hence, we recommend to use the smaller number of the probability grids for the data set with the large number of the observations.

     \section{ {\bf Other results from application} } 
  
\begin{figure}[!htb]
  \caption{Quantlets basis functions in simulation.
  \label{S2_QBE}}
\centering
\includegraphics[height=5.3in,width=5.7in]{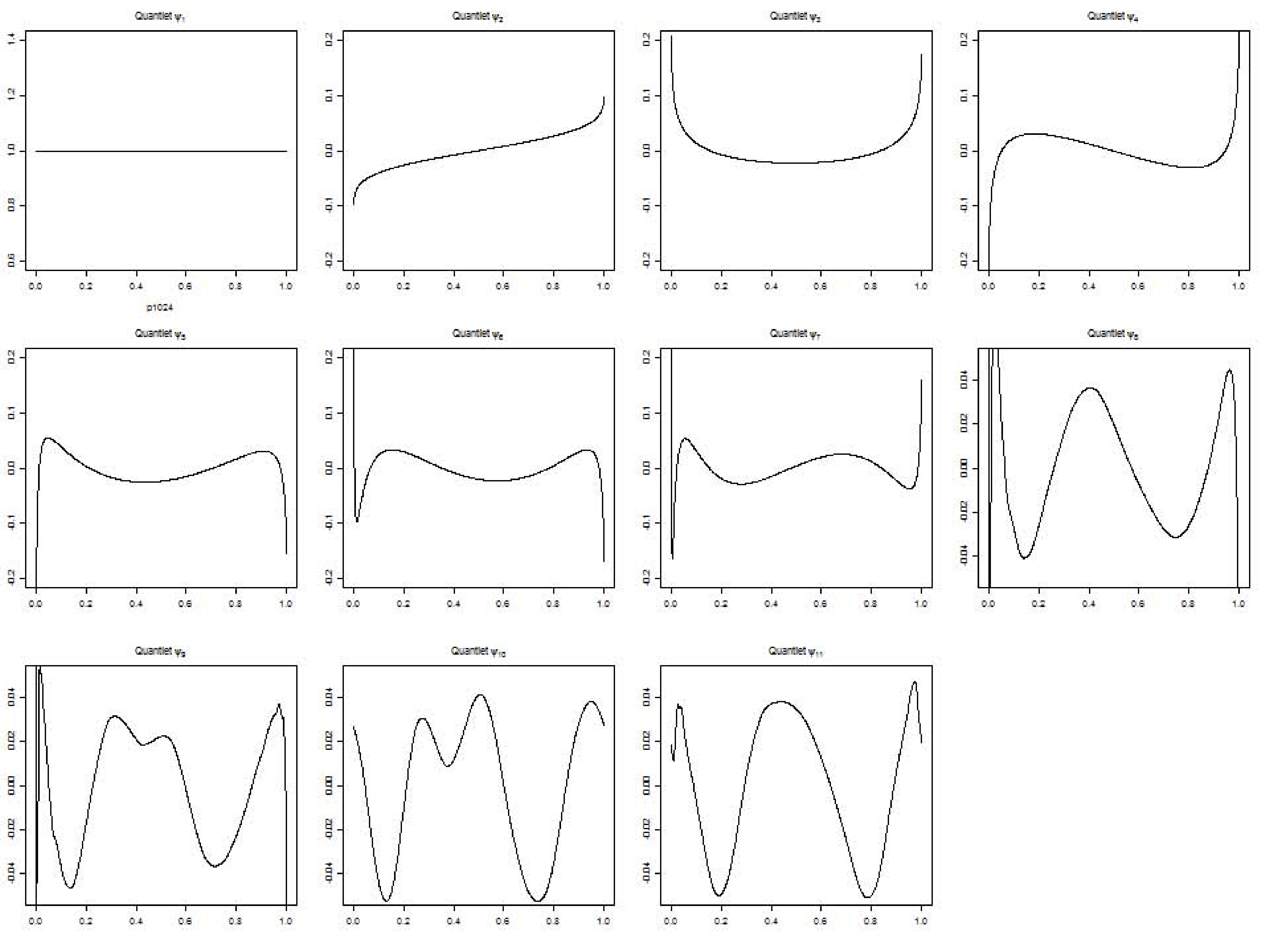}
\end{figure}

\begin{figure}[!htb]
 \caption{Principal Component basis functions in GBM application.
  \label{S5_PCBE}}
\centering
\includegraphics[height=5.3in,width=5.7in]{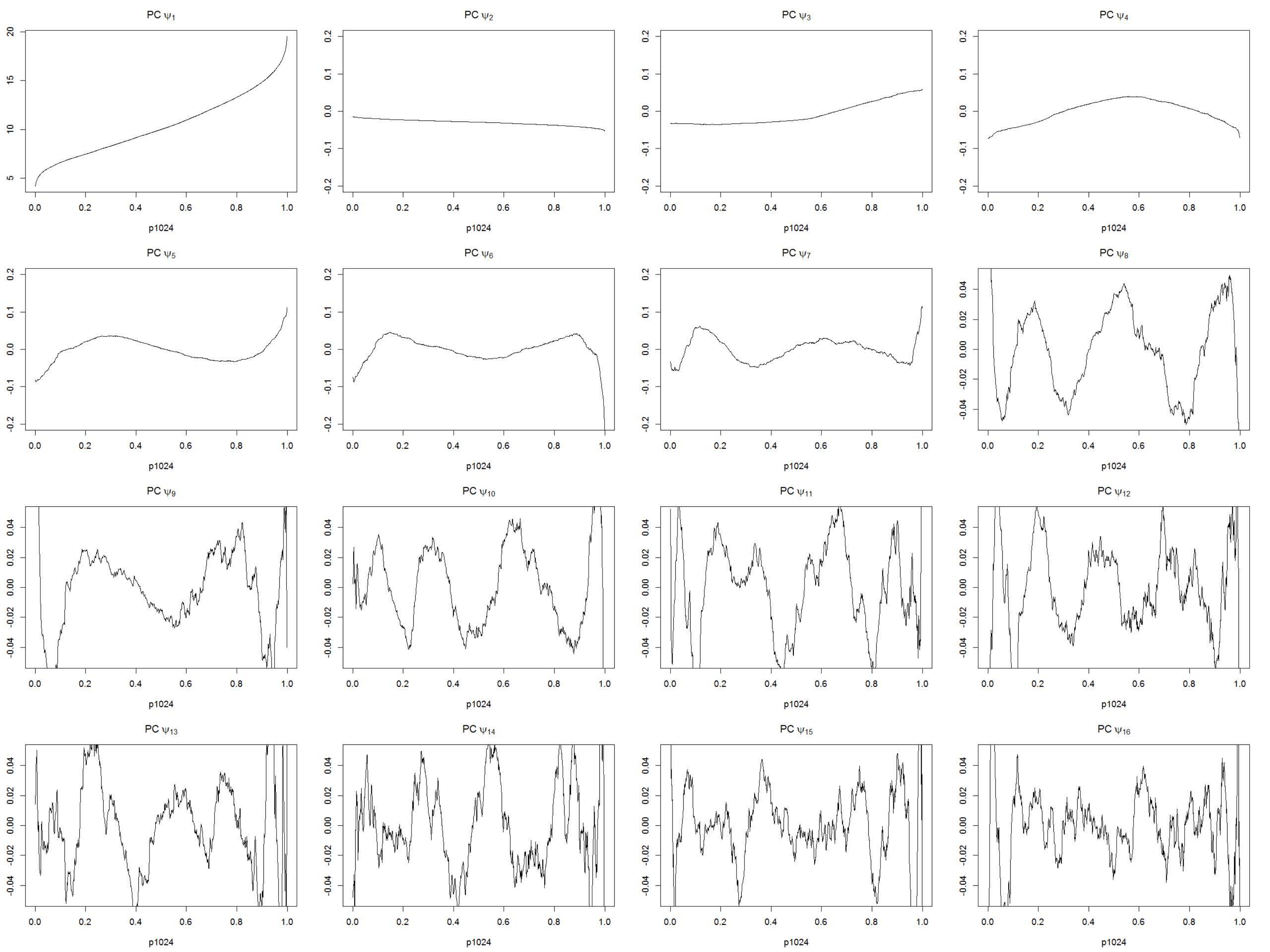}
\end{figure}

\begin{figure}[!htb]
 \caption{Wavelet denoising for 16 basis functions in GBM application: orthogonal basis (black) and
wavelet denoising basis (blue). 
  \label{S5_QBE_0}}
\centering
\includegraphics[height=5.3in,width=5.7in]{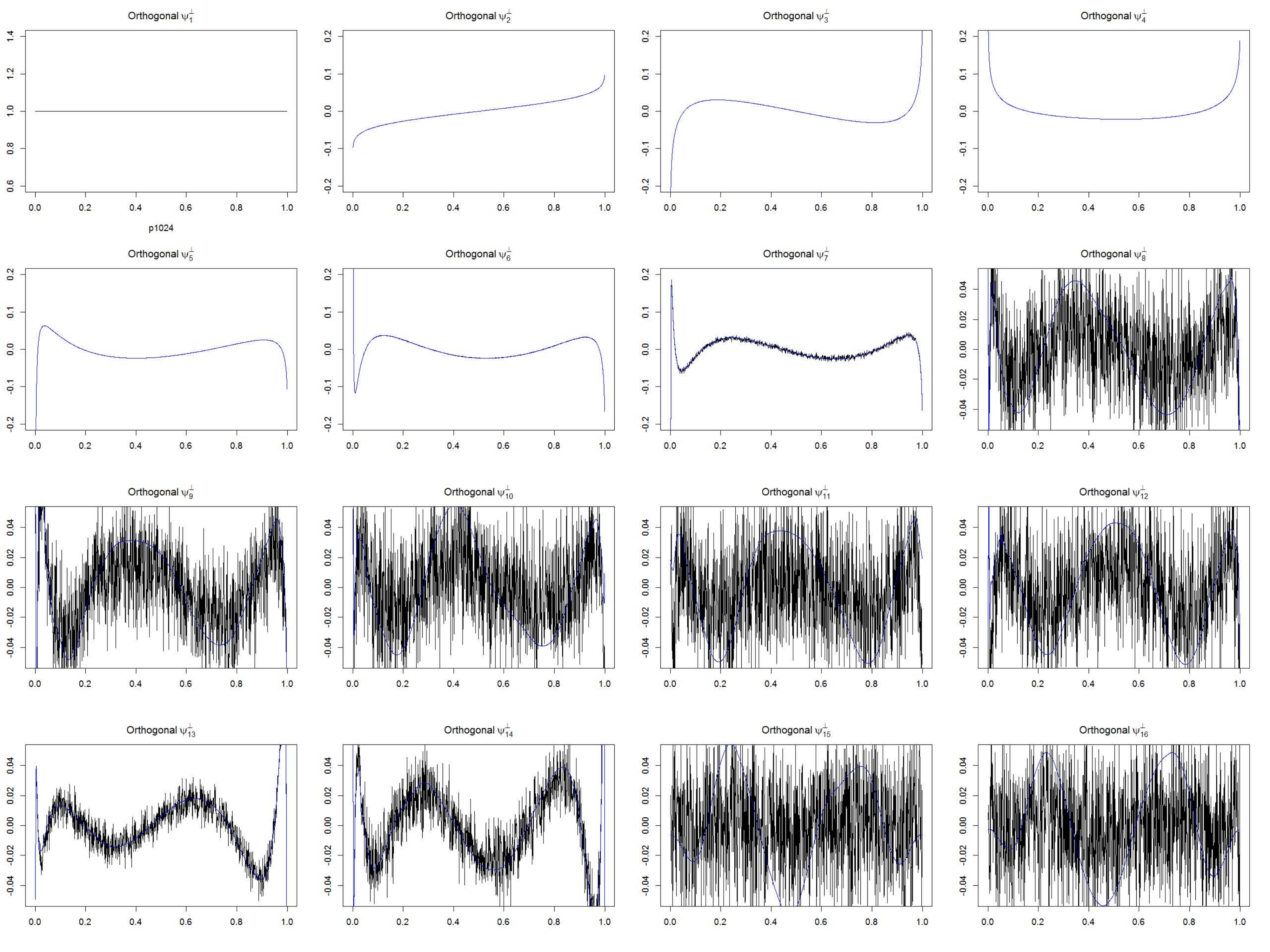}
\end{figure}

  There are additional results for the GBM study conducted in the main paper.
   Compared to principal components basis function (Figure~\ref{S5_PCBE}), the quantlets 
   (Figures~\ref{S2_QBE} and \ref{S5_QBE})
    have some level of interpretability in that the first two basis functions define the space of all Gaussian quantile functions ($\psi_1$, and $\psi_2$).  
    We see that in Figure~\ref{S5_QBE_0} orthogonal basis $\psi_k^{\perp}$ (black line) is wiggly up and down, compared to quantlets $\psi_k$ (blue line).
    Also, note that the next two quantlets for the GBM data seem to pick up on fundamental distributional characteristics like the kurtosis and skewness ($\psi_3$, and $\psi_4$).
For Gaussian data, only the first two basis functions will be needed, while comparing with dimensions $k=3, \ldots, K$ provides a measure of the degree of \textit{non-Gaussianity} in the distribution.

The summarizing Gaussianity score for the specific or reference group assessed by the relative sum is reported in Table \ref{S5_tabl_norm}. For instance, we see that the treatment group with the event time less than 12 months  can be summarized as the higher Gaussian score compared to its reference group. It was hard to explain the 
quantile trajectories of the male group or the group without mesenchymal status as a Gaussian quantile process because their scores explained by the normal quantile process   were not relatively high, which requires a nonparametric quantile process generated by mixed beta distributions to fully understand the entire quantile process.

The main results presented in the paper may depend on several modeling choices, containing the number of 
quantlets basis functions and determining the prior specification for $\nu$. Hence, we have conducted sensitivity
analysis under different modeling choices.
  Figures~\ref{S5_sc1} to \ref{S5_sc6} contain  
the posterior inference for functional coefficients for GBM data set:
for each covariate (6), the left panel includes posterior mean estimate, point and joint credible bands, GBPV in heading along with
SimBas less then $.05$ (orange line), 
 and the right panel includes predicted densities for the two levels of the covariate along with the posterior probability scores
   for the moment different testings, where Figure \ref{S5_sc4} presents the
   naive quantile functional regression approach.
   We also see that it does not produce different the results for varying $\nu$ in Figures \ref{S5_sc5} and \ref{S5_sc6}.
   Tables~\ref{S5_bigtab1} and \ref{S5_bigtab2} show
  specific results to assess sensitivity for a wide range of possible values of $K$ as well as different values of $\nu$, where they
include  global Bayesian p-values, run times along with area of the joint $95\%$ confidence intervals.  
Figure \ref{S5_sc7} depicts the functional coefficients if one naively applies regular (population) quantile regression methods across various quantiles p, and demonstrates that this approach gives nonsensical approaches for our application. 
As MCMC Diagnostic,
Figure~\ref{S5_T1_diag} contains
 Geweke’s diagnostic histograms \cite{geweke1991evaluating} for four models.  Under the null hypothesis of convergence, we would expect a uniform distribution of p-values.  We do not see any enrichment of small p-values in these histograms,
suggesting the chain converged.  The diagnostics are given for 
(A) model 1 (K=194), (B) model 2 (K=27), (C) model 3 (K=7), and model 4 (K=2). 

We lastly conducted  a sensitivity analysis  for lasso 
to see how selection of more or fewer dictionary elements via larger or smaller lasso parameters effects the ultimate number of quantlets.
The three panels of Figure~\ref{Re_select} 
show the common basis as the results from the choices including the large penalty (A), current penalty (B), and small penalty (C) of the  lasso in GBM data. We see that the path of concordance value was different from each case and the more sparse selection resulted in the smaller possible basis choices, and vice versa for the reason that the possible basis choice is represented by the number of the points in each panel. However, by the current near-lossless criteria (horizontal line), we can reduce this variability to 15, 27, and 38 basis functions for each case. Also, from Figures 11, 21 and 22 in Supplementary material, we see that there are not dramatic changes on the final results.

    \section{{ \bf Investigation of monotonicity} }
    
    By definition, quantile functions are monotone non-decreasing, since any decreases in the quantile function would correspond to negative probability densities.  
     There are a number of nonparametric smoothing methods in existing literature that impose monotonicity constraints on the functions, including  integrated splines \cite{ramsay1988monotone} (I-splines) and  convex splines \cite{meyer2008inference} (C-splines), which are adaptions of B-spline basis functions that enforce monotonicity.  A natural thought would be to consider utilizing basis sets like these for the quantile regression framework that could strictly enforce the monotonicity constraints.  We considered this, but chose to use quantlets instead for several reasons.

As an illustration in Subsection (\ref{sim1}), we generated empirical quantile functions from eight different parametric distributions and then fit C-spline, I-spline, and quantlet models to these data.  The IMSE is orders of magnitude smaller for quantlets than  I-splines or C-splines.  
We consider the flexibility of the basis set to capture the features of the data and the quantile functional regression coefficients $\beta_a(p)$ to be crucially important to this framework, so even a basis that constrains strict monotonicity may not be preferable if it lacks sufficient flexibility.
 
Second, from a modeling standpoint, in the quantile functional regression framework, monotonicity would have to be enforced for any possible combination of covariates $x_{a}, a=1, \ldots, A$,  a cumbersome and impractical constraint to impose.  

Third, since the quantlets are constructed from empirical quantile functions that are by definition monotone non-decreasing, we have found that in practice, our quantile functional regression framework tends to lead to virtually monotone predicted quantile functions for the various combinations of covariates.  While we would be concerned about a model producing gross non-monotonicities, we are not especially worried about very small magnitude non-monotonicities in the predicted values.  

It may be possible to adapt our quantlet basis in some manner to enforce strict monotonicity, but we leave that effort for future work.

 To investigate the degree of monotonicity afforded by the model, we construct predicted quantile functions for a broad range of covariate values (exhaustively if possible), and compute the degree of $\epsilon$-monotonicity, defined to be $P^\mathcal{M}_\epsilon(X)=\int_0^1 I[\widehat{Q}(p|X)-\text{max}_{p'<p}\{\widehat{Q}(p|X)\}>\epsilon] dp$ for some $\epsilon$ considered negligibly small in the context of the scale of $Y$ in the current data set.  We have found in our simulations and real data analyses that $P^\mathcal{M}_\epsilon \approx 0~ \forall X$, so it appears that for practical purposes, there is not a strong monotonicity problem in the models we have fit.  If $P^\mathcal{M}_\epsilon(X)$ is large for a given model, then one should carefully assess the model fit before scientifically interpreting its results.

We reported the empirical rates of the $\epsilon$-monotonicity as $1-n^{-1}\sum_{i=1}^{n}P^\mathcal{M}_\epsilon(X_i)$ for our simulation and GBM data sets in Table~\ref{S5_Mono}, where $n$ is the number of the possible levels of the predicted covariates, $X_i$.
We first generated $30$ additional predictors $u_{ij}$ from the uniform distribution defined on $(0,1)$ and replaced $\delta_{ij}$ by $u_{ij}$ for $j=2,3,4$, and $i=1,\dots,10$  in the simulation  
 and generated $82$ predictors as possible combinations of the discrete variables
  at the age evaluated by minimum, Q1, Q2, Q3, mean, or maximum ages in the GBM data. 
Based on the ranges of the observed data sets which were given as $(-20, 20)$ and 
 $(0, 100)$ for the simulation and GBM data, respectively, 
we set values for $\epsilon$ as shown in Table \ref{S5_Mono}. 
We see that the fitted quantile functions  based on our approach
show the monotonicity with the small scale of $\epsilon$ compared to the range of the original data set.
Figure \ref{S5_T2model3} 
shows that the predicted quantile functions
with bands for each level of the covariates in 
GBM dataset. We also see that they 
have $\epsilon$-monotonicity in that their quantile functions have
valid shape as the quantile function (nondecreasing shape).

 \section{\bf{ Software for implementation}} 
   
We provide description of the overall procedure to fit the quantile functional model and obtain inferential results
for the simulation and real application. 
We upload 
{\bf QFM.zip} file includes all the plots, estimates, and other inference results to reproduce works in this article. Among all files, we recommend to use the quantlets file, which  
produces the optimal {\it quantlets} basis function as the output for the input data set at a given probability grid of values under the options (\enquote{irregular} or \enquote{regular}), 
one that computes empirical quantiles based on the length of each observation and another that does it based on
 the common length across all observations.
The quantlets function requires
the glmnmet function in {\it glmnet} R package \cite{friedman2010regularization}
to figure out the union set of dictionary,
gramSchmidt function in {\it pracma} R package \cite{borchers2015pracma}
to obtain  the orthonormal set for the common basis set, and 
 wst function in {\it wavethresh} R package \cite{nason2010wavelet}  to utilize the non-decimated wavelet shrinkage method 
 after the set of the {\it overcomplete} dictionary was generated in the way described in the Section 2. 

Once we obtain the {\it quantlets} basis function, we can deal with it as the basis function and develop it
to fit the functional regression model. 
There are several possible ways to estimate the unknown parameters and obtain the posterior samples to produce the further inferential results.   
One possible way is to 
fit the quantlet-space functional regression model as part of the
Bayesian functional mixed model (BayesFMM) packages that have been developed in recent years
\cite{morris2006wavelet,zhu2011robust,zhu2012robust,meyer2015bayesian,zhang2016functional,Zhu2017robustf,lee2016semiparametric}.
We also mention WFMM executable as well as the BayesFMM packages,
 which is freely available at 
$$\text{https://biostatistics.mdanderson.org/SoftwareDownload,}$$ where
it does not need to formulate the random effect structure in the quantile functional regression model. 
To employ this,
we need to create the input file WFMM-input.mat which includes the empirical {\it quantlets} coefficients and design matrix structure. Such a file  will pipe into the WFMM software to fit our model.
There is a key commend to run the WFMM software in DOS window  as the following:
$$\text{wfmm WFMM-input.mat WFMM-output.mat $>$ WFMM-log.log}$$
Remark that the input file should be placed in the same directory that the commands are executed.
WFMM-output.mat will be produced by the above commend and contain 
the posterior samples of the quantile processes, which will be used for the further inference in R or Matlab
environments.

All our codes in {\bf QFM.zip} file are just for independent functional linear regression, while the FMM code can handle other structure including levels of random effects to model interfunctional correlation and nonparametric function.

\clearpage

\begin{table}
\caption{ Results for the simulation 1:
Empirical mean integrated squared error, $n^{-1}\sum_{i=1}^n( Q_{(\cdot)}(p_i)-\widehat {Q}_{(\cdot)}(p_i) )^2$ and monotonicity, $1-P^\mathcal{M}_\epsilon$ were computed based on
each basis representation for distributions, where
$N$, $T$, $G$, $D$, $M_1$, $M_2$, $M_3$, and $M_4$ indicate
normal, $t_{(1)}$, gamma, 
dirichlet,
 mixtures of $SN(-3.06,3.67,0)$ and $SN(9.11,7.89, -4)$  with $0.5$ probability, 
$SN(-7.1,2.4,0)$ and $SN(-3.11,7.89, 4)$  with $0.3$ and $0.7$ probabilities,
 $N(-2.5,2.5)$, $N(4, 3)$ and $N(9.5, 2.1)$ 
and  $N(-2.5,1.5)$, $N(4, 3.56)$ and $N(9.5, 1.1)$  
  with $0.3$, $0.5$ and $0.2$ probabilities
 distributions, respectively.
      \label{Basis_table}}
\begin{center}
\resizebox{0.80\columnwidth}{!}{

\begin{tabular}{ c | c | c c c c c c c c} \hline
 $$   &$$ 
& $\text{{\bf N}}$ &  $\text{{\bf T}}$ & $ \text{{\bf G}}$ & $\text{{\bf D}}$      
& $\text{{\bf M}}_1$ &  $\text{{\bf M}}_2$ & $ \text{{\bf M}}_3$ & $\text{{\bf M}}_4$  
  \\ \hline \hline
&$\text{{\bf Quantlets}}$ &	0.024 	&	0.001 	&	0.007 	&	0.004 	&	0.039 	&	0.050 	&	0.020 	&	0.022  \\
$\text{IMSE}$
&$\text{{\bf B-spline}}$ &	0.032 	&	0.029 	&	0.135 	&	0.057 	&	0.108 	&	0.166 	&	0.074 	&	0.094  \\
&$\text{{\bf I-spline}}$ &	0.698 	&	0.413 	&	0.599 	&	0.052 	&	2.283 	&	2.760 	&	1.004 	&	1.013  \\
&$\text{{\bf C-spline}}$	&0.203 	&	0.140 	&	0.414 	&	0.460 	&	0.548 	&	0.692 	&	0.343 	&	1.660  \\
\hline	
&$\text{{\bf Quantlets}}$ &	1.000 	&	1.000 	&	1.000 	&	1.000 	&	1.000 	&	1.000 	&	1.000 	&	1.000  \\
$\text{Monotonicity}$
&$\text{{\bf B-spline}}$ &	0.979 	&	0.996 	&	0.888 	&	0.976 	&	0.992 	&	0.973 	&	1.000 	&	1.000  \\
&$\text{{\bf I-spline}}$&	1.000 	&	1.000 	&	1.000 	&	1.000 	&	1.000 	&	1.000 	&	1.000 	&	1.000  \\
&$\text{{\bf C-spline}}$&	1.000 	&	1.000 	&	1.000 	&	1.000 	&	1.000 	&	1.000 	&	1.000 	&	1.000  \\
	\hline	
		\hline	
\end{tabular}
}
\end{center}
\end{table}      

 \begin{table}
\caption{Conditional moment statistics and 95\% confidence interval in simulation:
(A) naive quantile regression approach,
(B) naive quantile functional regression approach,
(C) principal component method, (D) {\it quantlet} space without sparse regularization,  and
(E) {\it quantlet} space with sparse regularization.
      \label{S2_Stat}    }
\begin{center}
\resizebox{0.98\columnwidth}{!}{   
\begin{tabular}{c   | c c c  c  c} \hline
 $\text{ {\bf True} }$   
 & $\text{ {\bf A} }$ &  $\text{{\bf B}}$ & $ \text{{\bf C}}$    & $ \text{{\bf D}}$    
 & $ \text{{\bf E}}$    
    \\ \hline
$\mu_{1}=1$  &0.93	(0.86,0.99) 
  &	1.20	(1.18,1.21)	&	1.2	(0.86,1.53)	&	1.19	(0.83,1.54)	&	1.18	(0.85,1.52)  
	\\
$\mu_{2}=3$  &2.71	(2.63,2.79) 
&	2.92	(2.91,2.94)	&	2.93	(2.61,3.24)	&	2.93	(2.57,3.29)	&	2.94	(2.60,3.30) 
\\
 $\mu_{3}=1$    &0.66	(0.59,0.72)
 &	0.99	(0.98,1.00)	&	0.99	(0.66,1.32)	&	1.00	(0.64,1.35)	&	0.99	(0.64,1.34) 
 \\
$\mu_{4}=3$ &2.59	(2.52,2.65)    
&	2.90	(2.88,2.91)	&	2.89	(2.56,3.22)	&	2.90	(2.54,3.25)	&	2.89	(2.54,3.25) 
\\	\hline	
$\sigma_1=5$	&4.92	(4.84,4.99)
&	4.97	(4.96,4.98)	&	4.97	(4.84,5.11)	&	4.97	(4.89,5.05)	&	4.97	(4.9,5.05) 
	\\
$\sigma_2=5$ 	&4.96	(4.85,5.06)
&	4.96	(4.95,4.97)	&	4.96	(4.83,5.11)	&	4.96	(4.89,5.03)	&	4.96	(4.88,5.04)
	\\
$\sigma_3=6.5$    &6.36	(6.27,6.46) 	
&	6.43	(6.42,6.44)	&	6.43	(6.29,6.56)	&	6.43	(6.35,6.51)	&	6.43	(6.36,6.51)
	\\
$\sigma_4=5$	&5.05	(4.94,5.14)
&	4.93	(4.92,4.95)	&	4.94	(4.82,5.07)	&	4.94	(4.86,5.01)	&	4.93	(4.86,5.01)
	\\
	\hline								
$\xi_{1}=0.00$  &-0.06	(-0.19,0.06)
&	0.01	(-0.01,0.02)	&	0.00	(-0.17,0.18)	&	0.01	(-0.21,0.22)	&	0.01	(-0.19,0.21)  \\
$\xi_{2}=0.00$  	 &-0.10	(-0.23,0.02) 
&	0.00	(-0.01,0.02)	&	0.00	(-0.16,0.17)	&	0.00	(-0.21,0.23)	&	0.00	(-0.21,0.22) 	\\
$\xi_{3}=0.00$   &-0.07	(-0.17,0.02)
	&	0.00	(-0.01,0.02)	&	0.00	(-0.13,0.13)	&	0.00	(-0.16,0.17)	&	0.00	(-0.16,0.17) 	\\
$\xi_{4}=-0.78$ &-0.91	(-1.11,-0.73)
 	&	-0.74	(-0.76,-0.73)	&	-0.74	(-0.95,-0.56)	&	-0.74	(-0.97,-0.52)	&	-0.74	(-0.96,-0.52)	\\	
\hline
\end{tabular}
}
\end{center}
\end{table}

 \begin{table}
\caption{Normality score of estimates for conditional subgroup in simulation. 
      \label{S2_Norm} }
\begin{center}
\resizebox{0.70\columnwidth}{!}{    
\begin{tabular}{c | c   | c } \hline 
   $\text{{\bf Group}}$
   &$\text{{\bf Estimate}}$    &$\text{ {\bf Percentage} ($95\%$ CI) }$   
    \\ \hline
 $(\xi, \omega, \alpha)=(1.0, 5.0, 0.0)$  &${\hat \beta_1(p)}$  & $ 68.8 \% ~(49.4,	88.4)$ \\
 $(\xi, \omega, \alpha)=(3.0, 5.0, 0.0)$  &${\hat \beta_1(p)}+ {\hat\beta_2(p)}$  &  $89.5 \% ~(78.5,	97.5 )$ \\
 $(\xi, \omega, \alpha)=(1.0, 6.5, 0.0)$  &${ \hat \beta_1(p)}+ {\hat\beta_3(p)}$  &  $81.9 \% ~(66.4, 94.9)$\\
 $(\xi, \omega, \alpha)=(9.1, 7.9, -4.0)$  &${\hat \beta_1(p)}+ {\hat\beta_4(p)}$  &  $34.7 \% ~(30.1,  39.0)$\\
\hline
\end{tabular}
}
\end{center}
\end{table}      

\begin{table}
\caption{Normality score of estimates for conditional subgroup in GBM application. 
      \label{S5_tabl_norm} }
\begin{center}
\resizebox{0.80\columnwidth}{!}{   
\begin{tabular}{c |c  |       c| c | c} \hline
   $$    &   $$     &$\text{{\bf Quantlet}}$  
   &   $$  
    &$\text{{\bf Quantlet}}$  
    \\    \hline
 $\text{{\bf Group}}$   &$\text{{\bf Treatment}}$     &$\text{ {\bf Percent} ($95\%$ CI) }$  
   &   $\text{{\bf Reference}}$   
    &$\text{ {\bf Percent} ($95\%$ CI) }$  \\  \hline
 $\text{{\bf Sex}}$  & $\text{{\bf Male}}$     & $63.3\% ~(40.0, 85.2)$   
 &$\text{{\bf Female}}$      & $77.9\% ~(59.2, 92.8)$    
  \\
 $\text{{\bf Age}}$  &$\text{{\bf 84 years}}$     & $65.3\% ~(44.7, 84.0)$   
&$\text{{\bf 60 years}}$    & $73.2\% ~(55.0, 89.1)$    
 \\
 $\text{{\bf DDIT3}}$  &   $\text{{\bf Yes}}$     & $70.9 \% ~(44.9,	91.7)$  
   &$\text{{\bf None}}$   & $69.9 \% ~(51.5,	86.6)$  
   \\    
 $\text{{\bf EGFR}}$  &    $\text{{\bf Yes}}$     & $71.4\% ~(49.6,	90.0)$ 
    &$\text{{\bf None}}$    & $71.3\% ~(50.2,	89.0)$ 
     \\ 
 $\text{{\bf Mesenchymal}}$  &  $\text{{\bf Yes}}$     & $78.4 \% ~(59.0,	94.0)$  
  &$\text{{\bf None}}$     & $61.3 \% ~(39.7,	82.3)$ 
   \\ 
  $\text{{\bf Survival status}}$  &   $\text{{\bf $\le$ 12 months}}$     & $80.6 \% ~(59.1,	96.2)$ 
     &$\text{{\bf  $>$ 12 months}}$    & $63.8 \% ~(44.6,	81.7)$ \\    
 \hline
\end{tabular}
}
\end{center}
\end{table}      


\begin{table}
\caption{GBM Results: Bayesian global p-values for quantlet models with various sizes of basis set $K$. 
      \label{S5_bigtab1}}
\resizebox{1.0\columnwidth}{!}{  
\begin{tabular}{ c c c c  c |c c c c c c c c} \hline  $$
& $K$      &$\rho^{0}$   &$\bar\rho$  &$\nu_0$
& $\text{{\bf Sex}}$ &  $\text{{\bf Age}}$ & $ \text{{\bf DDIT3}}$ & $\text{{\bf EGFR}}$  
& $\text{{\bf Mesenchymal}}$  &${\bf Survival_{12}}$    & $\text{{\bf Run time}}$  
  \\ \hline \hline
$\text{ Model~1}$	&	546	&.998  &1.000 &.006	&0.031	&	0.284	&	0.020	&	0.646	&	0.173	&	0.102	&	9.32	min	\\
$\text{ Model~2}$	&	194	&.997  &1.000 &.006	&	0.023	&	0.197	&	0.029	&	0.648	&	0.127	&	0.088	&	2.42	min	\\
$\text{ Model~3}$	&	101	&.997  &1.000 &.006	&	0.018	&	0.196	&	0.051	&	0.629	&	0.159	&	0.080	&	1.30	min	\\
$\text{ Model~4}$	&	66	&.996  &.999 &.006	&	0.017	&	0.271	&	0.018	&	0.646	&	0.188	&	0.106	&	46.20	sec	\\
$\text{ Model~5}$	&	50	&.996  &.999 &.006	&	0.032	&	0.242	&	0.048	&	0.698	&	0.229	&	0.092	&	39.13	sec	\\
$\text{ Model~6}$	&	44	&.996  &.999 &.006	&	0.027	&	0.239	&	0.023	&	0.657	&	0.175	&	0.097	&	33.11	sec	\\
$\text{ Model~7}$	&	42	&.994  &.999 &.006	&	0.005	&	0.122	&	0.006	&	0.648	&	0.058	&	0.038	&	35.60	sec	\\
$\text{ Model~8}$	&	31	&.993  &.999 &.006	&	0.008	&	0.177	&	0.009	&	0.627	&	0.116	&	0.061	&	26.19	sec	\\
$\text{ Model~9}$	&	27	&.990  &.998 &.006	&	0.016	&	0.168	&	0.012	&	0.605	&	0.087	&	0.067	&	21.83	sec	\\
$\text{ Model~10}$	&	23	&.989  &.997 &.006	&	0.036	&	0.242	&	0.014	&	0.684	&	0.128	&	0.076	&	18.78	sec	\\
$\text{ Model~11}$	&	19	&.989  &.997 &.006	&	0.038	&	0.297	&	0.035	&	0.668	&	0.221	&	0.132	&	15.31	sec	\\
$\text{ Model~12}$	&	15	&.988  &.997 &.006	&	0.019	&	0.226	&	0.020	&	0.683	&	0.103	&	0.111	&	12.89	sec	\\
$\text{ Model~13}$	&	13	&.981  &.996 &.006	&	0.041	&	0.274	&	0.042	&	0.796	&	0.302	&	0.143	&	12.93	sec	\\
$\text{ Model~14}$	&	10	&.981  &.996 &.006	&	0.036	&	0.294	&	0.027	&	0.694	&	0.218	&	0.113	&	11.97	sec	\\
$\text{ Model~15}$	&	9	&.981  &.996 &.006	&	0.006	&	0.152	&	0.007	&	0.605	&	0.128	&	0.048	&	11.81	sec	\\
$\text{ Model~16}$	&	8	&.964  &.993 &.006	&	0.027	&	0.239	&	0.025	&	0.640	&	0.129	&	0.104	&	10.65	sec	\\
$\text{ Model~17}$	&	7	&.962  &.993 &.006	&	0.007	&	0.147	&	0.006	&	0.607	&	0.084	&	0.063	&	9.12	sec	\\
$\text{ Model~18}$	&	5	&.860  &.974 &.006	&	0.014	&	0.160	&	0.009	&	0.561	&	0.096	&	0.063	&	0.022	sec	\\
$\text{ Model~19}$	&	2	&.858  &.966 &.006	&	0.014	&	0.053	&	0.006	&	0.494	&	0.067	&	0.042	&	0.006	sec	\\
$\text{ Naive}$	&	1024 &	 & & 	&	1.000	&	1.000	&	1.000	&	1.000	&	1.000	&	1.000	&	2.710	min	\\
$\text{ Model~9}$	&	27	& & &.01 &	0.016	&	0.169	&	0.012	&	0.607	&	0.088	&	0.068	&	22.25	sec	\\
$\text{ Model~9}$	&	27	& & &.0001 &	0.015	&	0.161	&	0.010	&	0.601	&	0.083	&	0.061	&	19.60	sec	\\
	\hline	
		\hline	
\end{tabular}
}									
\end{table}      

\begin{table}
\caption{GBM Results: Area of  the joint 95\% confidence intervals for quantlet models with various sizes of basis set $K$. 
      \label{S5_bigtab2}}
\resizebox{1.0\columnwidth}{!}{  
\begin{tabular}{ c c   c  c c |c c c c c c c c} \hline  $$
& $K$    &$\rho_0$  &$\bar\rho$  &$\nu_0$    
 & $\text{{\bf Intercept}}$  
& $\text{{\bf Sex}}$ &  $\text{{\bf Age}}$ & $ \text{{\bf DDIT3}}$ & $\text{{\bf EGFR}}$  
& $\text{{\bf Mesenchymal}}$  & ${\bf Survival_{12}}$   
  \\ \hline \hline
$\text{ Model~1}$	&	546	&.998  &1.000 &.006	&	17.373 	&	15.028 	&	32.168 	&	25.436 	&	15.919 	&	15.511 	&	14.707 	\\
$\text{ Model~2}$	&	194	&.997  &1.000	&.006&	17.127 	&	14.549 	&	29.665 	&	25.073 	&	14.940 	&	14.448 	&	14.056 	\\
$\text{ Model~3}$	&	101	&.997  &1.000	&.006&	16.680 	&	14.397 	&	30.862 	&	32.653 	&	15.015 	&	15.288 	&	14.498 	\\
$\text{ Model~4}$	&	66	&.996  &.999 &.006	&	18.154 	&	15.729 	&	33.493 	&	25.821 	&	15.295 	&	15.391 	&	15.665 	\\
$\text{ Model~5}$	&	50	&.996  &.999	&.006&	18.184 	&	15.772 	&	32.420 	&	30.657 	&	15.700 	&	15.801 	&	15.117 	\\
$\text{ Model~6}$	&	44	&.996  &.999 &.006	&	17.183 	&	15.654 	&	32.230 	&	26.074 	&	14.955 	&	15.221 	&	15.048 	\\
$\text{ Model~7}$	&	42	&.994  &.999 &.006	&	14.621 	&	12.503 	&	26.807 	&	22.012 	&	13.023 	&	12.269 	&	12.860 	\\
$\text{ Model~8}$	&	31	&.993  &.999	&.006&	15.948 	&	13.392 	&	29.069 	&	23.708 	&	13.981 	&	13.699 	&	13.566 	\\
$\text{ Model~9}$	&	27	&.990  &.998	&.006&	15.747 	&	13.035 	&	26.578 	&	22.607 	&	13.819 	&	13.807 	&	12.933 	\\
$\text{ Model~10}$	&	23	&.989  &.997 &.006	&	17.151 	&	15.533 	&	31.429 	&	24.489 	&	15.319 	&	14.935 	&	14.368 	\\
$\text{ Model~11}$	&	19	&.989  &.997 &.006	&	19.168 	&	17.205 	&	36.168 	&	28.363 	&	16.637 	&	16.385 	&	16.799 	\\
$\text{ Model~12}$	&	15	&.988  &.997	&.006&	17.176 	&	13.822 	&	30.512 	&	23.948 	&	14.784 	&	14.109 	&	14.297 	\\
$\text{ Model~13}$	&	13	&.981  &.996	&.006&	19.953 	&	16.687 	&	34.823 	&	28.207 	&	17.134 	&	17.192 	&	17.060 	\\
$\text{ Model~14}$	&	10	&.981  &.996	&.006&	19.087 	&	16.097 	&	33.559 	&	27.918 	&	16.082 	&	16.100 	&	16.433 	\\
$\text{ Model~15}$	&	9	&.981  &.996	&.006&	16.848 	&	13.388 	&	27.394 	&	22.433 	&	14.079 	&	14.100 	&	13.028 	\\
$\text{ Model~16}$	&	8	&.964  &.993	&.006&	16.922 	&	14.165 	&	30.335 	&	25.136 	&	14.903 	&	14.424 	&	14.194 	\\
$\text{ Model~17}$	&	7	&.962  &.993	&.006&	14.724 	&	12.679 	&	28.202 	&	22.669 	&	13.407 	&	12.965 	&	13.339 	\\
$\text{ Model~18}$	&	5	&.860  &.974	&.006&	15.679 	&	13.513 	&	27.982 	&	22.483 	&	13.584 	&	13.361 	&	13.288 	\\
$\text{ Model~19}$	&	2	&.858  &.966	&.006&	15.170 	&	12.845 	&	26.967 	&	21.397 	&	12.859 	&	12.525 	&	12.448 	\\
$\text{ Naive}$	&	1024&	& & &	26.013 	&	22.210 	&	47.030 	&	37.527 	&	22.680 	&	22.546 	&	21.837 	\\
$\text{ Model~9}$	&	& & 27&.01 	&	15.815 	&	13.093 	&	26.683 	&	22.705 	&	13.881 	&	13.865 	&	12.989 	\\
$\text{ Model~9}$	&	& &27&.0001	&	15.627 	&	12.934 	&	26.392 	&	22.435 	&	13.708 	&	13.702 	&	12.835 	\\
	\hline	
		\hline	
\end{tabular}
}									
\end{table}      

\begin{table}
\caption{ 
$\epsilon$-monotonicity of quantile functions for conditional subgroup.
      \label{S5_Mono} }
\begin{center}
\resizebox{0.7\columnwidth}{!}{    
        \begin{tabular}{c|c c |c c |c c}           
            \cline{1-7}                     
             & \multicolumn{2}{c| }{{\text{{\bf Simulation 1}}}}    
            & \multicolumn{2}{c| }{{\text{{\bf Simulation} 2}}}        
              & \multicolumn{2}{c }{{\text{{\bf GBM data}}}}         
                  \\ 
            \hline 
            \multicolumn{1}{c|}{}
              & $\epsilon=0.001$  &  $\epsilon=0.01$                   & $\epsilon=0.03$  &  $\epsilon=0.05$    
               & $\epsilon=0.1$ &  $\epsilon=0.5$  \\           \hline
 $\text{{\bf Naive}}$        &$25.8\%$     &$96.8\%$    &$0.0\%$     &$0.0\%$   &$0.0\%$   &$43.9\%$ \\
 $\text{{\bf PCA}}$          &$100.0\%$    &$100.0\%$    &$35.4\%$     &$83.8\%$   &$90.0\%$   &$93.9\%$ \\ 
 $\text{{\bf Quantlets}}$    &$100.0\%$    &$100.0\%$     &$100.0\%$     &$100.0\%$  &$93.9\%$   &$96.3\%$ \\            
                                                           \hline \hline   
        \end{tabular}
}
\end{center}
\end{table}      


   \begin{figure}[!htb]
   \caption{Quantlets basis functions  in GBM application.
  \label{S5_QBE}}
\centering
\includegraphics[height=5.3in,width=5.7in]{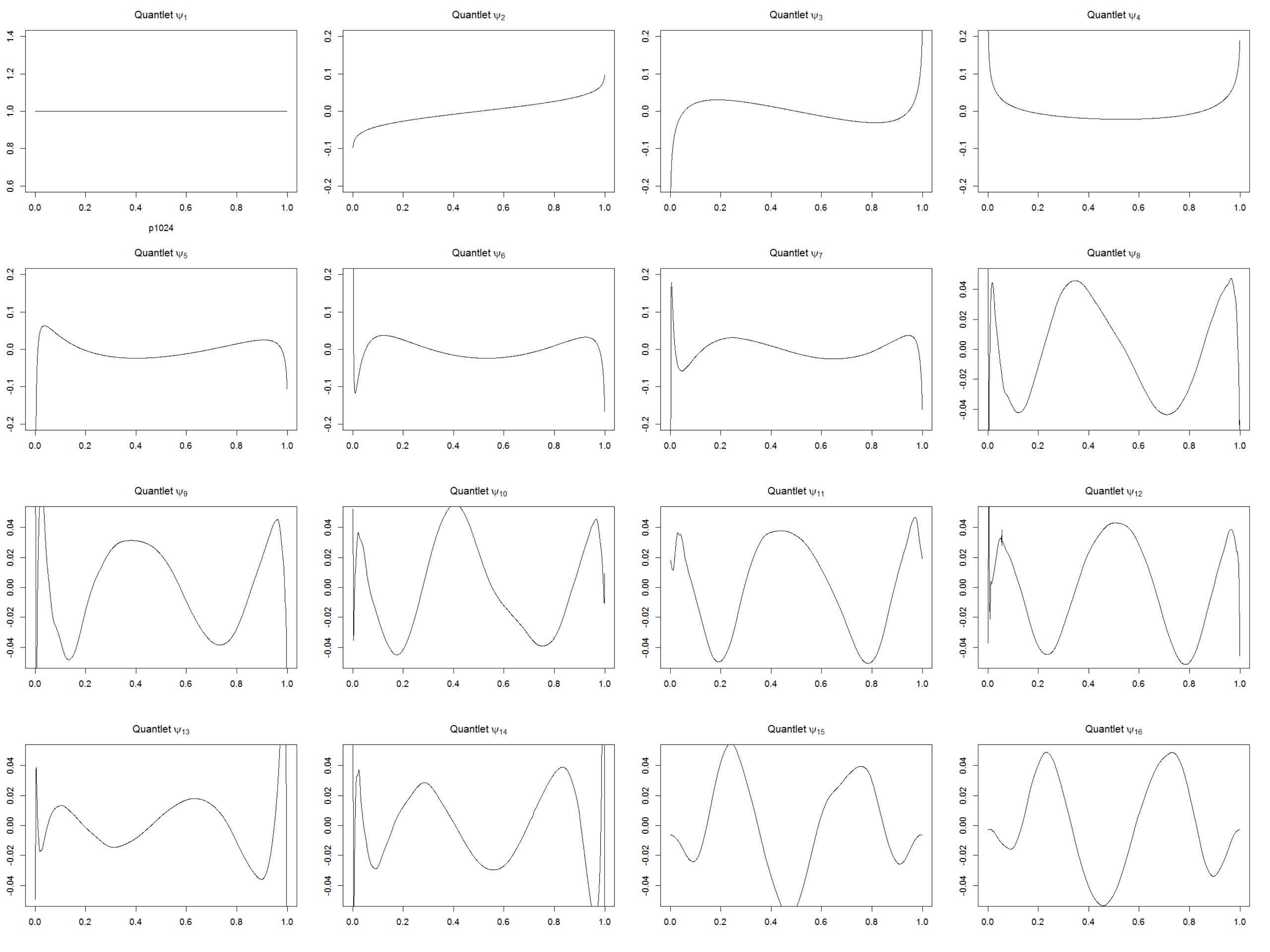}
\end{figure}         
         
  \begin{figure}[!htb]
  \caption{Simulation Results: Estimators and and $95\%$ joint credible intervals for $\beta_1(p)$ (black)
$\beta_2(p)$ (red), $\beta_3(p)$ (blue), $\beta_4(p)$ (green), 
their corresponding true coefficients (brown, orange, skyblue, and darkgreen, respectively),
and fitted values by {\it quantlets} (gray)   
are derived from the
 (A) naive quantile regression approach,
(B) naive quantile functional regression approach,
(C) principal component method, (D) {\it quantlet} space without sparse regularization, 
(E) {\it quantlet} space with sparse regularization, and (F) Gaussian {\it quantlet} space approach.
  \label{S2_Figure_2}}
\centering
\includegraphics[height=5.1 in,width= 5.2 in]{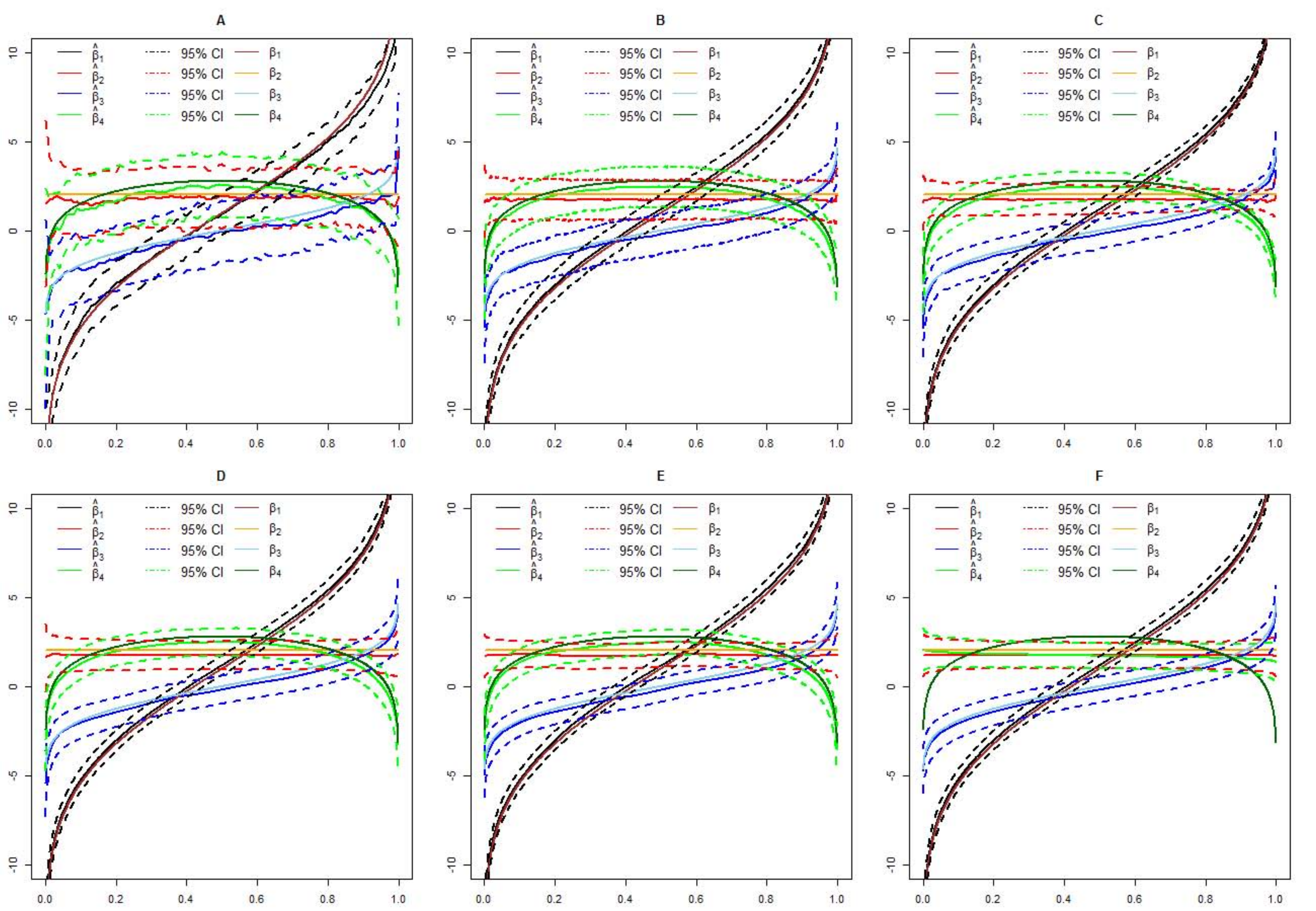}
\vspace{0.5cm} 
\end{figure}

 \begin{figure}[!htb]
 \caption{Intraquantile correlation estimated empirically (A) and assuming independence in the quantlet space (B)
in GBM application.
  \label{S5_ECOV}}
\centering
\includegraphics[height=2.3in,width=5.7in]{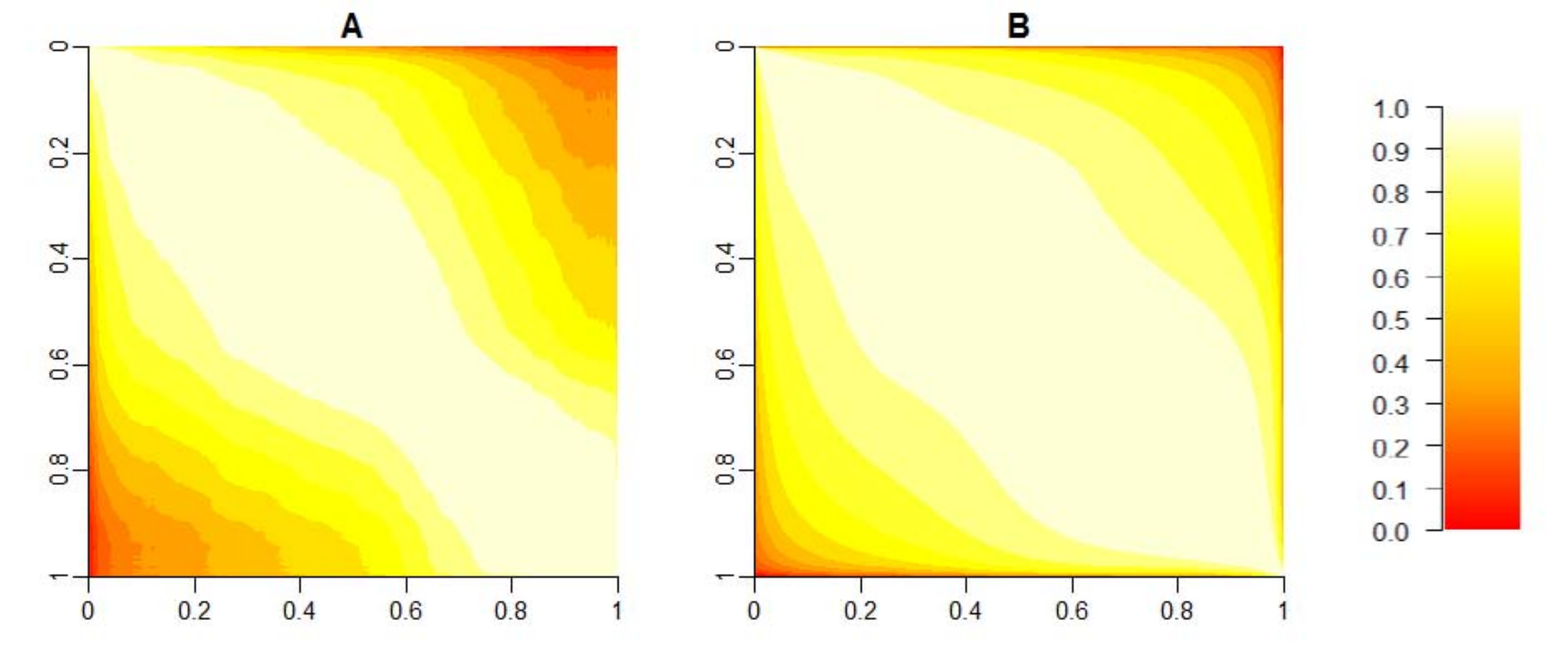}
\end{figure}

 \begin{figure}[!htb]
 \caption{Simulation Results: SimBaS for $\beta_3(p)$ (blue) and  $\beta_4(p)$ (green) at all $p \in \mathcal{P}$
are derived from the (B) quantile functional approach, (C) principal components method,
(D) {\it quantlet} space without sparse regularization,
and (E) {\it quantlet} space with sparse regularization,
 where vertical line (black) is significant level (0.05).
  \label{S2_Figure_3}}
\centering
\includegraphics[height=2.1in,width=5.2in]{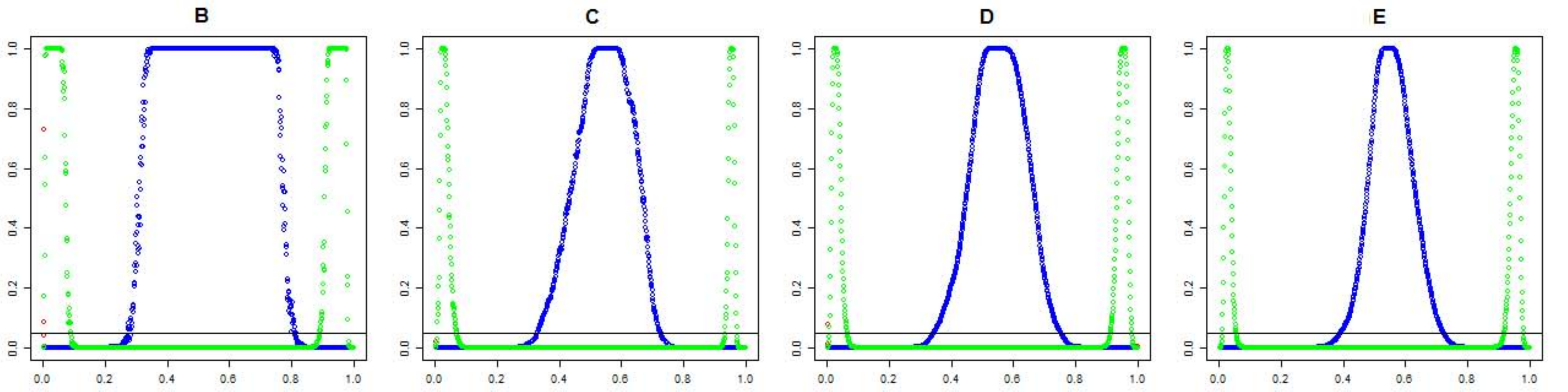}
\end{figure}

\clearpage

 \begin{figure}[!htb]
  \caption{Posterior inference for the model ($K=27$ and $\nu_0=.006$) in GBM application:
for each covariate (6), the left panel includes posterior mean estimate, point and joint credible bands, GBPV in heading along with
SimBas less then $.05$ (orange line), 
 and the right panel includes predicted densities for the two levels of the covariate along with the posterior probability scores
   for the moment different testings.
  \label{S5_sc1}}
\centering
\includegraphics[height=5.3in,width=5.7in]{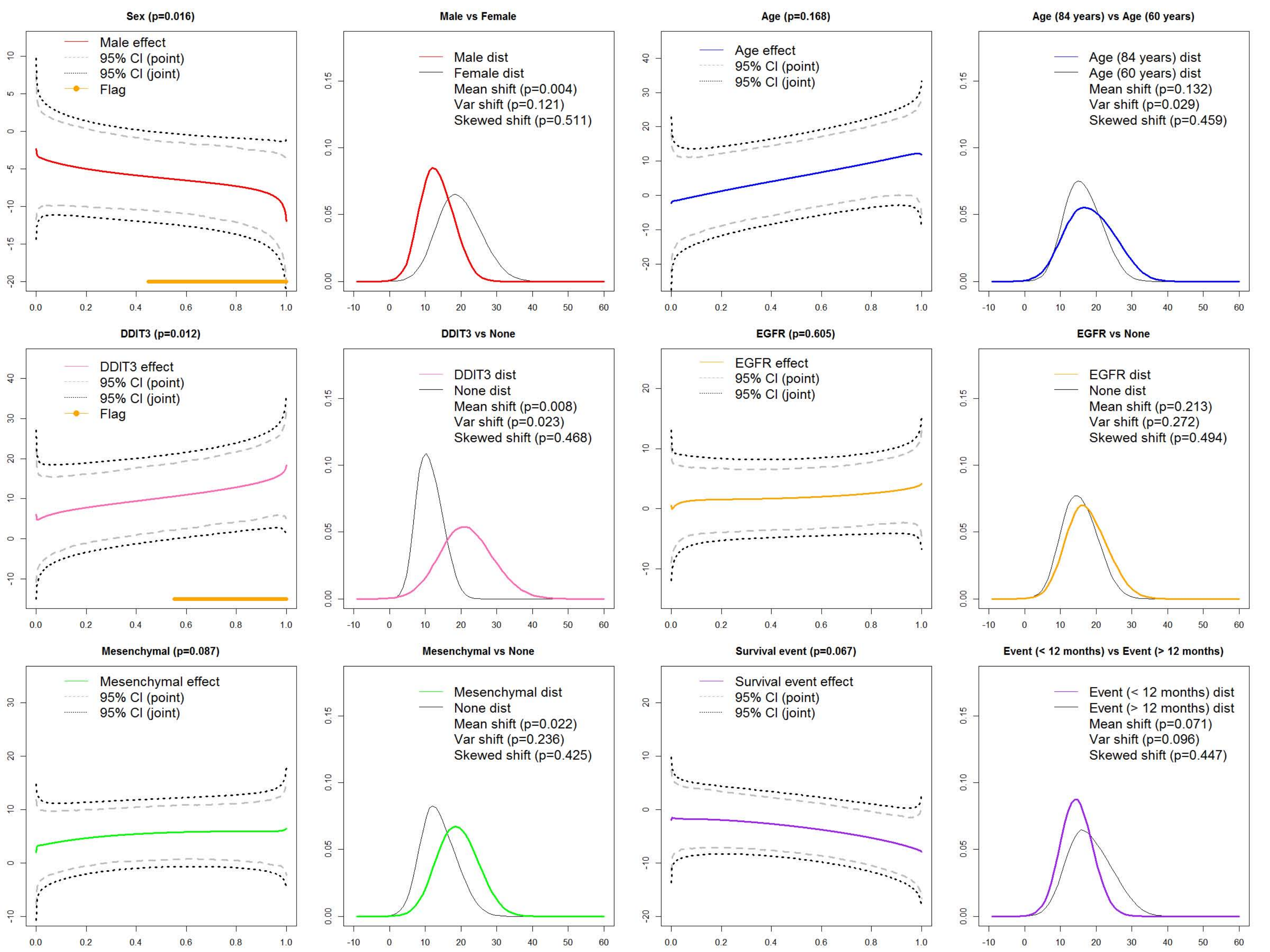}
\end{figure}

\clearpage
 \begin{figure}[!htb]
 \caption{Posterior inference for the model ($K=194$ and $\nu_0=.006$) in GBM application.
  \label{S5_sc2}}
\centering
\includegraphics[height=5.3in,width=5.7in]{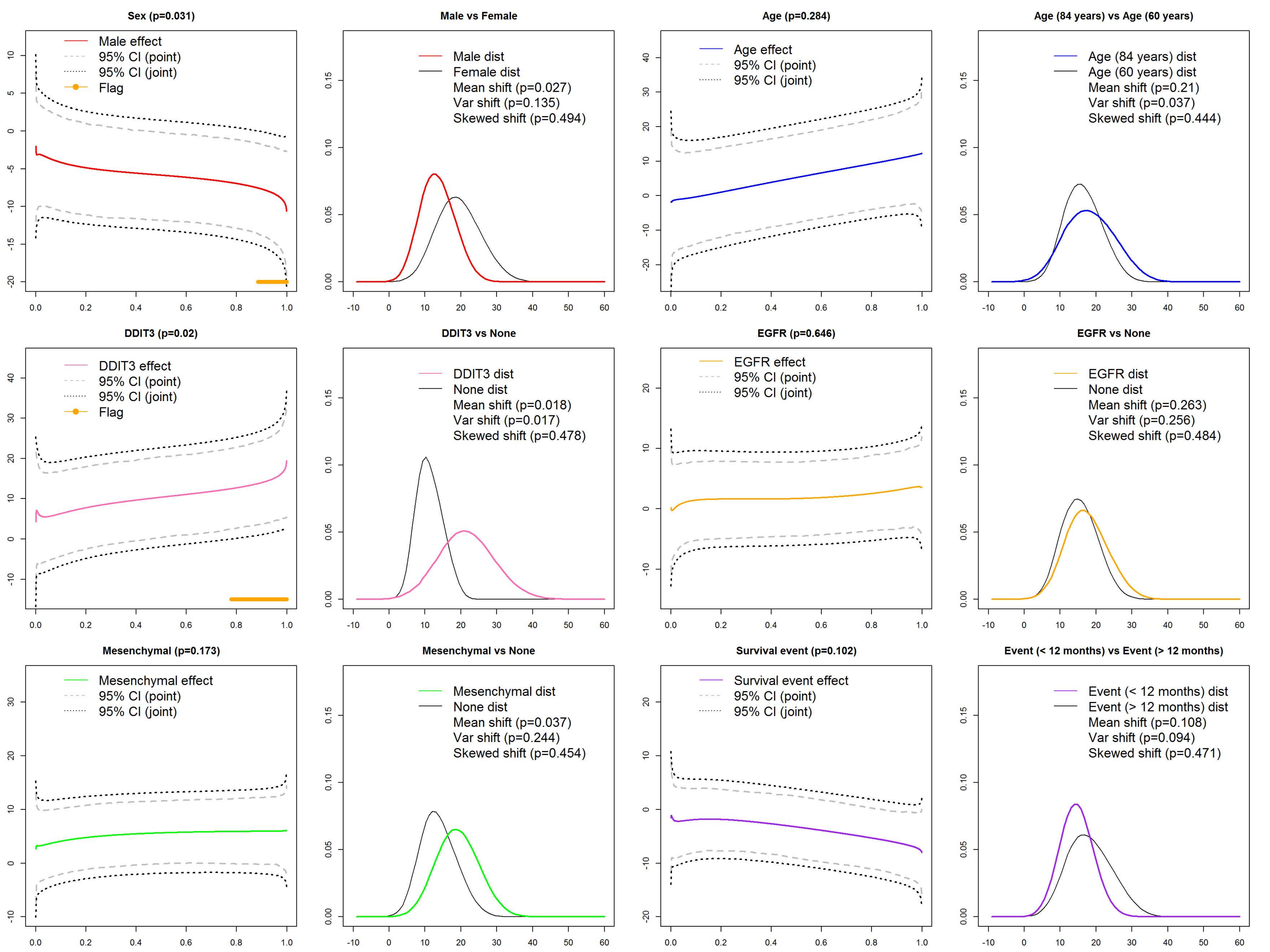}
\end{figure}

\clearpage
  \begin{figure}[!htb]
  \caption{Posterior inference for the model ($K=2$ and $\nu_0=.006$) in GBM application.
  \label{S5_sc3}}
\centering
\includegraphics[height=5.3in,width=5.7in]{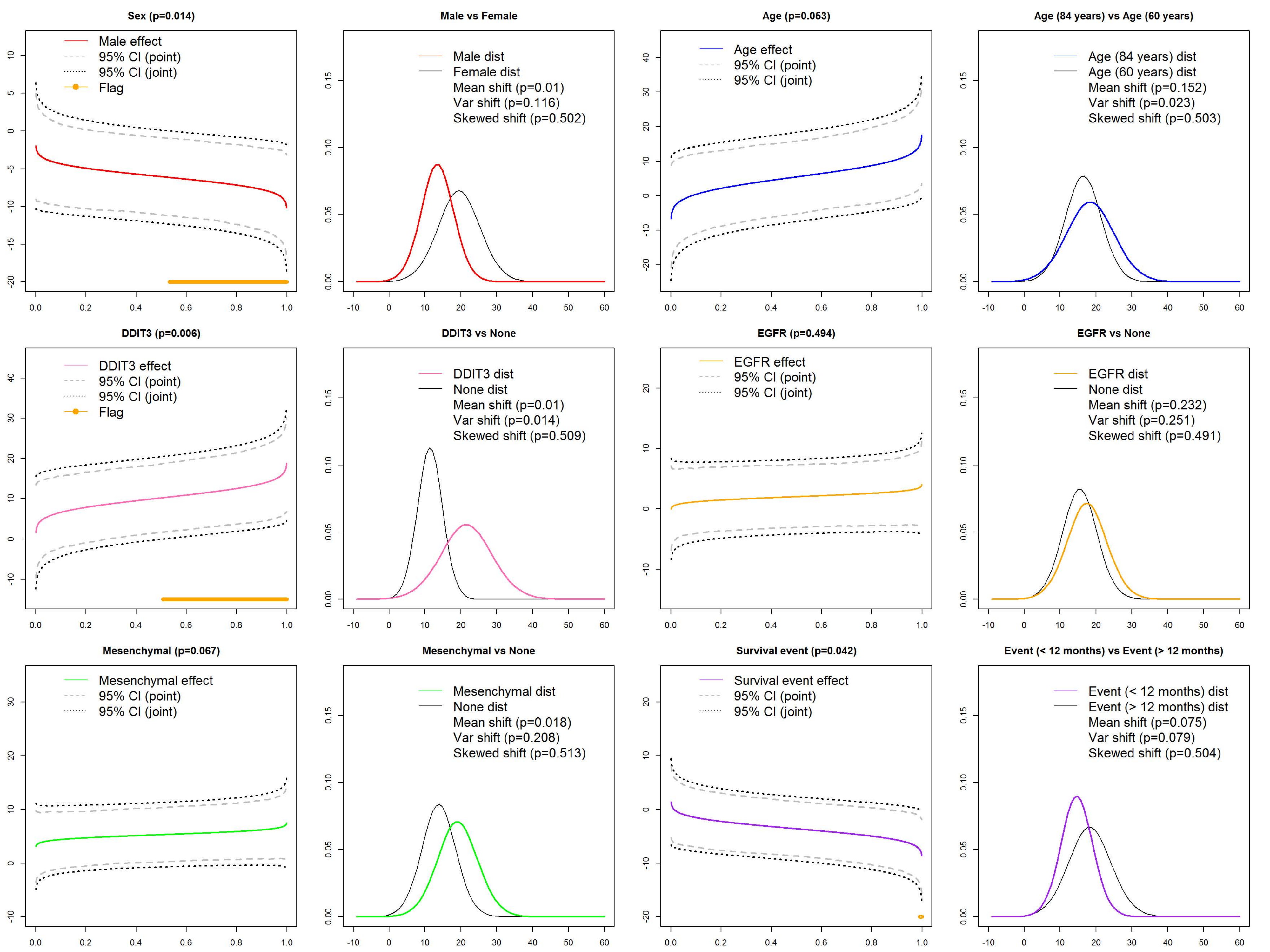}
\end{figure}        
       
         \clearpage  
   \begin{figure}[!htb]
   \caption{Inference from the naive quantile functional regression approach (separate functional regressions for each subject-specific quantile $p$) in GBM application.
  \label{S5_sc4}}
\centering
\includegraphics[height=5.3in,width=5.7in]{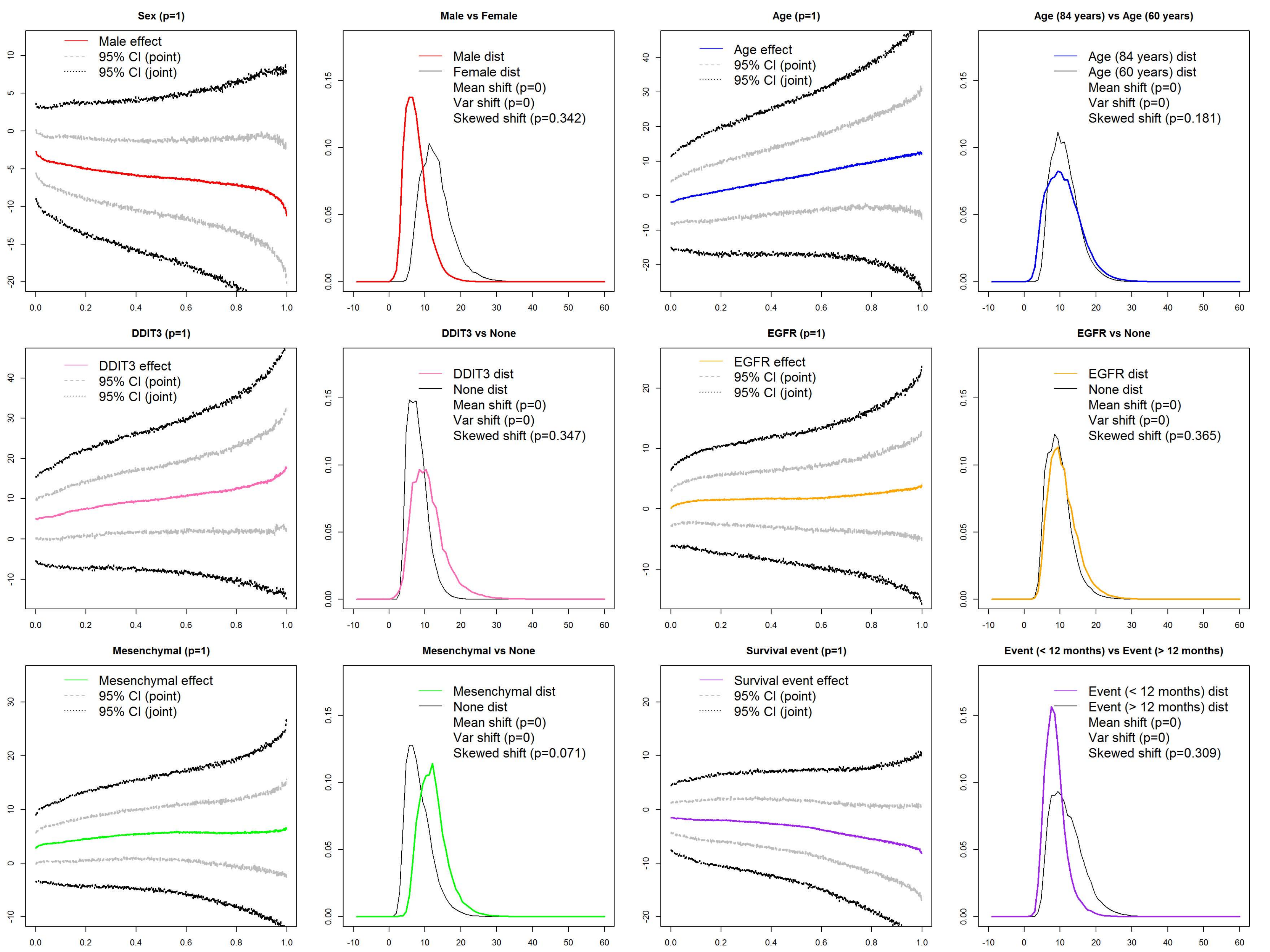}
\end{figure}

\clearpage
 \begin{figure}[!htb]
 \caption{Posterior inference for the model ($K=27$ and $\nu_0=.01$) for in GBM application.
  \label{S5_sc5}}
\centering
\includegraphics[height=5.3in,width=5.7in]{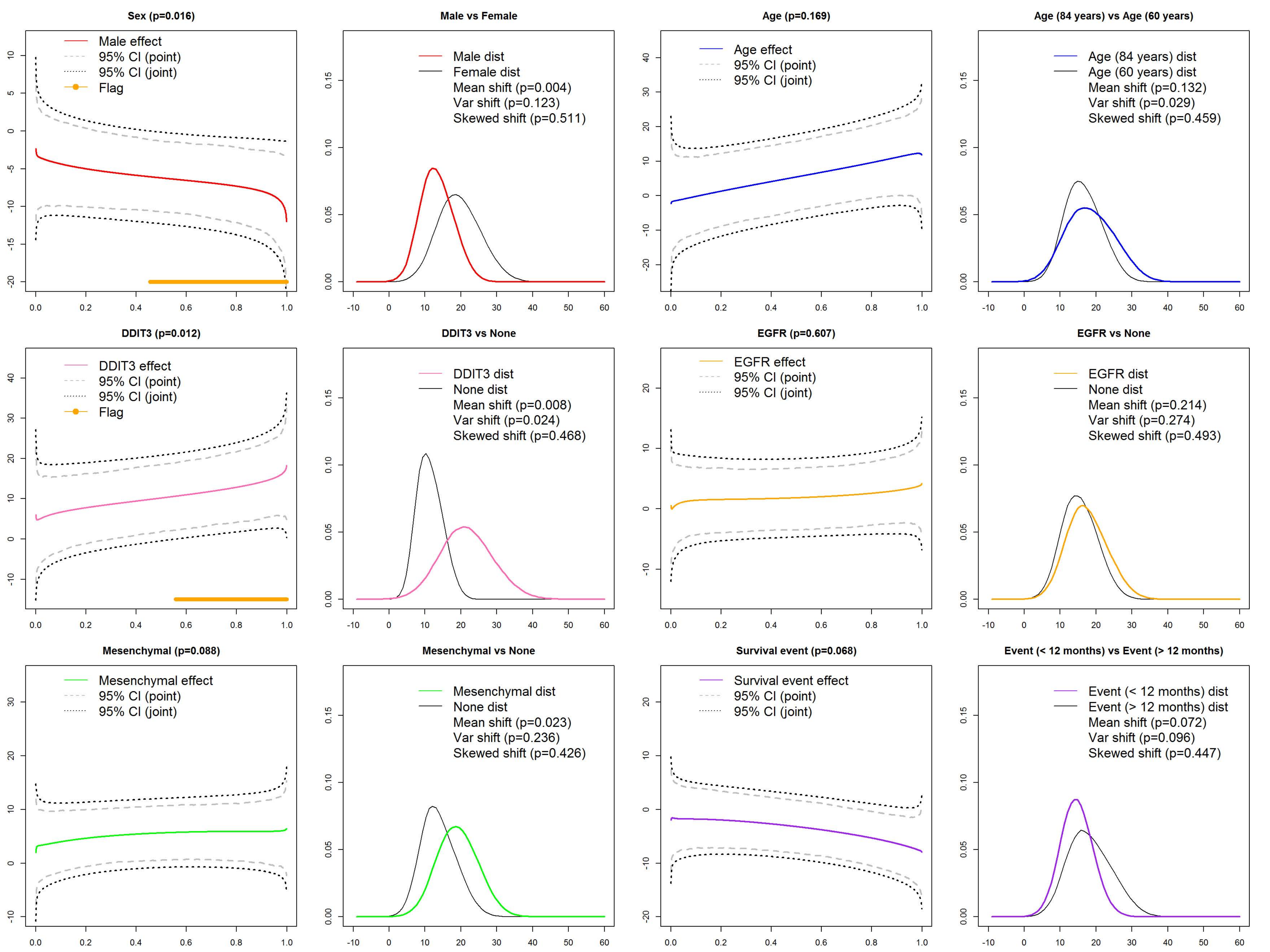}
\end{figure}

\clearpage
 \begin{figure}[!htb]
 \caption{Posterior inference for the model ($K=27$ and $\nu_0=.0001$) in GBM application.
  \label{S5_sc6}}
\centering
\includegraphics[height=5.3in,width=5.7in]{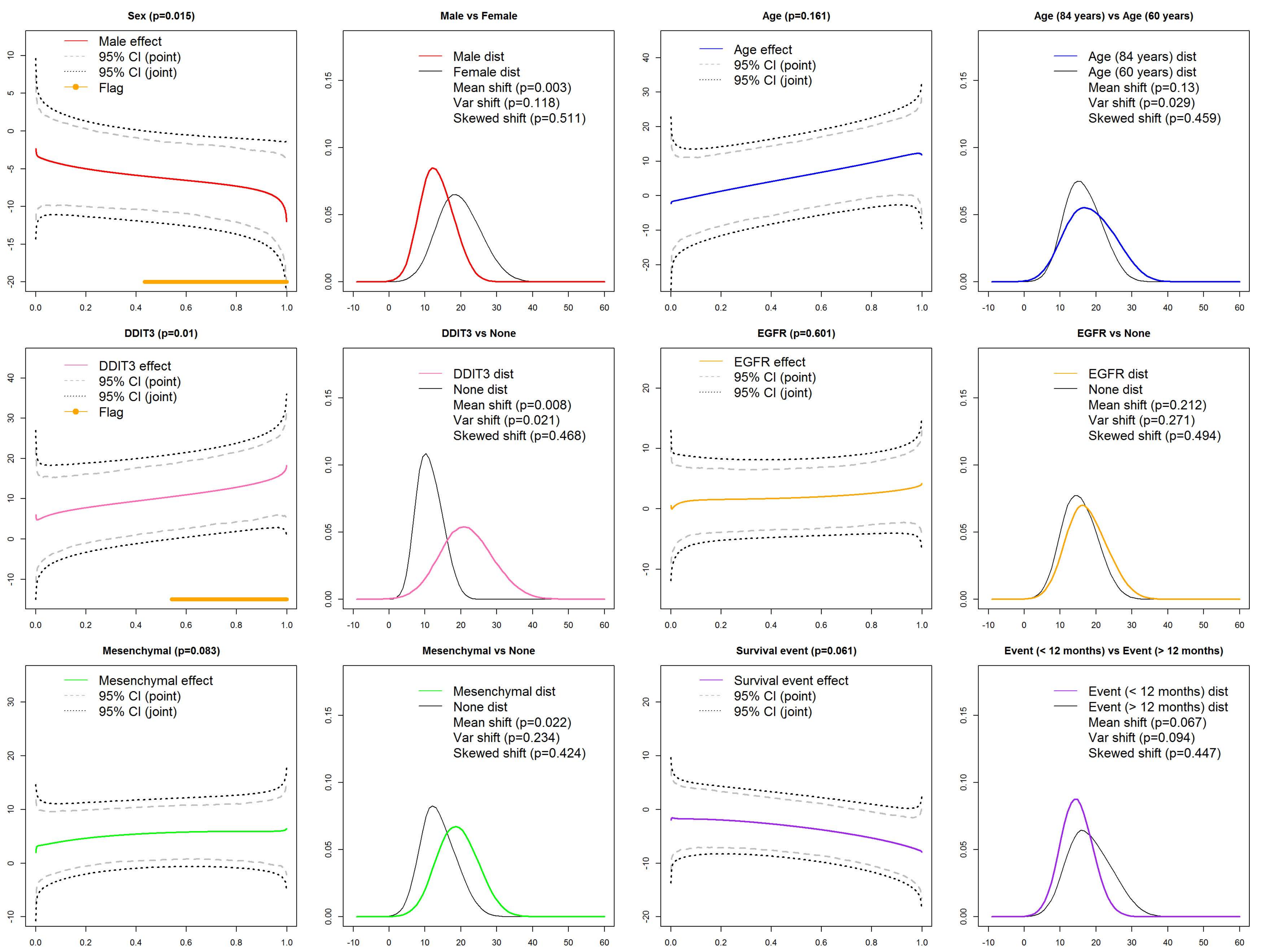}
\end{figure}

 \begin{figure}[!htb]
 \caption{Geweke’s diagnostic histograms  for four models in GBM application.  Under the null hypothesis of MCMC convergence, we would expect a uniform distribution in the p-values.  We see no enrichment of small p-values in these histogram, suggesting chain convergence.  Summaries are given for models
(A) model 1 (K=194), (B) model 2 (K=27), (C) model 3 (K=7), and model 4 (K=2). 
  \label{S5_T1_diag}}
\centering
\includegraphics[height=5.3in,width=5.7in]{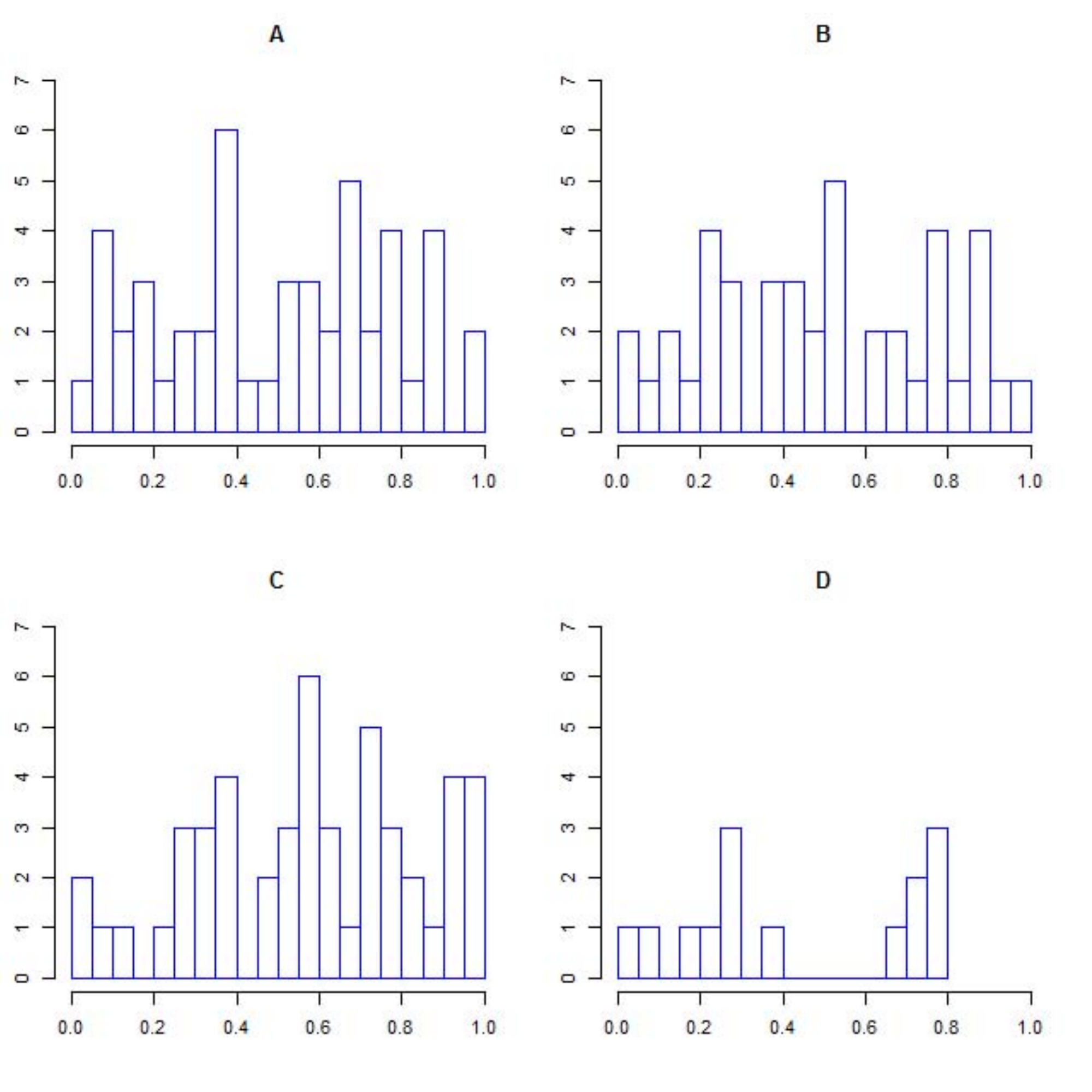}
\end{figure}

 \begin{figure}[!htb]
 \caption{Predicted quantile functions with the point and joint 95\% credible interval for the specific groups in GBM application: each row indicates the status of the sex and DDIT 3 
whereas each column reports at the summary value of the age (min, Q1, Q2, Q3, and max).
  \label{S5_T2model3}}
\centering
\includegraphics[height=5.3in,width=5.7in]{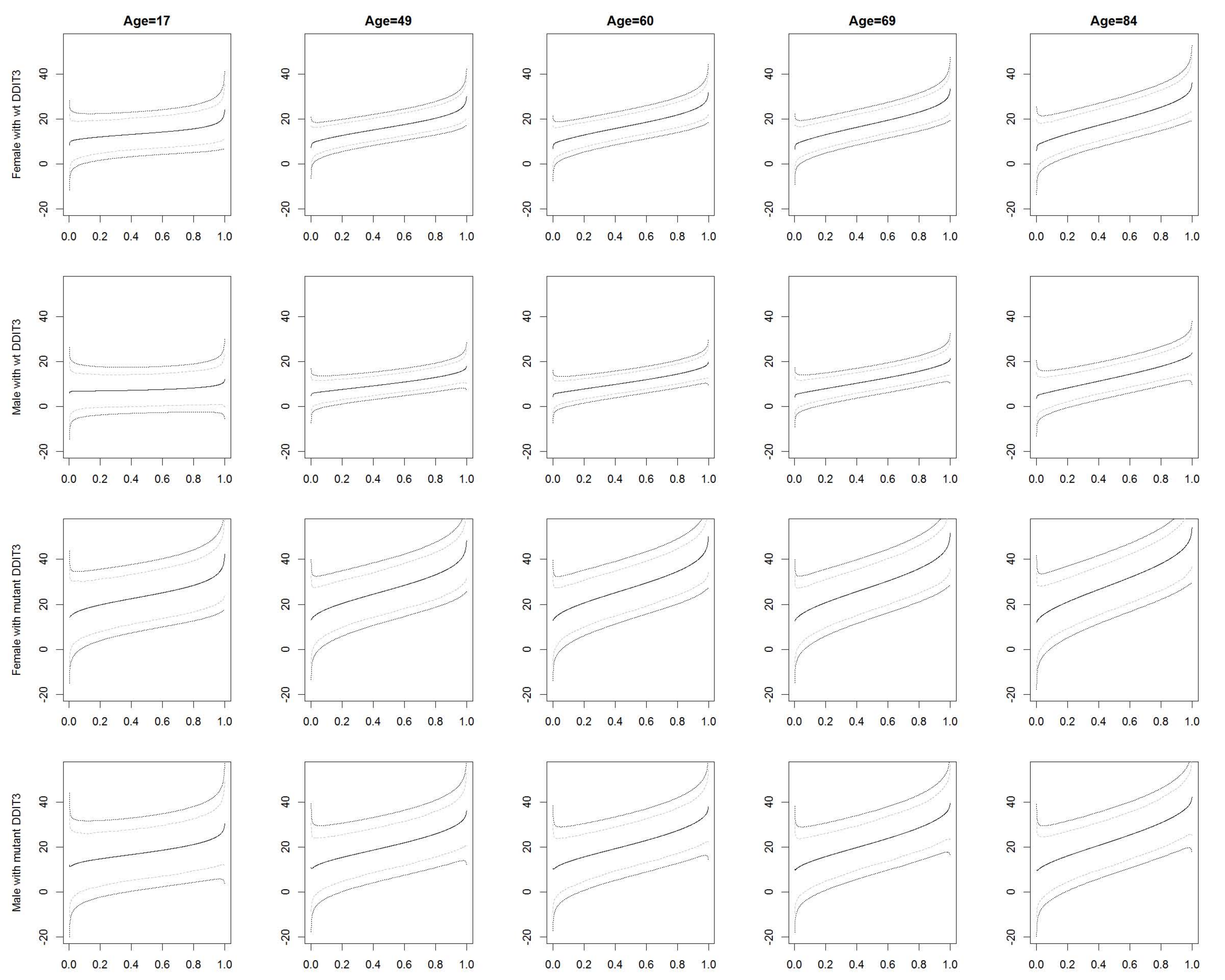}
\end{figure}

 \begin{figure}[!htb]
   \caption{Inference from the naive quantile regression method (separate classical quantile regressions for each $p$  in GBM application.
\label{S5_sc7}}
\centering
\includegraphics[height=5.3in,width=5.7in]{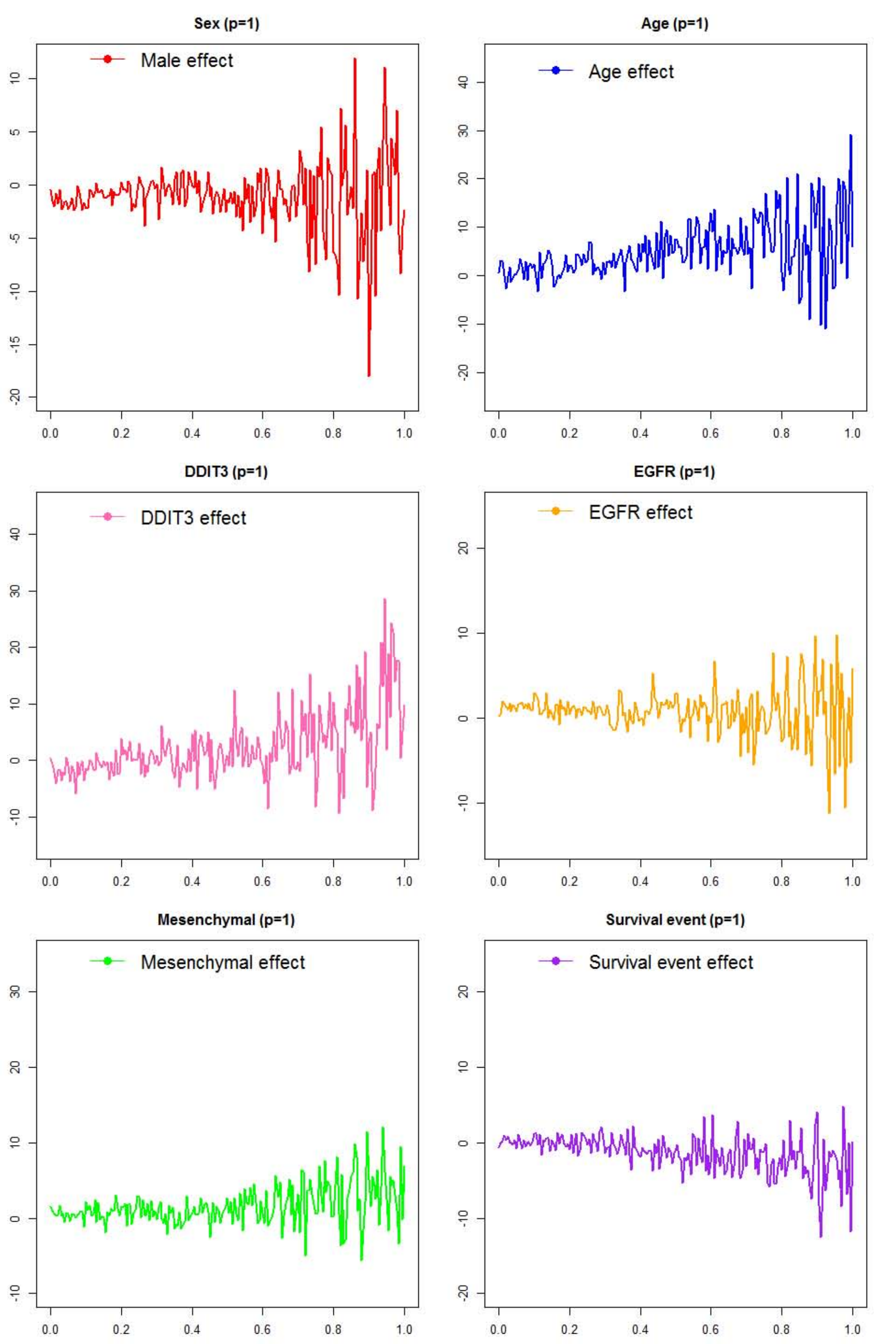}
\end{figure}

\clearpage
 \begin{figure}[!htb]
 \caption{Posterior inference for the model ($K=15$, $\lambda=10\lambda^{(c)}$  and $\nu_0=.006$) for in GBM application.
  \label{S5_finalsmall}}
\centering
\includegraphics[height=5.3in,width=5.7in]{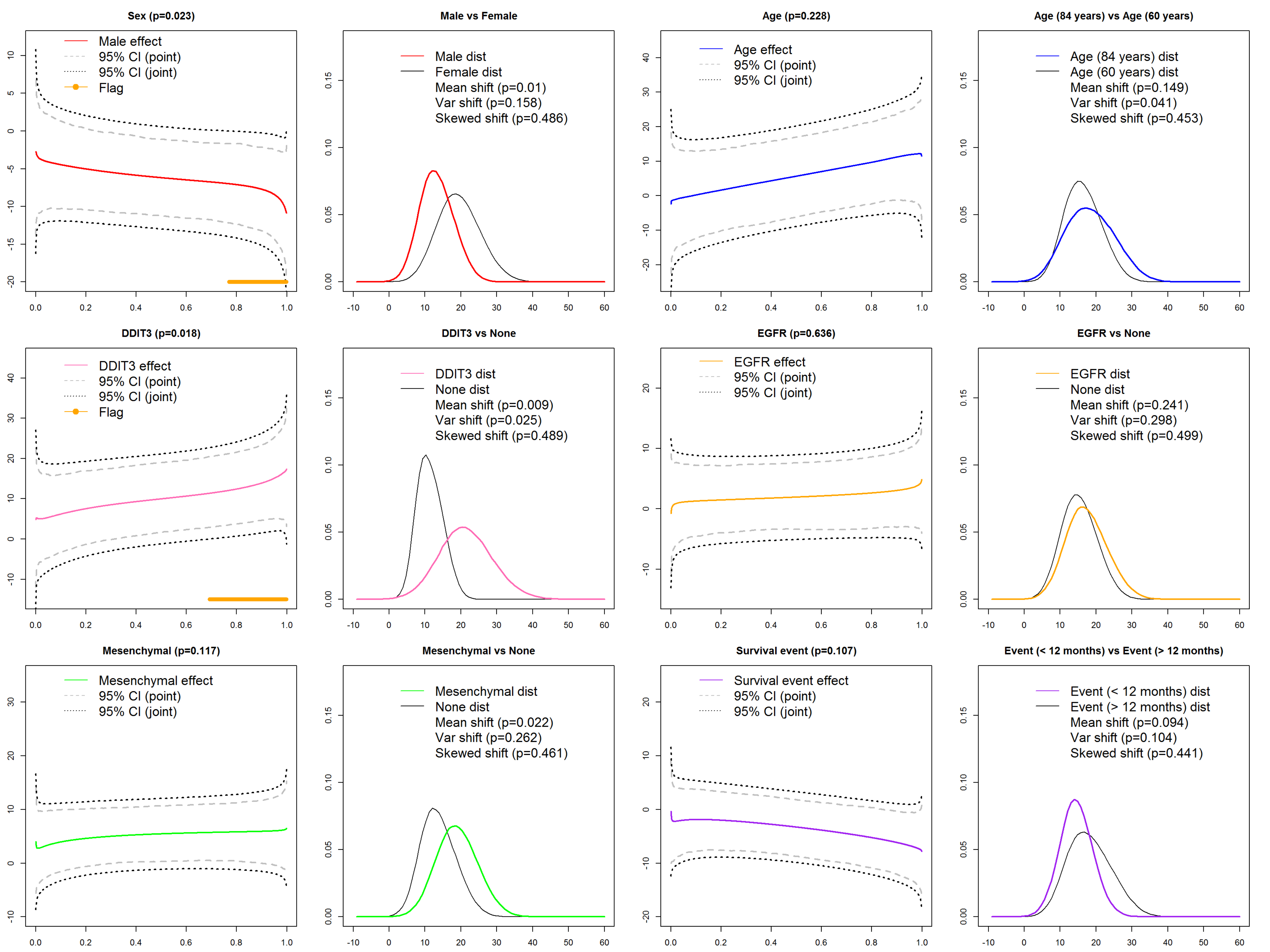}
\end{figure}

\clearpage
 \begin{figure}[!htb]
 \caption{Posterior inference for the model ($K=38$, $\lambda=.1\lambda^{(c)}$  and $\nu_0=.006$) for in GBM application.
  \label{S5_finalarg}}
\centering
\includegraphics[height=5.3in,width=5.7in]{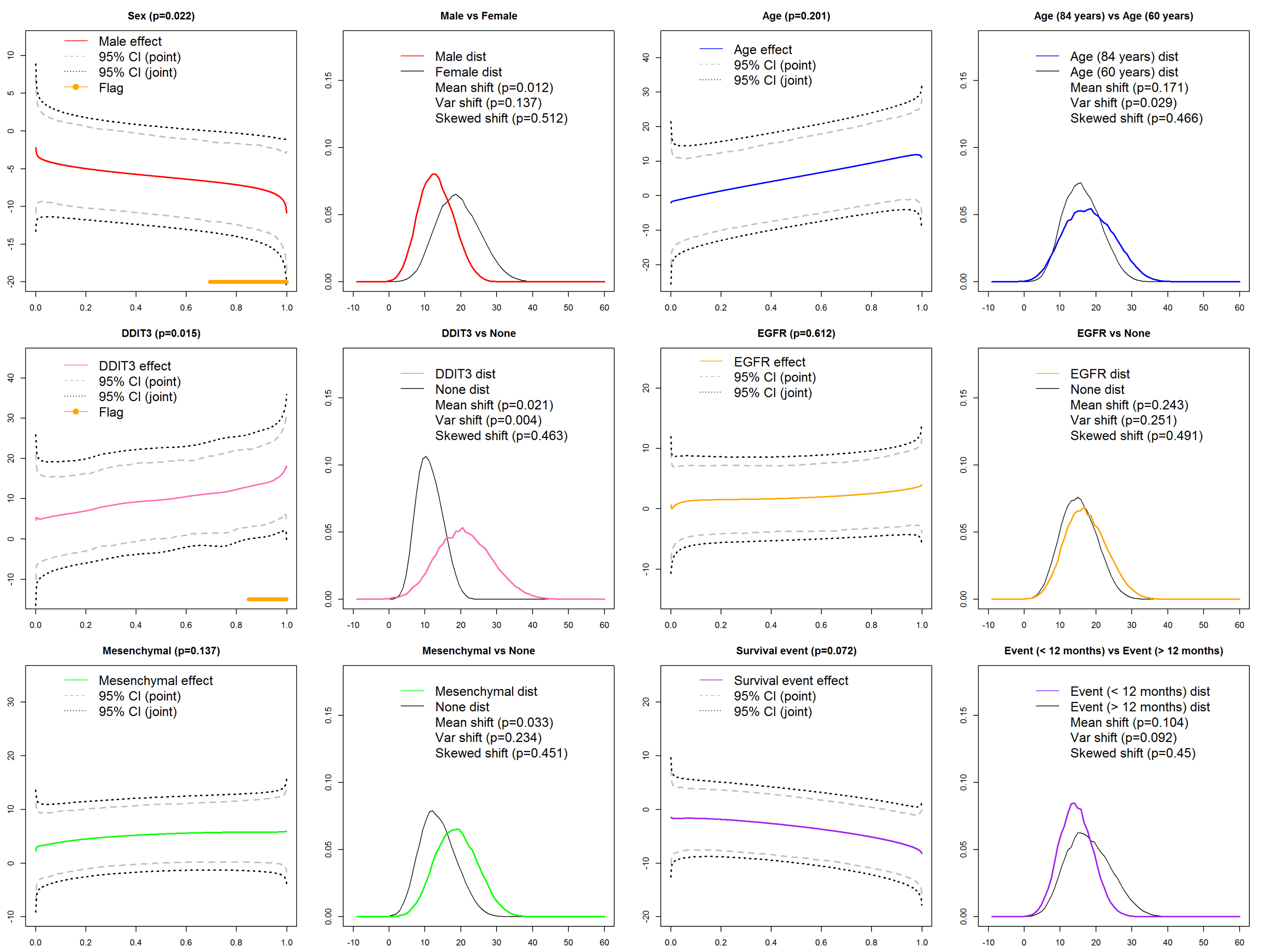}
\end{figure}

 \begin{figure}[!htb]
 \caption{Run time for the basis computation: $m_i$ indicates the number of the grid points. 
  \label{runtime}}
\centering
\includegraphics[height=3.0in,width=3.0 in]{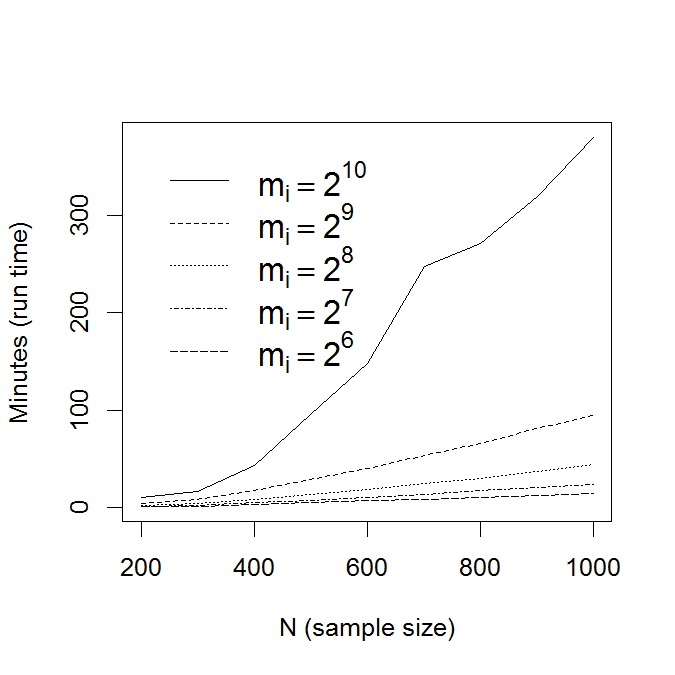}
\end{figure}

 \begin{figure}[!htb]
 \caption{ Near-lossless criterion for the different penalty ($\lambda$) of the lasso:  (A) large value of $\lambda$ , (B) current value of $\lambda$, and  (C) small value of $\lambda$, where  the large and small values are set to be 
$\lambda=10\lambda^{(c)}$ and $\lambda=0.1\lambda^{(c)}$ for the current penalty, $\lambda^{(c)}$.
  \label{Re_select}}
\centering
\includegraphics[height=1.9in,width=5.5in]{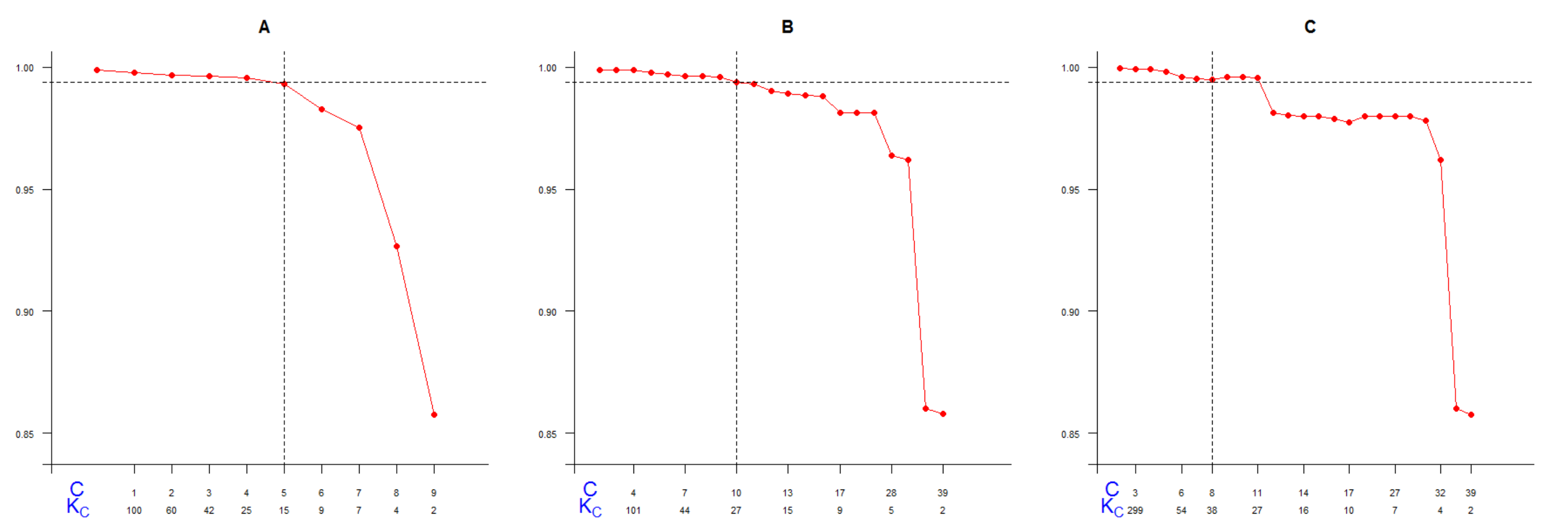}
\end{figure}

   \baselineskip 12pt             
\bibliographystyle{jasa_ECA}
\bibliography{JVHojin-2016-06-21}
\end{document}